\documentclass[aps,showpacs,prb,floatfix,twocolumn]{revtex4-1}
\usepackage{bm}
\usepackage{graphicx}
\usepackage{amsmath,amssymb}
\usepackage{float}
\usepackage{wrapfig}
\usepackage{layout}
\usepackage{color}
\usepackage[english]{babel}
\usepackage{pifont}
\newcommand{\cmark}{\ding{51}}
\newcommand{\xmark}{\ding{55}}
\usepackage{newtxtext,newtxmath}
\newcommand{\sss}{\mathrm{s}}
\newcommand{\ccc}{\mathrm{c}}
\newcommand{\ii}{\mathrm{i}}
\newcommand{\uu}{\mathrm{u}}
\newcommand{\cc}{\mathrm{c}}
\newcommand{\bb}{\mathrm{b}}

\newcommand{\dd}{\mathrm{d}}

\def\rightdef{=\mathrel{\mathop:}}

\usepackage{braket}
\begin{document}
\preprint{aps/}
\title{Instabilities in monoaxial chiral magnets under a tilted magnetic field}

\author{Yusuke Masaki}
\email{masaki@cmpt.phys.tohoku.ac.jp}

\selectlanguage{english}%

\affiliation{Department of Physics, The University of Tokyo, Bunkyo, Tokyo 113-0033,}
\affiliation{Research and Education Center for Natural Science, Keio University, Yokohama, Kanagawa 223-8521, }
\affiliation{Department of Physics, Tohoku University, Sendai, Miyagi 980-8578, Japan}
\date{\today}

\begin{abstract}
We thoroughly study the zero temperature properties of monoaxial chiral spin systems under a tilted magnetic field. 
The magnetic phase diagram includes two kinds of continuous phase transitions and one discontinuous phase transition. We clarify the properties of the phase transition in terms of the helical wave picture and the particle (soliton) picture and also in terms of the interaction between solitons. The interacting-soliton picture properly describes most of the discontinuous phase transition. In addition, we investigate several instabilities of the modulated structures such as an isolated soliton and the surface modulation,  since their instabilities should be important to the changes of the magnetic properties. For this purpose, we perform the analyses based on the energy landscape as well as the excitation spectrum; the former approach gives an intuitive interpretation. We clarify the mechanisms of the instabilities through these analyses and draw the stability lines of solitons in the magnetic phase diagram.
\end{abstract}
\maketitle

\section{Introduction}\label{sect-intro}
Two complementary pictures, particle and wave pictures are important to understand ordered phases with periodic structures\cite{Rosch2016}. In the former picture, the state can be regarded as an assembly of {\it emergent particles} (EPs), each of which can be characterized by a winding number and thus protected by topological stability. Examples are a vortex lattice state in type-II superconductors\cite{Abrikosov198811} and a skyrmion lattice (SkX) state\cite{BogdanovYablonskii1989,bogdanov1994thermodynamically,Muhlbauer2009,PhysRevLett.102.197202,NagaosaTokura2013} and a chiral soliton lattice (CSL) state\cite{Kishine2005,Kousaka2009,PhysRevLett.108.107202,Togawa2013,Togawa2016} in magnets and liquid crystals. The CSL state is described as an assembly of discommensurations proposed by McMillan\cite{McMillan1976}. On the other hand, they are sometimes described by a plane wave with a single or multiple wave number(s). Examples are a helical state of single-$q$ and a skyrmion lattice of triple-$q$ which appear in zero or low magnetic field applied to non-centrosymmetric magnets with the Dzyaloshinskii--Moriya interaction~(DMI), known as {\it chiral magnets}\cite{Dzyaloshinsky1958,Moriya1960}, or frustrated magnets\cite{Nagamiya1968,Okubo2012,Leonov2015,Hayami2016}. 

The two different pictures give two distinct continuous phase transitions (CPTs) between a nonuniform ordered phase and a disordered phase classified by de~Gennes\cite{DeGennes1975}: The instability-type transition is described in the wave picture, while the nucleation-type transition in the particle picture. 
They are summarized as follows: 
In the instability-type CPT, a mode with a {\it finite} wave vector drives the phase transition, i.e., the period of modulation remains {\it finite}.
The order parameter is a local quantity which can be infinitesimally small at the vicinity of the phase transition. 
The phase transition can be described by the Landau theory for such a small and local order parameter, and the fluctuations around the order parameter are important to the critical behavior, as is known as a textbook matter\cite{landau1980statphys}.
On the other hand, in the nucleation-type CPT, the order parameter is characterized by a topological number which counts the number of the EPs. 
At the phase boundary, the EP density becomes zero, i.e., the inter-EP-distance {\it diverges}. 
The particle picture describes the logarithmic criticality characteristic to the nucleation-type transition. 
Because of the topological stability of the EPs, the EP number in the ground state does not change by any instability with a negative eigenvalue of the Hessian against the fluctuations of the local quantity.
Examples of the nucleation-type CPTs other than those in the magnetic systems are the transition at the lower critical field of type-II superconductors\cite{DeGennes1966,Fetter1969}, and a transition between cholesteric and nematic liquid crystals\cite{deGennesLiquidCrystal}. These features are summarized in Table~\ref{table-CPT}.

\begin{table}[b]
\begin{tabular}{|c||c|c|}
\hline
CPT & instability-type & nucleation-type \\
\hline
condensate & wave of $\bm{q}$ mode & EP \\
\hline
period ($|\bm{q}|^{-1}$)& finite & $\to \infty$\\
\hline
order parameter & local and small & topological number\\
\hline\vspace{-0.3em}
against small fluctuations &  unstable  & topologically stable \\
(Hessian eigenvalue) &  (negative) & (non-negative) \\
\hline
hysteresis & \xmark & \cmark \\
\hline
\end{tabular}
\caption{Comparison between the two CPTs associated with periodic orders. The middle column summarizes the characters of the instability-type CPT, while the right one does those of the nucleation-type CPT. In the instability-type CPT, the Hessian has a negative eigenvalue for an instability mode.}
\label{table-CPT}
\end{table}

According to de~Gennes's textbook, the nucleation-type transition is accompanied by hysteresis even though it is a CPT\cite{DeGennes1975,deGennesLiquidCrystal}. 
One of the processes in which the number of EP changes is its nucleation or entrance at the surface, but a surface barrier for EP prevents this process. An example is the Bean--Livingston barrier in type-II superconductors\cite{Livingston1963,Bean1964}. The barrier vanishes at somewhat higher field than the lower critical field, up to which the Meissner state is maintained. The disappearance of the surface barrier may be related to the surface instability and important for changes of the topological charge. Actually, in the previous works\cite{Shinozaki2018,Masaki2018}, the authors attribute to the vanishing surface barrier the sharp jump observed in hysteresis loops of the magnetoresistance (MR) and magnetic torque (MT) measurements for micrometer-sized samples of Cr$_{1/3}$NbS$_{2}$ as explained later. 

Cr$_{1/3}$NbS$_{2}$ is a model material of monoaxial chiral magnets. There are intensive studies both theoretically and experimentally. In the absence of a magnetic field\cite{moriya1982evidence,MiyadaiKikuchiKondoSakkaAraiIshikawa1983}, it shows a helical state with its pitch of 48 nm along the $c$ axis, which we call the helical axis. 
The helical structure consisting of spins rotating in the $ab$ plane is robust because of the strong hard-axis anisotropy along the helical axis. When a magnetic field is applied, this system is a good playground for study of the two pictures mentioned above. 
Magnetic properties under the field have two regimes at zero temperature depending on the field direction\cite{moriya1982evidence,MiyadaiKikuchiKondoSakkaAraiIshikawa1983,KishineOvchinnikov2015,Togawa2016}.
A magnetic field perpendicular to the helical axis induces a chiral soliton lattice, where higher order harmonics are introduced: In the particle picture, each soliton can be regarded as the $2\pi$-domain wall of Bloch type, and its periodic array results in the soliton lattice. The phase transition from the soliton lattice to the uniform state is identified as the nucleation-type CPT\cite{dzyaloshinskii1964theory,dzyaloshinskii1965theory,dzyaloshinskii1965theory_2,Laliena2017}.

A field parallel to the helical axis induces a chiral conical state described in the wave picture, and the instability-type CPT to the uniform state is described on the basis of the Landau theory of a helical order parameter with uniform component parallel to the field\cite{MiyadaiKikuchiKondoSakkaAraiIshikawa1983,KishineOvchinnikov2015}. 
It is important to understand how these two distinct CPTs can be connected by changing the field angle\cite{Chapman2014,Laliena2016a}.  
Recently Laliena {\it et al.} have found a discontinuous phase transition (DPT) for the intermediate angles of two regimes and corresponding two multicritical points\cite{Laliena2016a}. The present author has used a linearization approach and clarified the origin of the DPT in Ref.~\onlinecite{Masaki2018}: The soliton interaction changes from the attractive one to the repulsive one at the multicritical point where  the nucleation type CPT and the DPT meet.

On the other hand, for micrometer-sized samples of Cr$_{1/3}$NbS$_{2}$, Togawa {\it et al.}\cite{Togawa2015} and Yonemura {\it et al.}\cite{Yonemura2017} recently found reproducible sharp jumps in large hysteresis loops by field sweep experiments of MR and MT measurements even for the nucleation-type CPT. Particularly in Ref.~\onlinecite{Yonemura2017}, they found the hystereses for various angles of tilted magnetic fields which include the nucleation-type CPT as well as DPT and  regarded them as evidences for chiral solitons. In my previous paper\cite{Masaki2018}, the authors discussed the necessary condition for the stability of the soliton and the surface instability mentioned above, in addition to the interaction properties of the solitons.  They corroborated that the theory of the surface barrier quantitatively explains the sharp jump in experiments. Therefore, it is also important to study instability processes when the nucleation-type CPT is concerned. As the sharp hysteresis was demonstrated by the surface instability, a large amount of changes can be caused by other instabilities of the topological structures. 

In this manuscript, we study the properties of the phase transitions and the instabilities of the monoaxial chiral magnet in the tilted magnetic field in terms of both particle and wave pictures. 
Since the Hessian remains positive for the nucleation-type CPT, 
not only the phase transitions but also the instabilities are important to the changes of physical properties (static properties such as the magnetization and transport properties, e.g., the MR, which is directly related to the presence and the number of chiral solitons).  
The remaining part of the present manuscript is structured as follows:
We briefly explain the model and the equation in Sect.~\ref{sect-model}. In Sect.~\ref{sect-pb-mcp}, we explain the linearization approach performed in Ref.~\onlinecite{Masaki2018} and study the properties of the multicritical points. We also show how the particle picture is effective by describing the DPT using the soliton interactions. In Sect.~\ref{sect-instability}, we completely investigate the instabilities of the modulated structures in this system. There are two other instabilities in addition to the surface instability: inflation instability and so-called $H_{0}$ line. The latter one is originally proposed in the skyrmion system\cite{Leonov2010}.  Finally we give the summary and some discussions in Sect.~\ref{sect-summary}. Some details of the calculations such as the details of expansion parameters and matrix elements are shown in Appendices~\ref{sect-landau-expansion} and \ref{sect-matrix-elements}. Appendix~\ref{sect-numerics} shows the details of our numerical conditions. In Appendices~\ref{sect-int-sol-surf}--\ref{sect-surface-twist-h0}, several details about the instabilities are described. We investigate, there, the interaction between an isolated soliton and the surface modulation (\ref{sect-int-sol-surf}), how the surface instability turns into the inflation instability (\ref{sect-turning-point}), the excitation spectrum when the surface modulation and an isolated soliton coexist (\ref{sect-inf-surface-soliton}), the continuum limit and the possibility of another mechanism for the $H_{0}$ line (\ref{sect-h0-Kdep-csg}), and the soliton picture of the surface modulation above the $H_{0}$ line (\ref{sect-surface-twist-h0}).  They are referred to also in the main text. Part of the present work has been published before in Ref.~\onlinecite{Masaki2018}.

The main results are summarized in Fig.~\ref{fig-hb-h0-phase}. In Fig.~\ref{fig-hb-h0-phase}(a), we provide the complete phase diagram including three different kinds of phase transitions, and the instability lines mentioned above are additionally shown. The properties of the phase transitions can be understood by using a linearization approach which separates the parameter region correspondingly into three regions as indicated by the red curve. 
It is also shown with the phase boundary in Fig.~\ref{fig-phase-diagram-critical}(a) for visibility.
In Fig.~\ref{fig-hb-h0-phase}(b), we show the schematic image of the spin structure in each region described within the linearization approach. Panels (i) and (ii) describe the tail structures of the soliton without and with oscillation, respectively. The former causes the repulsive interaction of solitons, while the latter does the attractive one. These interactions are attributed to the nucleation-type CPT and the DPT. Panel (iii) stands for the small helical wave structure rather than the soliton, and it is the order parameter of the Landau theory describing the instability-type CPT. The instability lines reduce the soliton region obtained on the basis of the linearization. 
The schematic images of instabilities are shown in Fig.~\ref{fig-hb-h0-phase}(c), where the solid curves are the initial spin profiles, while the dashed curves are the transient spin profiles of the instability processes. As discussed above, the surface instability (A) causes the soliton penetration from the surface. 
The inflation instability (B) corresponds to the winding process (the process for increase of the winding number), while the $H_{0}$ line (C) corresponds to the unwinding process (the process for decrease of the winding number), by pointing the spin to the $z$ direction (see the loops on the unit sphere and the light-blue dashed curves shown in panels (B) and (C) of Fig.~\ref{fig-hb-h0-phase}(c)). 
The parameter region surrounded by the inflation instability and the $H_{0}$ line gives the sufficient condition for the existence of the soliton in the bulk, which may be important to the manipulation of an isolated chiral soliton.

\section{Model and mean field equation}\label{sect-model}
In this paper, we only consider the zero temperature case of the monoaxial chiral magnets, where a spatial modulation of spins appears only in one direction. We take it as the $z$ axis. Since spins in each $xy$ plane are perfectly aligned, we consider the following energy functional of the spin chain along the $z$ direction for configuration $\vec{M}_{l} = \sum_{\mu = x,y,z}M_{l}^{\mu} \vec{e}^{\ \!\!\ \! \mu}$ with unit vectors in the spin space $\vec{e}^{\ \!\!\ \! \mu}$,\footnote{
We use the component representations of vectors for this basis as $\vec{M}_{l} = {}^{t}\! (M_{l}^{x},M_{l}^{y},M_{l}^{z})$. 
} and $xy$-layer index $l$:
\begin{align}
E[\{\vec{M}_{l}\}] 
&= -\sum_{l} \left[
J_{\parallel} \vec{M}_{l}\cdot\vec{M}_{l+1} + D \left(\vec{M}_{l}\times \vec{M}_{l+1}\right)^{z} \right.\nonumber \\ 
&\hspace{8em}\left.- \dfrac{K}{2} \left(M_{l}^{z}\right)^{2} +\vec{H}_{\mathrm{ex}}\cdot \vec{M}_{l}
\right]. \label{eq-energy-chain}
\end{align}
The total energy of the three dimensional system is given by $E N_{2\dd}$, while the energy density is given by $E/N_{z}$, where $N_{2d}$ and $N_{z}$ are, respectively, the number of spins in an $xy$ plane and on the chain.
We consider $|\vec{M}_{l}| =1$.
The summation index $l$ takes all of the sites on the chain under the appropriate boundary conditions discussed later and in Appendix.~\ref{sect-numerics}, and for the moment 
we define the lattice system of infinite length $l=-
\infty, \cdots,-1, 0, 1, 2, \cdots,\infty$. 
The first and second terms in Eq.~\eqref{eq-energy-chain} are the Heisenberg exchange interaction, and the DMI, respectively, on the nearest neighbor pairs along the helical axis. The third term is a single ion anisotropy energy, and when $K$ is positive, the helical axis is a hard axis. The fourth term describes the Zeeman energy due to the tilted magnetic field, denoted by $\vec{H}_{\mathrm{ex}}$, which has the $x$ and the $z$ component. 
The equation for a static configuration, $\vec{M}_{l}$, is given by the zero-torque condition. The effective field on site $l$, $\vec{H}_{l}^{\mathrm{eff}}$ is given by the variation of Eq.~\eqref{eq-energy-chain} with respect to $\vec{M}_{l}$, and the zero torque condition is equivalent to $\vec{M}_{l} \parallel \vec{H}_{l}^{\mathrm{eff}}$. At zero temperature, the modulus of the moment is always 1. Therefore, the equation can be summarized as follows:
\begin{align}
\vec{M}_{l} &= \hat{H}_{l}^{\mathrm{eff}} = \vec{H}_{l}^{\mathrm{eff}} / |\vec{H}_{l}^{\mathrm{eff}}|,\label{eq-moment}\\
\vec{H}_{l}^{\mathrm{eff}} & =\sum_{\mu = \pm1 }\left(J_{\parallel} \vec{M}_{l+\mu} + \mu D  \vec{M}_{l + \mu} \times \vec{e}^{\ \!\!\ \! z}\right) - KM_{l}^{z}\vec{e}^{\ \!\!\ \! z} + \vec{H}_{\mathrm{ex}}. \label{eq-field}
\end{align}
Equations~\eqref{eq-moment} and \eqref{eq-field} are the nonlinear coupled equations and determine all static spin profiles: isolated soliton state, soliton lattice state, conical state, and their mixtures. 

\section{Phase boundary and multicritical points}\label{sect-pb-mcp}
\begin{figure*}[t]
\begin{center}
\includegraphics[width = 0.95\hsize]{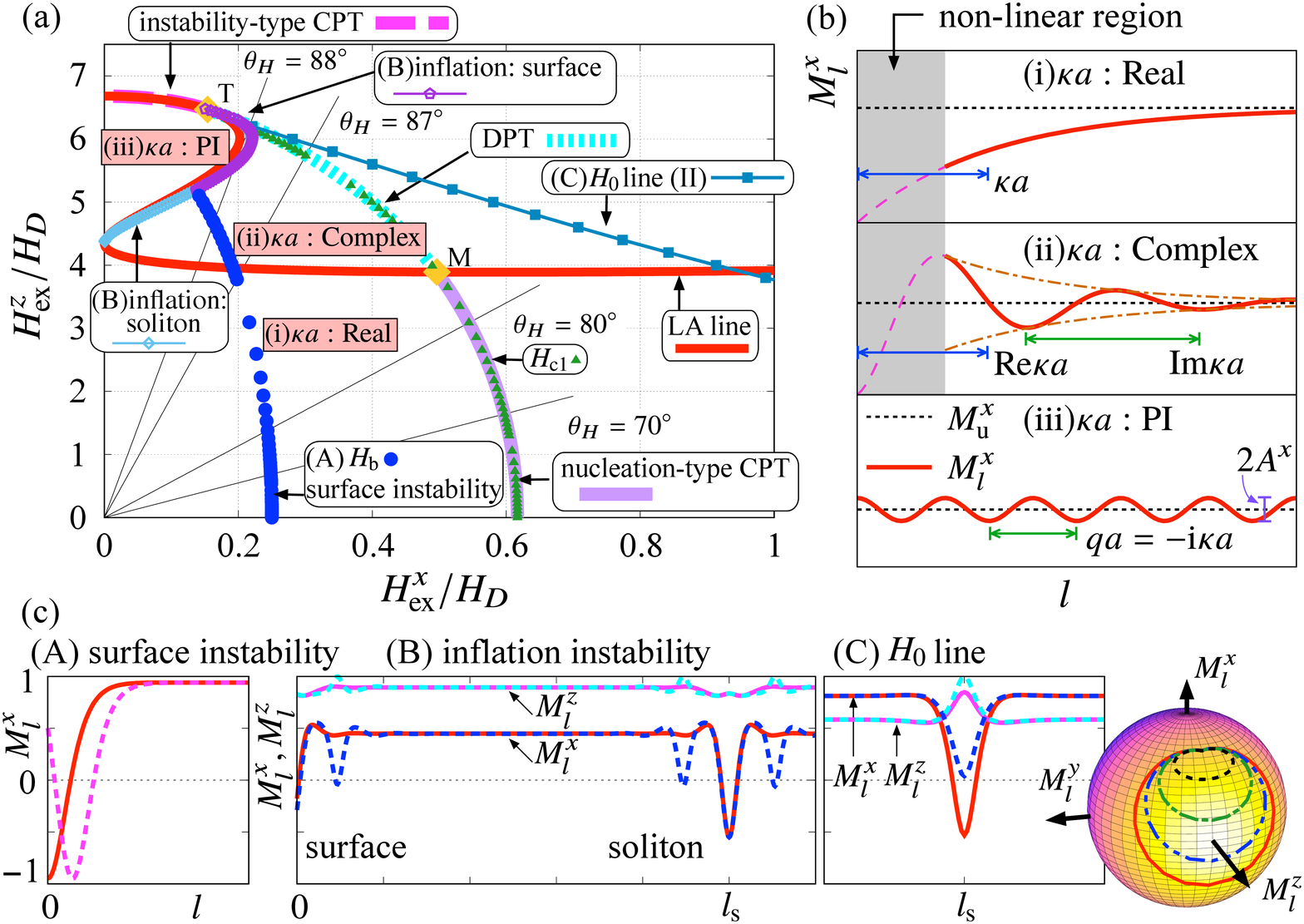}
\caption{(a) Phase diagram for realistic parameters: $D/J_{\parallel} = 0.16$ and $K/H_{D} = 5.68$. Two points denoted by ``T'' and ``M'' are the tricritical point and the multicritical point, respectively, and separate the phase boundary into the three segments: nucleation-type continuous phase transition line, the discontinuous phase transition line, and the instability-type continuous phase transition line. The red solid line labeled ``LA line'' is obtained using the linearization approach in Sect.~\ref{sect-LA}, and it divides the parameter space into the three regions: (i) real, (ii) complex, and (iii) pure imaginary (PI). The schematic images of these three cases are shown in panel (b). The black solid lines denote the angle of the tilted field for an eye guide. In addition, there are three kinds of instability field: (A) surface instability [Sect.~\ref{sect-surface-instability}] given by the barrier field $\vec{H}_{\bb}$ in the phase diagram, with blue circles, (B) two kinds of inflation-instability fields [Sect.~\ref{sect-div-inst}] with open symbols of pentagon and rhombus, which are labeled ``inflation: surface'' and ``inflation: soliton'', respectively, and (C) $H_{0}$ line [Sect.~\ref{sect-h0-line}] with solid squares. Their schematic images are shown in panel (c). For reference, we also show by small green triangles the field $\vec{H}_{\cc 1}$ at which the energy of a single soliton in the bulk is zero. 
(b) The schematic images of three cases considered in the linearization approach. The horizontal (vertical) axis stands for the coordinate along the chain (the $x$ component of the magnetic moment, $M_{l}^{x}$). In the gray regions in (i) and (ii), the nonlinear properties are important and the linearization is not valid. The red solid curves show $M_{l}^{x}$ and the black dotted lines show the uniform value $M_{\uu}^{x}$. The linearization is valid, when the difference between $M_{l}^{x}$ and $M_{\uu}^{x}$ is small, i.e. in the region far from the soliton center (out of the gray region) in the cases (i) and (ii) and when $A^{x}$ is sufficiently small in the case (iii). The real (imaginary) part of $\kappa$ stands for the spatial scale of the decaying length (the wavelength of the oscillation), as indicated by blue (green) bars.
(c) The schematic images of the three kinds of instabilities: (A) the surface instability, (B) the inflation instability, and (C) the $H_{0}$ line. 
The solid curves are the initial spin profiles, while the dashed curves are the transient spin profiles of these instability processes.
The first two instabilities increase the winding number, while the third one decreases. 
The surface instability causes the penetration of soliton from the surface. On the other hand, the inflation instability increases the winding number by changing the spin profiles around the surface ($l=0$) and/or the soliton center ($l=l_{\mathrm{s}}$).  At the $H_{0}$ line, the soliton unwinds to be the uniform state by pointing the spin to the helical axis at the soliton center $l_{\mathrm{s}}$. The winding/unwinding process is also illustrated  using the unit sphere on which the spin configuration $\{\vec{M}_{l}\}$ is given by the curve. The winding number changes when the loop on the sphere winds/unwinds the $M_{l}^{z}$ axis.
For inflation instability and the $H_{0}$ line, we show $M_{l}^{z}$ as well as $M_{l}^{x}$. The light-blue dashed curves touch $M_{l}^{z} = 1$ when the winding number changes in (B) and (C).
}
\label{fig-hb-h0-phase}
\end{center}
\end{figure*}
\begin{figure*}[t]
\begin{center}
\includegraphics[width=\hsize]{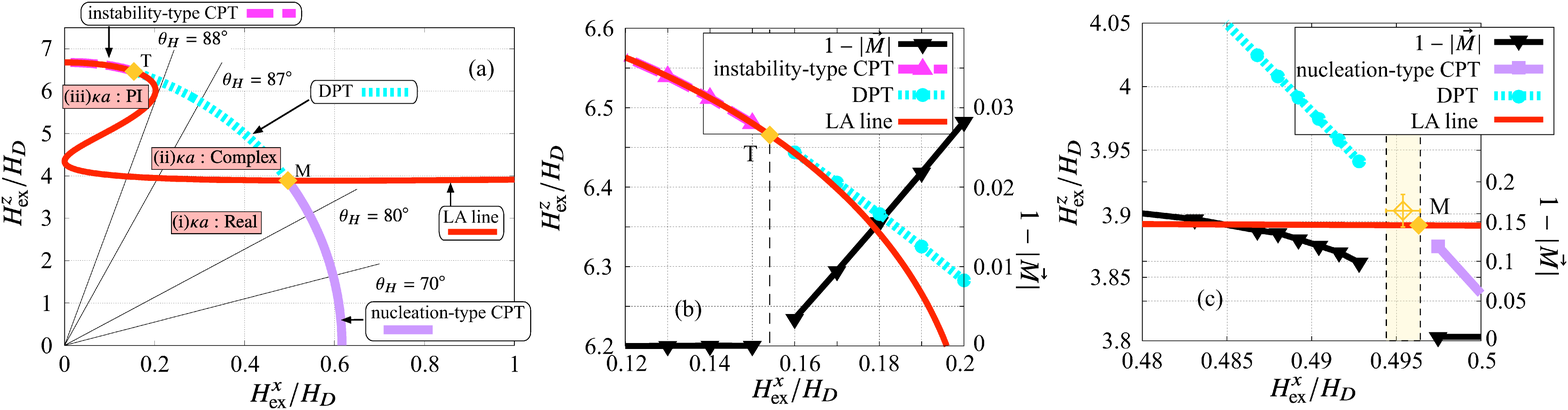}
\caption{(a) Phase diagram including three different kinds of phase boundaries and the LA line. The color plots are the same as in Fig.~\ref{fig-hb-h0-phase}(a). 
Magnified phase diagrams around (b) the tricritical point and (c) the multicritical point. The open rhombus in panel (c) is an estimated value of the multicritical point with error bars using the curve fitting. We show $1-|\vec{M}|$ with the uniform magnetization $\vec{M} = \sum_{l}\vec{M}_{l} / N_{z}$ as the limit from the low-field side to the phase boundary by inverted triangles, following Ref.~\onlinecite{Laliena2016a}. The values are indicated on the right vertical axis.}
\label{fig-phase-diagram-critical}
\end{center}
\end{figure*}
\subsection{Overview of the phase diagram}\label{sect-phase-diagram}
Let us start with overviewing the phase diagram of this system, 
and study it in detail in the following subsections. 
The phase boundary was first obtained in Ref.~\onlinecite{Laliena2016a} for a continuum model, but
we recalculate it using the lattice model with $N_{z} = 4000$ for the following analyses.
In Fig.~\ref{fig-phase-diagram-critical}(a), we show the phase diagram for the set of the realistic parameters of Cr$_{1/3}$NbS$_{2}$. The horizontal axis is the perpendicular component of the field $H_{\mathrm{ex}}^{x}/H_{D}$, while the vertical axis is the parallel component $H_{\mathrm{ex}}^{z}/H_{D}$. Here $H_{D} = 2[(J_{\parallel}^{2} + D^{2})^{1/2}-J_{\parallel}]$. The phase boundary is represented with several lines corresponding to the types of phase transition: instability-type CPT, DPT, and nucleation-type CPT. The two solid rhombuses on the phase boundary are the multicritical points. One is the tricritical point labeled T and the other is the multicritical point labeled M, which are  respectively referred to as TC1 and TC2 in Ref.~\onlinecite{Laliena2016a}. Here we regard 
point TC1 as a tricritical point in the sense that the point is determined by $a_{2} = 0$ and $a_{4}= 0$ with the coefficients in the Landau energy $E(\xi) = a_{0} + a_{2}\xi^{2} + a_{4}\xi^{4}$ for the amplitude $\xi$ of a certain order parameter. On the other hand, we call TC2 the multicritical point, following Ref.~\onlinecite{Schaub1985}. 
The parameter space of $H_{\mathrm{ex}}^{x}$ and $H_{\mathrm{ex}}^{z}$ is separated into the three region by the solid line colored with red, which is obtained by the linearization approach in Sect.~\ref{sect-LA}.
This line, which we call the LA line, is important to understand T and M, and the phase boundary, 
as seen in Figs.~\ref{fig-phase-diagram-critical}(a)--\ref{fig-phase-diagram-critical}(c).
Panels (b) and (c) are the enlarged images of Fig.~\ref{fig-phase-diagram-critical}(a) around T and M, respectively. 

The phase boundary and the LA line are also shown with instability lines in Fig.~\ref{fig-hb-h0-phase}(a). In addition to these phase structures, we add to Fig.~\ref{fig-hb-h0-phase}(a) several instability lines referred to as $H_{\bb}$, $H_{0}$-line, inflation-surface and inflation-soliton. They are instabilities of inhomogeneous structures such as an isolated soliton and a surface modulation, while the part of the LA line gives the instability of the uniform state. We will explain them in detail in Sect.~\ref{sect-instability}. 

\subsection{Linearization approach for soliton}\label{sect-LA}
In this subsection, we study a condition for the presence of a soliton solution. Since numerical search of a soliton solution takes time and is not easy to cover the whole parameter space, we try an alternative approach and analyze a small deviation from the uniform state on the basis of Eqs.~\eqref{eq-moment} and \eqref{eq-field} up to its linear order.

The phase diagram includes the two types of CPT: instability type for critical field $(H_{\mathrm{ex}}^{x},H_{\mathrm{ex}}^{z}) \approx (0,H_{\cc}^{z})$ and nucleation type for $(H_{\mathrm{ex}}^{x},H_{\mathrm{ex}}^{z}) \approx (H_{\cc}^{x},0)$.
Near the latter-type transition, solitons are nucleated and their interactions determine the critical properties at the transition. 
Since the soliton density is small near the transition, the interaction is mainly determined by the tail structure of an isolated soliton, and we will study it in the following.

Following Ref.~\onlinecite{Schaub1985}, let us consider an isolated soliton with its center at $l=0$ in the uniform background and assume its tail structure described with a finite real part of $\kappa$ as
\begin{align}
\vec{M}_{l}
\simeq\vec{M}_{\uu} + \vec{A}\exp(-\kappa x_{l})~\text{with}~x_{l} = l a ~(l\gg 1).\label{eq-asymptotic-form} 
\end{align}
When $\kappa$ has a finite imaginary part, we take the real part of this equation and do the same in the following. 
The schematic image of this profile for its $x$ component is shown in Fig.~\ref{fig-hb-h0-phase}(b). 
The uniform background should be determined for a given external field, and we can write it as $\vec{M}_{\uu} = (M_{\uu,\perp},0, M_{\uu,\parallel})$ with the normalization condition $|\vec{M}_{\uu}|^{2}= 1$.  Note that $M_{\uu}^{y} = 0 $ since $H_{\mathrm{ex}}^{y} = 0$.

The real part of $\kappa$ is the inverse decay-length of the soliton as indicated in panels (i) and (ii) of Fig.~\ref{fig-hb-h0-phase}(b), and when $\mathrm{Re}(\kappa a) > 0$, the deviation $\vec{A}e^{-\kappa x_{l}}$ is small for $l \gg 1$. We expand Eqs.~\eqref{eq-moment} and \eqref{eq-field} up to its linear order and examine the condition that the asymptotic form of a soliton exists. 
Note that this is a necessary condition for the existence of an isolated soliton. Its sufficient condition will be discussed in Sect.~\ref{sect-h0-line} by considering its core structure.

The second term in Eq.~\eqref{eq-asymptotic-form} describes not only a soliton tail but also another spin structure. 
When $\kappa$ is pure imaginary $\kappa = \ii q$, it corresponds to a distorted conical structure with wave number $q$ [see panel (iii) of Fig.~\ref{fig-hb-h0-phase}(b)]. This solution can be also examined by the same approach when the amplitude is small, $|\vec{A}|\ll 1$. We call the order whose leading term is described by Eq.~\eqref{eq-asymptotic-form} with $\kappa = \ii q$ the {\it distorted conical order}.

Equations~\eqref{eq-moment} and \eqref{eq-field} are expanded up to first order with respect to the second term of Eq.~\eqref{eq-asymptotic-form}, in which either $\vec{A}$ or $\exp(-\kappa x_{l})$ is considered to be sufficiently small. The normalization condition $|\vec{M_{l}}|^{2} = 1$ imposes the following relation between $A^{x}$ and $A^{z}$:
\begin{align}
\vec{M_{\uu}}\cdot \vec{A} = M_{\uu,\perp} A^{x} + M_{\uu,\parallel}A^{z} =0. \label{eqA1A3}
\end{align}
Defining the following quantities:
\begin{align}
H_{\uu}^{x} &= 2 J_{\parallel}M_{\uu,\perp} +H_{\mathrm{ex}}^{x},\\
H_{\uu}^{z} &= 2 J_{\parallel}M_{\uu,\parallel} -KM_{\uu,\parallel}+H_{\mathrm{ex}}^{z},\\
H_{\uu} &= \sqrt{(H_{\uu}^{x})^{2} + (H_{\uu}^{z})^{2}},
\end{align}
we substitute Eq.~\eqref{eq-asymptotic-form} into Eq.~\eqref{eq-field}:
\begin{align}
H_{l}^{\mathrm{eff},x}&= H_{\uu}^{x}
+[2J_{\parallel}\mathrm{ch}(\kappa a)A^{x} + 2D\mathrm{sh}(\kappa a)A^{y}]e^{-\kappa x_{l}},\\
H_{l}^{\mathrm{eff},y}&= [2J_{\parallel}\mathrm{ch}(\kappa a)A^{y} - 2D\mathrm{sh}(\kappa a)A^{x}]e^{-\kappa x_{l}},\\
H_{l}^{\mathrm{eff},z}&= H_{\uu}^{z}
+[2J_{\parallel}\mathrm{ch}(\kappa a) -K ]A^{z}e^{-\kappa x_{l}}, \\
|\vec{H}_{l}^{\mathrm{eff}}|^{2}&\simeq H_{\uu}^{2} + 2\left\{
H_{\uu}^{x}[2J_{\parallel}\mathrm{ch}(\kappa a)A^{x} + 2D\mathrm{sh}(\kappa a)A^{y}]\right.
\nonumber \\
&\hspace{5em}\left.+H_{\uu}^{z}[2J_{\parallel}\mathrm{ch}(\kappa a)-K]A^{z}\right\}e^{-\kappa x_{l}}.
\end{align}
Note that $x_{l+1} - x_{l} = a$. In the last equation, we retain the terms up to first order in the deviation. For convenience, we write $\cosh$ and $\sinh$ as $\mathrm{ch}$ and $\mathrm{sh}$, respectively.
The coupled linear-equations are derived from the mean field equation \eqref{eq-moment} using complementary relation \eqref{eqA1A3}. 
Its $x$ component reads
\begin{align}
M_{\uu,\perp} &+ A^{x}e^{-\kappa x_{l}}\nonumber  \\
&\simeq\dfrac{H_{\uu}^{x}}{H_{\uu}}
+\dfrac{(H_{\uu}^{z})^{2}}{H_{\uu}^{2}}
\dfrac{
[2J_{\parallel}\mathrm{ch}(\kappa a)A^{x} + 2D\mathrm{sh}(\kappa a)A^{y}]e^{-\kappa x_{l}}}{H_{\uu}}\nonumber \\
&\hspace{5em}-
\dfrac{H_{\uu}^{x}H_{\uu}^{z}[2J_{\parallel}\mathrm{ch}(\kappa a)-K]A^{z}e^{-\kappa x_{l}}}{H_{\uu}^{3}},\label{eq-x-comp}
\end{align}
and the $y$ component does
\begin{align}
 A^{y}e^{-\kappa x_{l}}
&\simeq\dfrac{[2J_{\parallel}\mathrm{ch}(\kappa a)A^{y} - 2D\mathrm{sh}(\kappa a)A^{x}]e^{-\kappa x_{l}}} 
{H_{\uu}}. \label{eq-y-comp}
\end{align}
The equation for the $z$ component can be similarly written. The zeroth order terms for the $x$  and the $z$ component lead to the equation for the uniform component 
\begin{align}
\vec{M}_{\uu} = \vec{H}_{\uu}/H_{\uu}\to
\left(1-\dfrac{K}{H_{\mathrm{ex}}^{z}}M_{\uu,\parallel}\right)M_{\uu,\perp } - \dfrac{H_{\mathrm{ex}}^{x}}{H_{\mathrm{ex}}^{z}}M_{\uu,\parallel}= 0 \label{eq-uniform-comp}
\end{align}
with $|\vec{M}_{\uu}| = 1$.
When $H_{\mathrm{ex}}^{z}=0 $, $M_{\uu,\perp} = 1$ and $M_{\uu,\parallel}=0$. When $K=0$, $M_{\uu,\perp(\parallel)} = H_{\mathrm{ex}}^{x(z)}/H_{\mathrm{ex}}$. Note that the relation \eqref{eqA1A3} can be written as $\vec{H}_{\uu}\cdot\vec{A}=0$.
The first order terms in the equation for the $z$ component are equivalent to those in Eq.~\eqref{eq-x-comp} through Eq.~\eqref{eqA1A3}.
We obtain the coupled equation for $(A^{x},A^{y})$
\begin{align}
\begin{pmatrix}
2J_{\parallel}\mathrm{ch}(\kappa a)-KM_{\uu,\perp}^{2}-H_{\uu} 
& M_{\uu,\parallel}^{2}2D\mathrm{sh}(\kappa a)\vspace{0.5em}\\
-2D\mathrm{sh}(\kappa a) 
& 2J_{\parallel}\mathrm{ch}(\kappa a)-H_{\uu}
\end{pmatrix}
\begin{pmatrix}
A^{x} \\A^{y}
\end{pmatrix}
=0.
\end{align}
The condition that the nontrivial mode exists is given by the null determinant of the coefficient matrix, i.e., 
\begin{align}
A \cosh^{2}(\kappa a) + B \cosh (\kappa a) + C =0, \label{eq-quadratic}
\end{align}
where $A=4(J_{\parallel}^{2} + D^{2}M_{\uu,\parallel}^{2})$,
$B = - 2J_{\parallel}(2H_{\uu}+KM_{\uu,\perp}^{2})$, 
$C= -4D^{2}M_{\uu,\parallel}^{2} +H_{\uu}(H_{\uu}+KM_{\uu,\perp}^{2})$.
This can be regarded as the quadratic equation with respect to $\cosh(\kappa a)$. The discriminant of the quadratic equation and its solution
\begin{align}
\mathsf{D} =  \dfrac{B^{2}}{4} - AC,~\cosh (\kappa a) = \dfrac{-B\pm \sqrt{B^{2}-4AC}}{2A}.
\end{align}
classify values of $\kappa a$ into the three cases: (i) real, (ii) complex, and (iii) pure imaginary. Their boundaries given by $\mathsf{D} = 0$ are displayed by the red line in Figs.~\ref{fig-hb-h0-phase}(a) and \ref{fig-phase-diagram-critical}(a). We will explain the details of the phase diagram in the following subsections.
When $\mathsf{D}$ is positive, $\cosh \kappa a$ is real, which means $\kappa a$ is either real or pure imaginary. The condition for real (pure imaginary) $\kappa a$ is given by $\cosh\kappa a \ge (\le) 1$.
When $\mathsf{D} $ is negative, $\kappa a$ is complex. The complex $\kappa a$ means that the tail of the soliton has the structure of damped oscillation [see panel (ii) of Fig.~\ref{fig-hb-h0-phase}(b)]. We will show the typical profiles and that they are related to the interaction potential of solitons in Sect.~\ref{sect-multicritical}.
The pure imaginary $\kappa a$ means that the trial form we assumed has no decaying solution but oscillation of a single-$q$ mode as indicated in panel (iii) of Fig.~\ref{fig-hb-h0-phase}(b). As we see in the next subsection, this regime is described by a wave structure with helical pitch $q\equiv -\ii \kappa$ and roughly speaking the Fourier amplitude with $q$ is the order parameter in the Landau theory.

\subsection{Instability-type phase transition}\label{sect-inst}
In the previous subsection, we determined the region of solutions with pure imaginary $\kappa = \ii q$ and its boundary $\mathsf{D} = 0$. 
In this region, an isolated soliton cannot exist, and the ordered phase is described by the distorted conical spin structure. The line of $\mathsf{D}=0$ is the instability-field line of the uniform state. In this subsection, we study the phase boundary between the distorted conical phase and the disordered phase and determine it by more rigorous criterion on the basis of energy analysis. Instead of the microscopic energy functional~\eqref{eq-energy-chain}, 
we analyze the Landau energy of the distorted conical order parameter.  
Following Ref.~\onlinecite{Schaub1985}, we expand the spin structure using the order parameter $\xi$ as
\begin{align}
\begin{pmatrix}
M_{l}^{x} \\ M_{l}^{y}
\end{pmatrix}
\simeq 
\begin{pmatrix}
M_{c}^{x} \\  0
\end{pmatrix}
+ \sum_{n=1}^{3}\xi^{n} 
\begin{pmatrix}
\sigma_{n,x} \cos (nq x_{l}) \\
\sigma_{n,y} \sin (nq x_{l})
\end{pmatrix}, \label{eq-expand-1} \\
M_{c}^{x} \simeq M_{\uu,\perp} + \alpha_{M} \xi^{2} + \beta_{M}\xi^{4},~
q \simeq q_{c} + \alpha_{q} \xi^{2} + \beta_{q} \xi^{4}. \label{eq-expand-2}
\end{align}
Here $\sigma_{n=\{1,2,3\},\mu=\{x,y\}}$ with the normalization $\sigma_{1,x}^{2} + \sigma_{1,y}^{2} = 1$, $\alpha_{M,q}$, $\beta_{M,q}$, and $q_{c}$ are the parameters to be determined for minimizing the energy for a given $\xi$. The terms consisting of the single $q$ mode and its self-interactions should be enough to describe the instability-type phase transition, but we make a few remarks. First, the fourth-order harmonic terms $\cos 4qx_{l}$ and $\sin 4qx_{l}$ are not necessary, because we do not consider the case $qa = 0$, $\pi/2 $, or $\pi$, and their contributions with other $qa$ vanish after site summation. In a similar way, an initial phase $\phi$ which appears by shifting $qx_{l} \to qx_{l} + \phi = q y_{l}$, is not necessary either since $\sum_{l} \cos (n q y_{l}) = 0$ when $qa$ is neither $0$ nor $\pi/n$ for $n=1,\cdots,4$. Second, the odd-order terms in $\xi$ do not appear in $M_{c}$ and $q$. We can confirm this as follows: First we do not expand $M_{c}$ and $q$ with respect to $\xi$, and calculate the energy density. As we see later, it depends only on even-order terms in $\xi$. We obtain $\xi$ dependence of $M_{c}$ and $q$ from the minimization condition of the energy with respect to $M_{c}$ and $q$, and the equations include the even-order terms in $\xi$. Thus we can expand $M_{c}$ and $q$ as Eqs.~\eqref{eq-expand-2}. Physically, this means that transformation $\xi\to -\xi$ corresponds to shifting of the spin structure as $x_{l} \to x_{l} + \pi/q$, and it does not change the uniform magnetization.
These considerations on commensurate wave number $q$ are needed for lattice models in contrast to  continuum models such as a model in Ref.~\onlinecite{Schaub1985}.

Another difference from the study~\cite{Schaub1985} is the presence of the third component $M_{l}^{z}$, but this is only technical at zero temperature, and the normalization condition determines it as $M_{l}^{z} = [1-(M_{l}^{x})^{2} - (M_{l}^{y})^{2}]^{1/2}$. We obtain the expansion of $M_{l}^{z}$ up to order $\xi^{4}$ as
\begin{align}
M_{l}^{z} &\simeq M_{c}^{z}  +( \xi \sigma_{1,z} + \xi^{3}\sigma_{1,z}^{\prime})\cos qx_{l} \nonumber \\&+ (\xi^{2} \sigma_{2,z} +\xi^{4}\sigma_{2,z}^{\prime})\cos 2qx_{l} + (\xi^{3} \sigma_{3,z})\cos 3qx_{l} \label{eq-expand-3}
\end{align}
with $M_{c}^{z} = M_{\uu,\parallel} + \alpha_{M,z} \xi^{2} + \beta_{M,z}\xi^{4}$.
The parameters are related to those of the $x$ and the $y$ component, and the relations are given in Appendix~\ref{sect-landau-expansion}. 
Note that the term with $4q$ modulation is not necessary for the same reason as the above. 

We substitute these forms \eqref{eq-expand-1} and \eqref{eq-expand-3} into Eq.~\eqref{eq-energy-chain} and write the Landau energy density up to fourth order in $\xi$ as $E(\xi)/N_{z} = a_{0} + a_{2} \xi^{2} +a_{4}\xi^{4}$. Spatially dependent terms vanish owing to the summation over $l$. The detail of the calculation is referred to as Appendix~\ref{sect-landau-expansion}, and here we write down the coefficients. The first term in the Landau energy, $a_{0}$, includes $\vec{M}_{\uu}$, which has been already determined in the above linearization approach, and its form is given by 
$a_{0} = (\vec{H}_{\uu} + \vec{H}_{\mathrm{ex}})\cdot \vec{M}_{\uu}/2$. 
The stationary condition of $a_{0}$ with respect to $\vec{M}_{\uu}$ is equivalent to Eq.~\eqref{eq-uniform-comp}. The coefficient for $\xi^{2}$ is given by 
\begin{align}
a_{2} &= 
\dfrac{1}{2}\left[
-J_{\parallel}\cos q_{c} a + \left(\dfrac{H_{\uu}^{z}}{2M_{\uu,\parallel}} + \dfrac{K}{2}\right)\right]\left(1 + \dfrac{M_{\uu,\perp}^{2}\sigma_{1,x}^{2}}{M_{\uu,\parallel}^{2}}\right)\nonumber \\
&- D \sigma_{1,x}\sigma_{1,y} \sin q_{c} a - \dfrac{K}{4},
\end{align}
which includes $q_{c}$, $\sigma_{1,x}$ and $\sigma_{1,y}$ with the relation $\sigma_{1,x}^{2} + \sigma_{1,y}^{2}  = 1 $. They are determined by the stationary condition of $a_{2}$. We can show that $a_{2} = 0$ is equivalent to $\mathsf{D} = B^{2} /4- A C = 0 $ which is obtained using the linearization approach in the previous section. These details are shown in Appendix~\ref{sect-landau-expansion}. 

Next, we obtain the coefficient $a_{4}$ in the Landau energy.
The coefficient $a_{4}$ in the Landau energy depends on 
$\sigma_{2,\mu=\{x,y,z\}}$, $\sigma_{1,z}^{\prime}$, and $\alpha_{M}$,
while $\vec{M}_{\uu}$, and 
\{$q_{c}$, $\sigma_{1,x}$, $\sigma_{1,y}$\} have been determined by the stationary conditions of $a_{0}$ and $a_{2}$, respectively. 
The other  parameters in Eqs.~\eqref{eq-expand-1} and \eqref{eq-expand-2} are determined from the higher order coefficients. 
We obtain $a_{4}$, with $\sigma_{2}^{2} = \sum_{\mu=x,y,z}\sigma_{2,\mu}^{2}$, as
\begin{align}
a_{4} &= J_{\parallel}\left[
\sigma_{2}^{2}\sin^{2} q_{c}a  + \sigma_{1,z}\sigma_{1,z}^{\prime} ( 1- \cos q_{c}a)\right] \nonumber \\
&\hspace{2em} -D \sigma_{2,x} \sigma_{2,y} \sin 2q_{c} a -\dfrac{K}{2}\left(
\alpha_{M}^{2} + \dfrac{\sigma_{2,x}^{2} + \sigma_{2,y}^{2}}{2}
\right) \nonumber \\
&\hspace{2em}+ \dfrac{H_{\mathrm{ex}}^{z} }{2M_{\uu,\parallel}}\left(\alpha_{M}^{2} + \alpha_{M,z}^{2}  + \dfrac{\sigma_{2}^{2}
}{2}  + \sigma_{1,z}\sigma_{1,z}^{\prime}\right). \label{eq-a4}
\end{align}
Parameters $\sigma_{2,x}$, $\sigma_{2,y}$, and $\alpha_{M}$ should be determined through the stationary condition of $a_{4}$ with respect to them. The derivatives of $a_{4}$ are shown in Appendix~\ref{sect-landau-expansion}, from which we can easily obtain $\alpha_{M}$, $\sigma_{2,x}$, and $\sigma_{2,y}$ and find $a_{4}$ on the basis of Eq.~\eqref{eq-a4} for a given $\vec{H}_{\mathrm{ex}}$.
The condition $a_{4} = 0$ and $a_{2} = 0$ gives the tricritical point, and its value will be shown in Sect.~\ref{sect-tricritical}.

Near the phase boundary of the instability-type transition, there is the parameter region where $M_{l}^{x}$ becomes non-negative (roughly given by $M_{c}^{x} \gtrsim \xi \sigma_{1,x}$ in Eq.~\eqref{eq-expand-1}). In such a region, the winding number is zero, and the character of the ordered state is helical order rather than the assembly of the solitons.

\subsection{Tricritical point}\label{sect-tricritical}
First we consider the tricritical point, T ($\vec{H}_{\mathrm{c}}$), and the phase boundary of CPT ($H_{\mathrm{ex}}^{x}  < H_{\mathrm{tri}}^{x}$).
The numerically obtained phase boundary of this CPT is consistent with the analytical line based on the instability-type phase transition. 
Thus we conclude that the phase transition in this region is the instability-type continuous one which can be described by the Landau theory for the distorted conical  order. 
The tricritical point appears when $a_{2} = 0$ and $a_{4} = 0$, and it is obtained as $(H_{\mathrm{tri}}^{x},H_{\mathrm{tri}}^{z})\simeq 
(0.1541, 6.466) H_{D}$.
The value is in good agreement with the numerical calculation as shown in Fig.~\ref{fig-phase-diagram-critical}(b); the quantity $1-|\vec{M}|$ stays zero on the instability-type CPT line and becomes finite on the line being the other side of the tricritical point. Here, $1-|\vec{M}|$ is obtained by approaching to the transition line in the ordered phase.
We make a remark on the discontinuity which depends linearly on $H_{\cc}^{x} - H_{\mathrm{tri}}^{x}$ with the $x$ component of DPT field $H_{\cc}^{x}$, as shown in Fig.~\ref{fig-phase-diagram-critical}(b). We assume that the coefficient of $\xi^{6}$ in the Landau expansion $a_{6}> 0$, and $a_{4}$ changes its sign at the tricritical point along the phase boundary as $ a_{4} = \bar{a}_{4} (H_{\mathrm{tri}}^{x}-H_{\cc}^{x} )/H_{\mathrm{tri}}^{x}$ with $\bar{a}_{4} > 0$. The discontinuity of $\xi^{2}$ is given by $-a_{4}/(2a_{6}) =[\bar{a}_{4}/(2a_{6})] (H_{\cc}^{x}-H_{\mathrm{tri}}^{x} )/H_{\mathrm{tri}}^{x}$. The leading order in $\xi$ of the total magnetization, defined by $\vec{M} = \frac{1}{N_{z}}\sum_{l} \vec{M}_{l} \simeq (M_{c}^{x}, 0, M_{c}^{z}) $, is calculated from the expansion forms \eqref{eq-expand-1}--\eqref{eq-expand-3} and $M_{c}^{z}$, and we obtain the linear dependence on $H_{\cc}^{x} - H_{\mathrm{tri}}^{x}$ as
\begin{align}
1- |\vec{M}| \approx 
 \dfrac{M_{\uu,\parallel}^{2} + M_{\uu,\perp}^{2}\sigma_{1,x}^{2}}{4M_{\uu,\parallel}^{2}} \dfrac{\bar{a}_{4}}{2a_{6}}
 \dfrac{H_{\cc}^{x}-H_{\mathrm{tri}}^{x} }{H_{\mathrm{tri}}^{x}}.
\end{align}

\subsection{Multicritical point}\label{sect-multicritical}
\begin{figure*}[t]
\begin{center}
\includegraphics[width = 0.95\hsize]{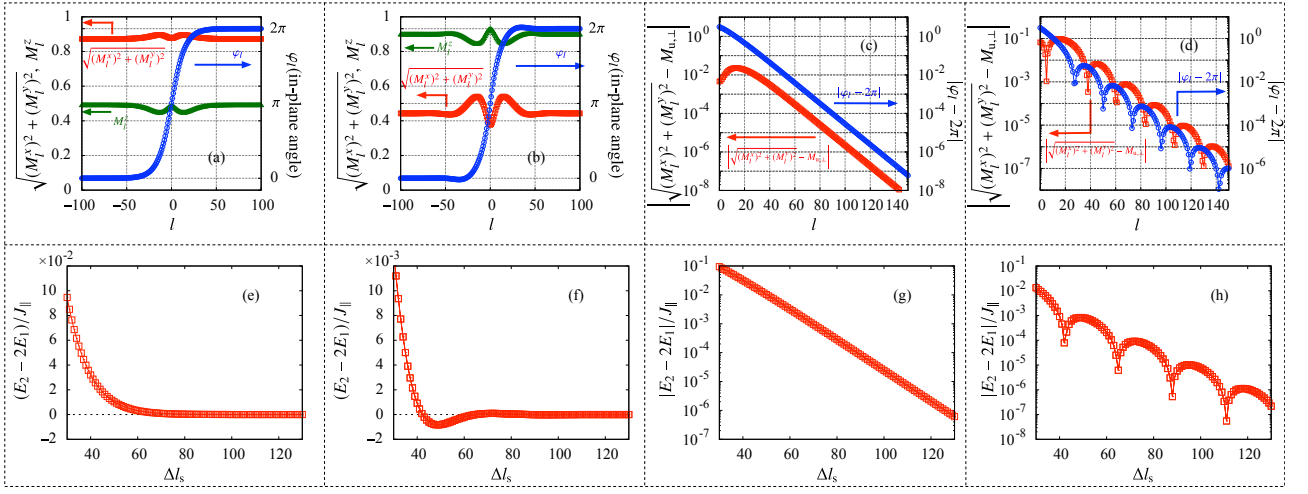}
\caption{(a), (b) Typical spin structures of isolated solitons. In both panels, red squares, green triangles, and blue circles with lines stand for the in-plane spin amplitude defined by$[(M_{l}^{x})^{2}+(M_{l}^{y})^{2}]^{1/2}$, $M_{l}^{z}$, and in-plane angle ($\varphi_{l}=\tan^{-1}(M_{l}^{y}/M_{l}^{x})$) for (a) the field in the repulsive region of the phase diagram, $\vec{H}_{\mathrm{ex}}=(0.4, 0.0, 3.0)H_{D}$, and (b) the field in the attractive region $\vec{H}_{\mathrm{ex}}=(0.3, 0.0, 5.7)H_{D}$. (c), (d) Deviations of in-plane amplitude (red squares, left axis) and phase (blue circles, right axis) from the asymptotic values in log scale of (a) and (b), respectively. (e), (f) Interaction energy of two solitons as a function of intersoliton distance $\Delta l_{\mathrm{s}}$, which is the variable of horizontal axes. Each vertical axis is the two soliton energy measured from $2E_{1}$ with the single soliton energy $E_{1}$, and the unit of the energy is $J_{\parallel}$. Each single soliton profile is given the top panels (a) and (b), respectively. The detail including the definition of energy is described in the text. (g), (h) The logarithmic plots of the absolute values of the panels (e) and (f), respectively.}
\label{fig-spin-interaction}
\end{center}
\end{figure*}

Next we see how we can understand the multicritical point, M. The phase boundary crosses the LA line at M in Figs.~\ref{fig-phase-diagram-critical}(a) and \ref{fig-phase-diagram-critical}(c). The LA line is defined by the null discriminant and separates the parameter space into the three regions on the basis of the type of $\kappa a$. For a complex $\kappa a$, the tail of an isolated soliton decays with oscillation. For a real $\kappa a $, the tail  exponentially decays without oscillation.

We confirm these tail structures for each representative field value in Fig.~\ref{fig-spin-interaction}. 
The panels (a) and (b) show the spin structures for the isolated solitons for real and complex $\kappa a$, respectively. The field values are shown in the caption. These isolated soliton solutions are obtained under the periodic boundary condition (PBC) for sufficiently large systems, and here we set $N_{z}$ = 2000. We impose an additional condition for the in-plane angle $\varphi_{l=0} = \pi$ to fix the soliton position. 
In panel (a), the in-plane amplitude defined by $[(M_{l}^{x})^{2} + (M_{l}^{y})^{2}]^{1/2}$ shows a small dip at the soliton center and once increases with distance, finally approaching the asymptotic value from above. On the other hand, in panel (b), the amplitude shows damped oscillation and hence has local maxima and minima also for $l \ge 50$. We can see such a difference also in the in-plane angle: In panel (a), the angle increases monotonically, while not monotonically but with oscillation in panel (b). We show the oscillation structures in the in-plane amplitude and angle, respectively, in Fig.~\ref{fig-spin-interaction}(d) more clearly. They are shown as the absolute values of the deviation from the asymptotic values in logarithmic scale. For comparison, we show those for a real $\kappa a$ in Fig.~\ref{fig-spin-interaction}(c).

For these two field values, we calculate the interaction energy of two isolated solitons\cite{Jacobs1980,Schaub1985,Leonov2010,Leonov2018,Masaki2018}. To eliminate the system size dependence, we measure the energy of $w$-soliton (winding) state from the energy of the uniform state $\vec{M}_{l} = \vec{M}_{\uu}$: $E_{w} = E - E_{\mathrm{uni}}$ with $E_{\mathrm{uni}} = - N_{z} (J_{\parallel}|\vec{M}_{\uu}|^{2} + \vec{H}_{\mathrm{ex}}\cdot \vec{M}_{\uu} - K (M_{\uu}^{z})^{2}/2 )$.
We define the interaction energy of two isolated solitons with distance $\Delta l_{\mathrm{s}}$ as 
\begin{align}
E_{\mathrm{int}} (\Delta l_{\mathrm{s}})= E_{2} (\Delta l_{\mathrm{s}}) - 2 E_{1}, \label{eq-int-e}
\end{align}
We construct the two soliton state under the PBC and  $\varphi_{l=0} = \varphi_{l=\Delta l_{\mathrm{s}}} = \pi$. 
We set $N_{z}=500$ to calculate $E_{\mathrm{int}} (\Delta l_{\mathrm{s}})$, which is sufficiently large for the interaction potential to decay well. We show $E_{\mathrm{int}} (\Delta l_{\mathrm{s}})$ in Fig.~\ref{fig-spin-interaction}(e) for $\vec{H}_{\mathrm{ex}}/H_{D}=(0.4, 0.0, 3.0)$ and in Fig.~\ref{fig-spin-interaction}(f) for  $\vec{H}_{\mathrm{ex}}/H_{D}=(0.3, 0.0, 5.7)$. Their absolute values in the logarithmic scale are also shown in Figs.~\ref{fig-spin-interaction}(g) and \ref{fig-spin-interaction}(h), respectively, as well as spin profiles.

Figure~\ref{fig-spin-interaction}(e) shows that the interaction is repulsive, namely the energy becomes lower as $\Delta l_{\mathrm{s}}$ increases. The nucleation-type CPT can be understood through the following emergent-particle picture: When the field is lower than the field denoted by $\vec{H}_{\mathrm{c}1}$ at which $E_{1} = 0$, adding solitons into the system lowers the energy owing to negative single soliton energy, which contrasts with the positive interaction energy. The competition between them determines the number of solitons in the system. The condition can be approximated as the minimization of the energy density under the PBC:  
\begin{align}
\min_{\Delta l_{\mathrm{s}}}[(E_{1} + E_{\mathrm{int}}(\Delta l_{\mathrm{s}}))/\Delta l_{\mathrm{s}}]
~\text{with} ~\Delta l_{\mathrm{s}} = N_{z} / w. \label{eq-nuc-cond}
\end{align}
We can neglect more than two body interactions near the phase boundary. The negative $E_{1}$ approaches zero with the increasing field, and finally $w=1$ is achieved and the critical field $\vec{H}_{\mathrm{c}}$ is given by the condition $E_{1} = 0$, i.e. $\vec{H}_{\mathrm{c}} = \vec{H}_{\mathrm{c}1}$.  In the thermodynamic limit, the winding number density defined by $\bar{w} = \lim_{N_{z}\to \infty} w/N_{{z}}$ changes from $(2\pi)^{-1}\tan^{-1}(D/J_{\parallel})$ to 0 continuously. For a fixed $H_{\mathrm{ex}}^{z}$, we explain the diverging behavior of the period at  $H_{\mathrm{ex}}^{x} \lesssim H_{\cc}^{x}$. First we approximate the single soliton energy and the interaction energy, respectively,  as 
\begin{align}
E_{1} = \varepsilon_{1} (H_{\mathrm{ex}}^{x} - H_{\cc}^{x})/H_{\cc}^{x},~E_{\mathrm{int}} = \varepsilon_{\mathrm{int}} e^{-\kappa a \Delta l_{\mathrm{s}}}.
\end{align}
Using the condition~\eqref{eq-nuc-cond}, i.e.,
$\varepsilon_{1} (H_{\mathrm{ex}}^{x} - H_{\cc}^{x})/H_{\cc}^{x} + \varepsilon_{\mathrm{int}}(\kappa a\Delta l_{\mathrm{s}} + 1 )e^{-\kappa a\Delta l_{\mathrm{s}}} = 0$, we obtain a finite intersoliton distance, which is a period of the soliton lattice for $0< H_{\cc}^{x} -H_{\mathrm{ex}}^{x} \ll H_{\cc}^{x}$ and correspondingly for $\kappa a\Delta l_{\mathrm{s}} \gg 1$:
\begin{align}
\Delta l_{\mathrm{s}} \approx \dfrac{1}{\kappa a} \ln\left(\dfrac{H_{\cc}^{x}}{H_{\cc}^{x}-H_{\mathrm{ex}}^{x}}\right).
\end{align}
This critical behavior with logarithmic divergence is consistent with the result of chiral sine-Gordon model $L_{\mathrm{CSL}}/ a \approx [4/(\pi Q_{0} a)] \ln [H_{\cc}^{x}/(H_{\cc} - H_{\mathrm{ex}}^{x})]$ with $Q_{0} = D/(J_{\parallel} a)$ at $H_{\mathrm{ex}}^{z} = 0$, where $\kappa a \to (\pi/4)(D/J_{\parallel})$ for $H_{\mathrm{ex}}^{x}\to H_{\cc}^{x}$.

On the other hand, the situation in Fig.~\ref{fig-spin-interaction}(f) is more complicated; the interaction potential steeply decays also in this case but with oscillation which comes from the oscillation of each soliton profile. As a consequence, $E_{\mathrm{int}}(\Delta l_{\mathrm{s}})$ has local minima. 
Let us define $\Delta l_{\mathrm{s},\min}$, at which the negative global minimum is taken. In Fig.~\ref{fig-spin-interaction}(f), $\Delta l_{\mathrm{s},\min}\sim50$. How does the oscillation in $E_{\mathrm{int}}(\Delta l_{\mathrm{s}})$ change the above mechanism? 
Even at $\vec{H}_{\mathrm{ex}} = \vec{H}_{\mathrm{c}1}$, i.e., $E_{1} = 0$, the energy density $E_{\mathrm{int}}(\Delta l_{\mathrm{s}})/\Delta l_{\mathrm{s}}$ is minimized at $\Delta l_{\mathrm{s}} ( < \Delta l_{\mathrm{s},\min})$, which corresponds to the soliton lattice state. More precisely, $\Delta l_{\mathrm{s}} ( < \Delta l_{\mathrm{s},\min})$ minimizes the total energy density when $E_{1} + E_{\mathrm{int}}(\Delta l_{\mathrm{s},\min}) < 0$. For $\vec{H}_{\mathrm{ex}}$ such that $0 < E_{1} < - E_{\mathrm{int}}(\Delta l_{\mathrm{s},\min})$, the soliton lattice state with finite intersoliton distance is more favored than the uniform state, though the isolated soliton state has higher energy than the uniform state. When $E_{1} + E_{\mathrm{int}}(\Delta l_{\mathrm{s},\min}) = 0$, we can show that the total energy density is minimized at $\Delta l_{\mathrm{s},\min}$ ( i.e., the minimum is 0). The DPT point is given by the field at which the energy of the soliton lattice is the same as that of the uniform state. Therefore, the DPT point can be evaluated by $\vec{H}_{\mathrm{ex}}$ such that $E_{1} + E_{\mathrm{int}}(\Delta l_{\mathrm{s},\min}) = 0$, in the simplified picture based on the two body interaction.
\begin{figure}[t]
\begin{center}
\includegraphics[width = 0.95\hsize]{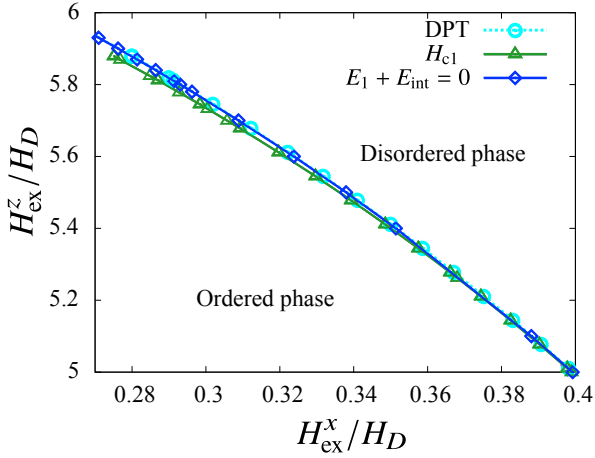}
\caption{Phase boundary of DPT. Light-blue circles, green triangles, and blue rhombuses represent the DPT line, $\vec{H}_{\cc1}$, and the line determined by the condition  $E_{1} + E_{\mathrm{int}}(\Delta l_{\mathrm{s},\min}) = 0$. See the text for their definitions.}
\label{fig-e1-eint}
\end{center}
\end{figure}

The evaluated values of DPT field such that $E_{1} + E_{\mathrm{int}}(\Delta l_{\mathrm{s},\min}) = 0$ are shown in Fig.~\ref{fig-e1-eint} with the blue rhombuses. 
In Fig.~\ref{fig-e1-eint}, we also show the DPT line (light-blue circles) and $\vec{H}_{\cc1}$ (green triangles), defined by $E_{1} = 0$. 
$\vec{H}_{\cc1}$ underestimates the phase boundary of the DPT and is in the ordered state, though it is the phase boundary when the phase transition is the continuous one of the nucleation-type. 
In contrast to $H_{\cc1}$, the curve with blue rhombuses agrees quite well with the DPT line, and thus the emergent particle picture is effective. 

Now we theoretically know that the sign of the interaction between solitons determines whether the phase transition is a continuous one of nucleation type or a discontinuous one. According to the numerical results shown in Fig.~\ref{fig-phase-diagram-critical}(c), the LA line crosses a phase boundary expected between the two end points, the light-blue circle and the purple square, and thus the calculated data are consistent with the linearization approach.
This approach is helpful in determining  the multicritical point M with high accuracy. M is a critical point at which the sign of the soliton interaction changes. Since the phase boundary of the nucleation-type CPT is the same as $\vec{H}_{\mathrm{c}1}$, we can identify M as the crossing point of $\vec{H}_{\mathrm{c}1}$ and the LA line. Its value is $(H_{\mathrm{multi}}^{x},H_{\mathrm{multi}}^{z}) \simeq (0.4963, 3.891 )H_{D}$. 
Note that $\vec{H}_{\cc1}$ can be calculated much more easily than the DPT line. 

\begin{figure}[tb]
\begin{center}
\includegraphics[width = 0.95\hsize]{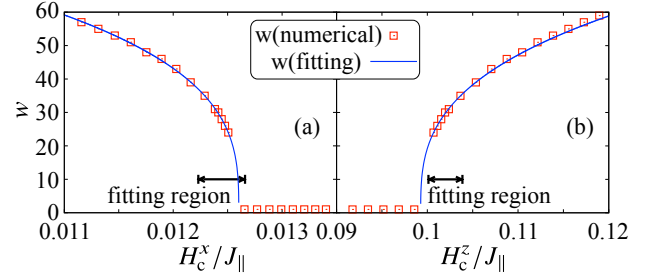}
\caption{Curves obtained through the fitting method around the multicritical point M.  The black arrows represent the region which we use for the curve fitting. The curve fitting is performed with respect to (a) the $x$ component and (b) the $y$ component.}
\label{fig-fitting}
\end{center}
\end{figure}

The point M can be evaluated also using the numerical data about the discontinuous jump in the winding number $w$, at the DPT. We perform curve fitting with a function $w = f_{\mu}[ (H_{\cc}^{\mu} - H_{\mathrm{multi}}^{\mu})/J_{\parallel}]^{e_{\mu}} (\mu = x,z)$ along the phase boundary $(H_{\cc}^{x},H_{\cc}^{z})$, where $f_{\mu}$, $H_{\mathrm{multi}}^{\mu}$, and $e_{\mu}$ are fitting parameters.
Here $e_{\mu}$ is a critical exponent of the discontinuity $w$ around M. We obtain the fitting parameters, as shown in Table~\ref{tab-fitting}, and the obtained fitting curves agree with the raw data as shown in Figs.~\ref{fig-fitting}(a) and \ref{fig-fitting}(b).
\begin{table}[t]
\begin{center}
\begin{tabular}{|c|c|c||c|c|c|}\hline
parameters & values & errors&parameters & values & errors \\ \hline
$H_{\mathrm{multi}}^{x}/H_{D}$ & 0.4954 & 0.0010 & $H_{\mathrm{multi}}^{z}/H_{D}$&  3.902&0.013 \\ \hline
$f_{x}$ & 468 & 164 &$f_{z}$ &  216&50 \\ \hline
$e_{x}$ & 0.321&0.046 &$e_{z}$ & 0.335 &0.046 \\\hline
\end{tabular}
\caption{Fitting parameters.}
\label{tab-fitting}
\end{center}
\end{table}
As seen from both Table~\ref{tab-fitting} and Fig.~\ref{fig-phase-diagram-critical}(c), two evaluated values of M are consistent within the error-bars. 

We mention the critical behavior of $1-|\vec{M}|$ near the multicritical point M. 
Since an intersoliton distance $\Delta l_{\mathrm{s}}$ is large, we can expect that 
$1-|\vec{M}| \approx w|\Delta \vec{M}|$, 
where $\Delta \vec{M} = N_{z}^{-1}\sum_{l}(\vec{M}_{\uu} - \vec{M}_{l})$ with a single soliton profile $\vec{M}_{l}$. The magnetization reduction owing to a single soliton, $\Delta \vec{M}$, does not change at around M,
and thus the field dependence may be almost the same as that of $w$, i.e.,
\begin{align}
1-|\vec{M}| \propto \left|\dfrac{H_{\cc}^{\mu} - H_{\mathrm{multi}}^{\mu}}{ H_{\mathrm{multi}}^{\mu}}\right|^{e_{\mu}}~\text{for}~\mu=x,z.
\end{align}
The critical behaviors of $1-|\vec{M}|$ and $w$ are different from the linear one, and in that sense, M is different from the conventional tricritical point.

\section{Instabilities of surface modulation and soliton}\label{sect-instability}
In the previous section, we show that a part of the phase transition is nucleation-type. There are many states with positive eigenvalues of the Hessian, and  hysteresis due to this kind of stability appears frequently also in the real material\cite{Togawa2015,Tsuruta2016,Yonemura2017,Mito2018}. 
In other words, because of the topological stability, the physical quantities for a state which is not the most stable spin structure are easily realized when the nucleation type is concerned. Actually, it was corroborated that the theory of the surface instability quantitatively explains the sharp jump appearing in the hysteresis process observed  in experiments for the micrometer-size sample of Cr$_{1/3}$NbS$_{2}$ in my previous papers~\cite{Shinozaki2018,Masaki2018}. Therefore, it is also important to study instability processes, as a process where a large amount of changes can be realized. In addition, when manipulation of solitons is considered, the condition for the existence of solitons is also important, which is described by instabilities of an isolated soliton, see Sect.~\ref{sect-div-inst} and Sect.~\ref{sect-h0-line}. 
In this section, we show three kinds of instability: surface instability, inflation instability, and an instability of soliton which results in the so-called $H_{0}$ line.  
Here we  study instabilities using excitation spectra of spin waves, and energy landscapes. They are complementary ways: The excitation spectrum allows us a systematic study of the instability. On the other hand, an intuitive picture of the instability mode is given by the energy landscape for a proper horizontal axis as the disappearance of the local minimum structure. Before dealing with specific problems, we first derive the eigenequation to obtain the excitation spectrum in the presence of a modulated structure as a static solution\cite{Muller2016}~(see also the supplemental material of Ref.~\onlinecite{Masaki2018}, where we consider it for the model in three dimensions). In the following, we only focus on the uniform mode in the plane perpendicular to the helical axis, which is valid for the system with the single DMI.

Let us introduce a new spin coordinate system $\vec{M}_{l} = \sum_{\mu=x,y,z}\tilde{M}_{l}^{\mu}\vec{\tilde{e}}_{l}^{\ \!\!\ \!\mu}$. The new basis set is spatially dependent, and it is determined by the static configuration $\vec{M}_{\mathrm{s},l} = \!\cos\varphi_{\mathrm{s},l}\sin\theta_{\mathrm{s},l}\vec{e}^{\ \!\!\ \!x} + \sin\varphi_{\mathrm{s},l}\sin\theta_{\mathrm{s},l}\vec{e}^{\ \!\!\ \!y}+ \cos\theta_{\mathrm{s},l}\vec{e}^{\ \!\!\ \!z}$, where subscripts $\sss$ denote {\it static}. 
The new $z$ direction points to the local magnetic moment of the equilibrium configuration: $\tilde{M}_{\mathrm{s},l}^{z} = \vec{\tilde{e}}_{l}^{\ \!\!\ \!z} \cdot \vec{M}_{\sss,l} = 1$. In addition, the set is chosen to satisfy the relation $\vec{\tilde{e}}_{l}^{\ \!\!\ \!x} \cdot \vec{e}^{\ \!\!\ \!z} = 0$. 
We write down the energy functional up to second order of $\{\tilde{M}_{l}^{x}\}$ and $\{\tilde{M}_{l}^{y}\}$ using the hermitian matrix $(\mathcal{K}_{l,m}^{\mu,\nu})^{*} = \mathcal{K}_{m,l}^{\nu,\mu}$ in the form 
\begin{align}
E(\{\vec{M}_{l}\})\simeq E(\{\vec{M}_{\mathrm{s},l}\}
) + \dfrac{1}{2}\sum_{l,m}\sum_{\mu,\nu = x,y}
\tilde{M}_{l}^{\mu} \mathcal{K}_{l,m}^{\mu,\nu}\tilde{M}_{m}^{\nu}, \label{eq:hessian}
\end{align} 
where we use $\tilde{M}_{l}^{z} \simeq 1 - \sum_{\mu = x,y}(\tilde{M}_{l}^{\mu})^{2}/2$.
The first order terms of $\{\tilde{M}_{l}^{x}\}$ and $\{\tilde{M}_{l}^{y}\}$ vanish owing to the equilibrium condition of $\{\varphi_{\sss,l}\}$ and $\{\theta_{\sss,l}\}$. The details of the matrix elements are written in Appendix~\ref{sect-matrix-elements}. The equation of motion is given by the Landau Lifshitz equation $\frac{\dd \vec{M}_{l}}{\dd t} = -\vec{M}_{l}\times \vec{H}_{l}^{\mathrm{eff}}$ with $ \vec{H}_{l}^{\mathrm{eff}} = -\frac{\partial \mathcal{H}}{\partial \vec{M}_{l}}$ given by Eq.~\eqref{eq-field}. By retaining the terms up to  first order in $\{\tilde{M}_{l}^{\mu = x,y}\}$ and performing the Fourier transform in time, the equation of motion now reads 
\begin{align}
-\ii \omega_{n} 
\begin{pmatrix}
\tilde{M}_{n,l}^{x} \\
\tilde{M}_{n,l}^{y}
\end{pmatrix}
=\sum_{m}\begin{pmatrix}
-\mathcal{K}_{l,m}^{yx} &
-\mathcal{K}_{l,m}^{yy} \\
\mathcal{K}_{l,m}^{xx} &
\mathcal{K}_{l,m}^{xy}
\end{pmatrix}
\begin{pmatrix}
\tilde{M}_{n,m}^{x}\\
\tilde{M}_{n,m}^{y}\\
\end{pmatrix}, \label{eq-bog}
\end{align}
where $n$ is a label of the eigenstates. For later convenience, we define the spatial profile of the amplitude  
for the spin wave of the $n$ th mode
\begin{align}
\tilde{A}_{n,l} = \sqrt{(\tilde{M}_{n,l}^{x})^{2} + (\tilde{M}_{n,l}^{y})^{2}}, \label{eq-amp-wf}
\end{align}
and the dimensionless eigenenergy $\epsilon_{n} = \omega_{n}/H_{D}$. It should be noted that the eigenequation has the same properties as the Bogoliubov equation for the Bose system, for which instabilities are intensively studied on the basis of its excitation spectrum. The correspondence is clearer by performing the following map: $\tilde{M}_{l}^{x} + \ii \tilde{M}_{l}^{y} \sim \sqrt{2S}u_{l}$ and $\tilde{M}_{l}^{x} - \ii \tilde{M}_{l}^{y} \sim \sqrt{2S}v_{l}$ and seeing the eigenequation for $u_{l}$ and $v_{l}$. The map reminds us of the Holstein--Primakov transformation for quantum spins up to the lowest order.

First we solve the mean field equation to obtain the static profile $\vec{M}_{\mathrm{s},l}$ and then investigate the excitation modes from $\vec{M}_{\mathrm{s},l}$ by numerically diagonalizing Eq.~\eqref{eq-bog} and obtain the excitation spectrum $\epsilon_{n}$ and eigenvectors $\{\vec{\tilde{M}}_{l}\}$. The PBC or the open boundary condition (OBC) given by $\vec{M}_{l=-1} = \vec{M}_{l=N_{z}} = \vec{0}$ for $l = 0, \cdots, N_{z}-1$ is properly used. In order to exclude the surface twist structure at around $l=N_{z}-1$ for the OBC, we use sites $l=0,\cdots, N_{z}/2-1$ for calculation of excitation spectrum.  Thereby, we can approximately deal with a semi-infinite system with boundary at $l=0$. The boundary condition for the diagonalization is correspondingly given by $\tilde{M}_{l=-1,N_{z}/2}^{x,y} = 0$. Note that the condition at $N_{z}/2$ gives finite size effects, but the effects on the localized modes are negligible and those on the extended modes are not very important in the following. 

\subsection{Surface instability and surface barrier}\label{sect-surface-instability}
When we decrease the perpendicular component of the field $H_{\mathrm{ex}}^{x}$, the surface instability occurs, which allows the emergence of solitons from the surface. It is interpreted as the vanishing surface barrier of a soliton, as discussed in Refs.~\onlinecite{Shinozaki2018,Masaki2018} using energy landscape for the soliton coordinate. It may be noted that the vortex coordinate was used to study the Bean--Livingston barrier for type-II superconductors\cite{Bean1964}.
The energy landscape has the local maximum structure around the surface which prevents the solitons from penetrating into the system. The local maximum structure can be regarded as the surface barrier and remains down to the barrier field $\vec{H}_{\bb}$, which is lower than the transition field.

The vanishing surface barrier causes drastic increase of soliton number, which is observed as sharp jumps in MR and MT measurements. However, in the former discussion we neglect the effects on the energy landscape of a modulation of the spin structure near the surface called surface twisted structure\cite{Du2013,IwasakiMochizukiNagaosa2013a,Rohart2013, Sampaio2013, Wilson2013,Meynell2014}. Here we study the surface instability from this point of view and show that it does not affect the barrier field $\vec{H}_{\bb}$. We also show excitation spectra and energy landscapes similar to those shown in the previous papers\cite{Shinozaki2018,Masaki2018} for readability. In addition, we discuss the soliton bound around the surface when the soliton interaction is attractive, which can be an evidence for the attractive soliton interaction and thus for the DPT.

\begin{figure}[t]
\begin{center}
\includegraphics[width = 0.95\hsize]{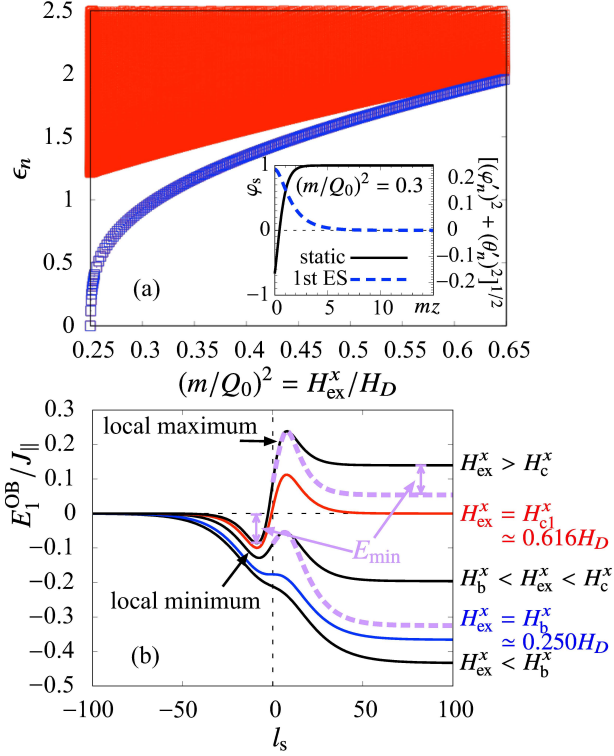}
\caption{(a) Energy spectrum for the continuum model as a function of $H_{\mathrm{ex}}^{x}/H_{D}$. The bound state with energy below the continuum is indicated by blue squares. The inset shows, at $H_{\mathrm{ex}}^{x}/H_{D} = 0.3$, the in-plane spin angle for the static state $\varphi_{\mathrm{s}}$ using the black solid curve, and the amplitude for the spin wave profile of the first excited state using the blue broken curve. (b) Energy profiles of an isolated soliton. Several curves correspond to different values of $H_{\mathrm{ex}}^{x}$. We set $H_{\mathrm{ex}}^{z} = 0$. 
The purple dashed lines describe the energy profiles calculated by taking account of the surface modulation. They agree quite well with the results without surface modulation near the surface barrier. Far from the surface, a spin profile of the dashed  curves includes a single soliton and the surface twisted structure and thus the energy is lower than the energy indicated by the corresponding solid curve by the energy of the surface twisted energy, which is the value of the local minimum in the solid curve. The energy profiles shown with red and blue colors correspond to the cases of $H_{\mathrm{ex}}^{x}= H_{\cc}^{x}$ and $H_{\bb}^{x}$, respectively. The precise values of the field are $0.2H_{D}, H_{\bb}^{x},0.4H_{D},H_{\cc}^{x},0.8H_{D}$ from the bottom, where $H_{D} = 2[(J_{\parallel}^{2} + D^{2})^{1/2}-J_{\parallel}]$. }
\label{fig-surface-barrier-1}
\end{center}
\end{figure}
We start with the excitation spectrum at $H_{\mathrm{ex}}^{z} = 0$. Figure~\ref{fig-surface-barrier-1}(a) shows the field dependence of the spectrum on the basis of chiral sine-Gordon model. The definition and formulation of the chiral sine-Gordon model are summarized in Appendix~\ref{sect-h0-Kdep-csg}. The static configuration is given by\cite{Meynell2014,Shinozaki2018}
\begin{align}
\varphi_{\mathrm{s}}(z) = 4\tan^{-1}[e^{m(z-z_{\mathrm{s}})}] \label{eq-single-soliton}
\end{align}
with
\begin{align}
z_{\mathrm{s}} = -\dfrac{1}{m}
\left|\cosh^{-1}\left(\dfrac{2m}{Q_{0}}\right)\right| \equiv z_{\mathrm{s,min}}. \label{eq-surface-twist}
\end{align}
The surface barrier vanishes when $z_{\mathrm{s,min}} = 0$, namely $m/Q_{0} = 1/2$.  This is also confirmed in Fig.~\ref{fig-surface-barrier-1}(a). The excitation spectrum consists of the continuum part represented by red color and the isolated branch with the blue symbols. The branch is for a surface bound spin wave, whose energy becomes zero at $m/Q_{0} = 1/2$. In this calculation, we take account of the surface modulations and reproduce $\vec{H}_{\bb}$ in Ref.~\onlinecite{Shinozaki2018}, which does not include the effects of the surface modulation on the surface barrier.

\begin{figure*}[t]
\begin{center}
\includegraphics[width = 0.95\hsize]{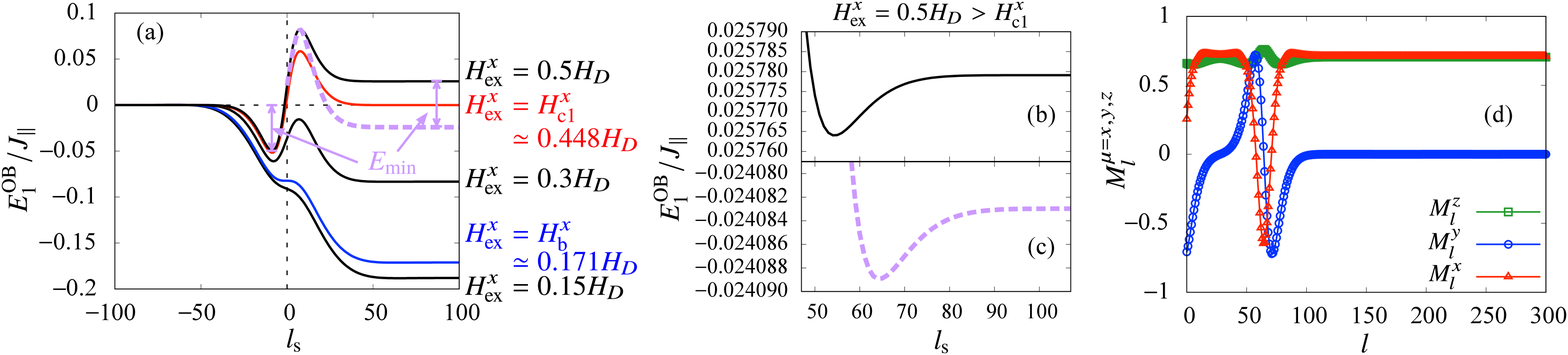}
\caption{(a) Energies of a soliton when the soliton interaction is attractive for several values of $H_{\mathrm{ex}}^{x}/H_{D}$. We use $H_{\mathrm{ex}}^{z} = 4.5 H_{D}$. Here $H_{\cc1}^{x} \simeq 0.448H_{D}$ is not the critical field but the field at which the single soliton energy vanishes. The middle panels (b) and (c) show the first local minima from the surface, which are not clear in panel (a) at $H_{\mathrm{ex}}^{x}/H_{D} = 0.5$. Panel (d) shows the spin profile for the surface modulation and the soliton which is bound to the local minimum shown in panel (c). }
\label{fig-surface-barrier-2}
\end{center}
\end{figure*}
We compare the energy landscapes of an isolated soliton near the surface with and without the surface modulation. The calculation methods are detailed in Appendix~\ref{sect-int-sol-surf}, where the calculation in the absence (presence) of the surface modulation is referred to as Method~I (Method~II).
The energy profiles are shown as functions of the soliton center $l_{\mathrm{s}}$ in Fig.~\ref{fig-surface-barrier-1}(b) for several values of $H_{\mathrm{ex}}^{x}$ when  $H_{\mathrm{ex}}^{z} = 0$. For $H_{\mathrm{ex}}^{z} = 0$, $H_{\bb}^{x} \simeq 0.250 H_{D}$ and $H_{\cc}^{x}\simeq 0.616 H_{D}$. The solid curves are obtained without surface modulation for $H_{\mathrm{ex}}^{x}= 0.2H_{D}$, $H_{\bb}^{x}$, $0.4H_{D}$, $H_{\cc}^{x}$, and $0.8H_{D}$ from the bottom for both positive and negative $l_{\mathrm{s}}$. Note that the negative $l_{\mathrm{s}}$ describes the situation where the soliton center is outside of the system, but its tail appears inside the system. These energy profiles have local maximum and local minimum structures at $l_{\mathrm{s}} = l_{\mathrm{s},\max} > 0$ and $l_{\mathrm{s},\min} = -l_{\mathrm{s},\max}$, respectively, for $H_{\mathrm{ex}}^{x} > H_{\bb}^{x}$. When $D/J_{\parallel} \ll 1$, $l_{\mathrm{s},\min}$ is approximated by $z_{\mathrm{s},\min}/a$. The local minimum structure easily traps a soliton at $l_{\mathrm{s},\min}$, and its tail structure in the system ($l \ge 0$) describes the surface modulated spin structure, which is the static configuration considered above [Eqs.~\eqref{eq-single-soliton} and \eqref{eq-surface-twist}, and see also Fig.~\ref{fig-spin-profile}]. 
The dashed curves are obtained by taking account of the surface modulation for $H_{\mathrm{ex}}^{x}= 0.4H_{D}$, and $0.8H_{D}$. We remark that this method is available only for $H_{\mathrm{ex}}^{x} > H_{\bb}^{x}$ and $l_{\mathrm{s}} \ge 0$. Compare black and purple curves for $l_{\mathrm{s}}\ge 0$. For $0\le l_{\mathrm{s}} \lesssim l_{\mathrm{s},\max}$, the difference appears to be small for both field values. When the soliton is deeply inside the system ($l_{\mathrm{s}} \gg l_{\mathrm{s},\max}$), 
the energy difference is given by the energy at $l_{\mathrm{s},\min} (< 0)$ denoted by $E_{\min} (< 0)$ in Fig.~\ref{fig-surface-barrier-1}(b). This implies that the interaction between soliton and surface disappears for sufficiently large distance, and the total energy is given by the sum of the energies for a single soliton and the surface modulation.  As seen from Fig.~\ref{fig-surface-barrier-1}(b), the energy barrier for a soliton inside the system becomes higher owing to the surface modulation, while that for a virtual soliton is not affected very much. For an in-between distance, it is necessary to take account of their interaction, which is discussed in Appendix~\ref{sect-int-sol-surf}.

We next study the surface barrier when the soliton interaction is attractive. 
For $H_{\mathrm{ex}}^{z} = 4.5 H_{D}$, we show the energy landscapes for $H_{\mathrm{ex}}^{x} = 0.15H_{D}$, $H_{\bb}^{x}$, $0.3H_{D}$, $H_{\cc1}^{x}$, and $0.5H_{D}$ from the bottom in Fig.~\ref{fig-surface-barrier-2}(a).
The whole structure of $E_{1}(l_{\mathrm{s}})$ [Fig.~\ref{fig-surface-barrier-2}(a)] is quite similar to the repulsive case, while there is small oscillation [(b) and (c)]. 
In the attractive case, the phase transition is discontinuous. 
For convenience, we use $H_{\cc1}^{x}$ at which the energy of the single soliton in the bulk is zero, instead of the  transition field at which the DPT occurs. 
The oscillation structures in the energy profiles are associated with the oscillation of the asymptotic behavior and there are tiny local minimum structures, though they are not visible in this scale in Fig.~\ref{fig-surface-barrier-2}(a). To visualize the local minimum structure, we show the enlarged profiles around the first local minimum in Fig.~\ref{fig-surface-barrier-2}(b) or \ref{fig-surface-barrier-2}(c). Their  presence is independent of whether the surface modulation is present or not, though the position of the local minimum changes. We expect that the local minimum structure attracts a metastable isolated  soliton to the surface, which can be an evidence of the attractive interaction. The binding energy of the local minimum in panel (c) is approximated as $5\times 10^{-6} J_{\parallel}$ per chain. We consider the $ab$ plane of $1 \mathrm{\mu m}\times1\mathrm{\mu m}$, which reads $N_{2\dd} = 4\times10^{6}$. Noting that $J_{\parallel}/k_{\mathrm{B}} \sim 10 \mathrm{K}$, the binding energy of the soliton in the three dimensional system is on the order of $200 \mathrm{K}$, which implies the possibility of the experimental observation. Panel (d) shows the soliton bound to the surface in the presence of the surface modulation.

\subsection{Inflation instability}\label{sect-div-inst}

\begin{figure}[t]
\begin{center}
\includegraphics[width = 0.95\hsize]{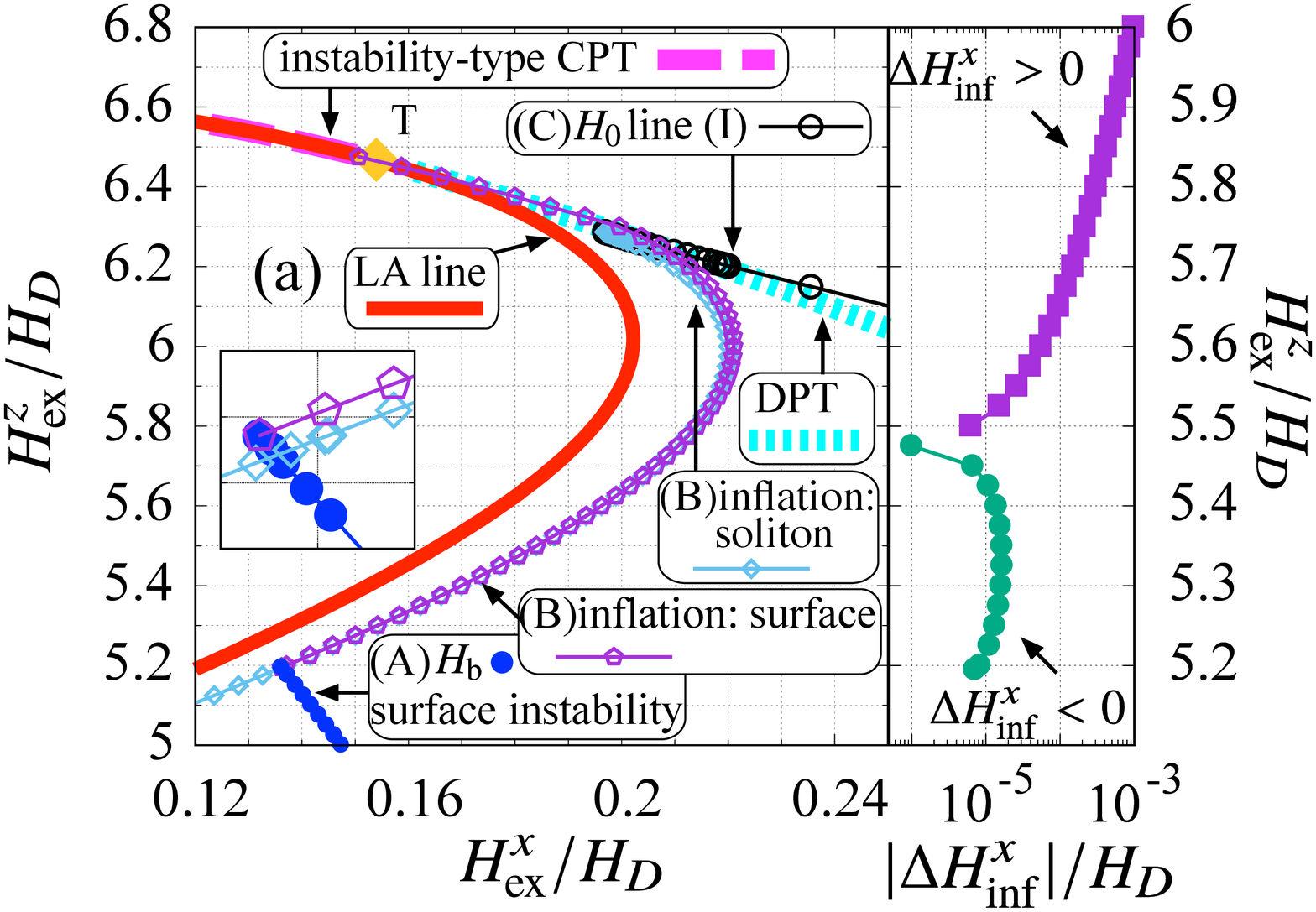}
\caption{(a) Phase diagram around the inflation instabilities. The inset shows the vicinity of the point where $\vec{H}_{\bb}$ and $\vec{H}_{\mathrm{inf,sur}}$ meet. The horizontal axis ranges from $H_{\mathrm{ex}}^{x}/H_{D} = 0.13614$ to $0.13616$ and the vertical axis ranges from $H_{\mathrm{ex}}^{z}/H_{D} = 5.19455$ to $5.19485$. The same symbols are used as in the main plot. (b) The absolute values of $\Delta H_{\mathrm{inf}}^{x}/H_{D}$ are shown on the horizontal axis, as a function of $H_{\mathrm{ex}}^{z}/H_{D}$ on the vertical axis, where $\Delta H_{\mathrm{inf}}^{x} \equiv H_{\mathrm{inf,sur}}^{x} - H_{\mathrm{inf, sol}}^{x}$ is the difference between the two inflation-instability fields. We use the square (circle) symbols for $\Delta H_{\mathrm{inf}}^{x} > 0$ ($\Delta H_{\mathrm{inf}}^{x} < 0$). }
\label{fig-inflation1}
\end{center}
\end{figure}

\begin{figure}[t]
\begin{center}
\includegraphics[width = 0.95\hsize]{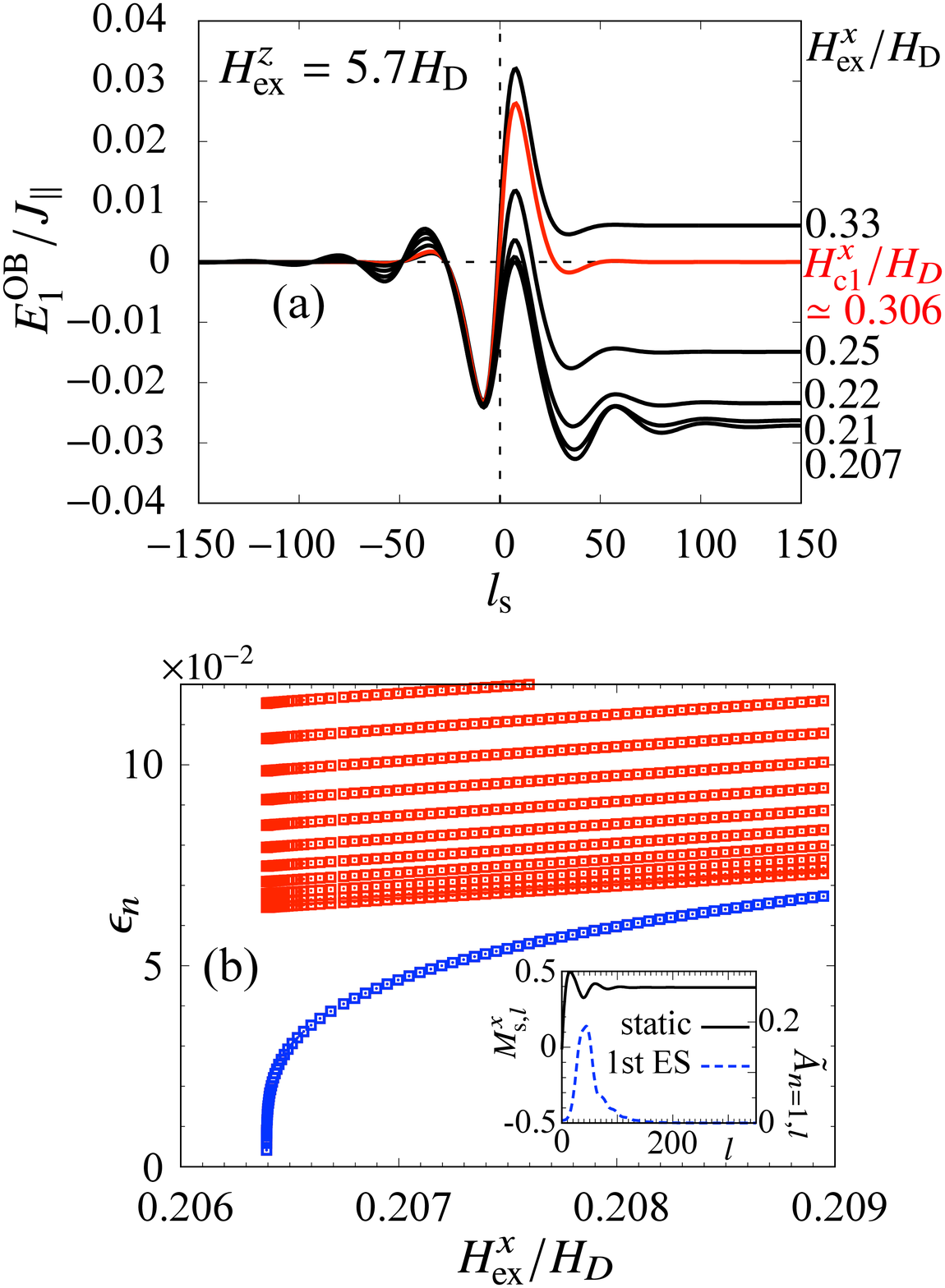}
\caption{
(a) Energies of the single soliton as functions of the soliton center for several values of $H_{\mathrm{ex}}^{x}/H_{D}$ as indicated in the panel.  
The red curve is for $\vec{H}_{\mathrm{c}1}$. (b) Energy spectra at $H_{\mathrm{ex}}^{z}/H_{D} = 5.7$. The blue symbols indicate the eigenenergies of the surface bound state. The inset shows the $x$ component of the magnetic moment for the static profile, $M_{\mathrm{s},l}^{x}$ using the black curve, and the amplitude for the first excited state, $\tilde{A}_{n=1,l}$ using the blue dashes curve at $H_{\mathrm{ex}}^{x}/H_{D} = 0.2064$. 
}
\label{fig-inflation2}
\end{center}
\end{figure}

Here we discuss the inflation instability. When the magnetic field changes from the region of complex $\kappa a$ to the pure imaginary region across the LA line [see Fig.~\ref{fig-hb-h0-phase}(a)], $\mathrm{Re}(\kappa a)$ vanishes. Hence a part of the LA line implies the instability that the soliton size diverges. However, we remark that this instability line has been determined using the linear approximation, and the pure imaginary region stands for the evolution of the distorted conical order from the uniform state instead of the soliton size divergence. As we see below, the soliton-size divergence occurs at slightly outside of the dome of the pure imaginary region, and we call this instability an inflation instability. There are two kinds of the inflation instabilities: One is the inflation of the soliton, and the other is the inflation of the surface modulation. These two instability fields are almost the same. The inflation instability of the soliton gives a part of the sufficient condition for the presence of the isolated soliton. The field values of the inflation instabilities are shown in Fig.~\ref{fig-inflation1}(a). In the following, we clarify the inflation instabilities in detail. 

First, we show the phase diagram around the complex and the pure imaginary region of $\kappa a$ in Fig.~\ref{fig-inflation1}(a). There are two curves labeled ``inflation: soliton'' and ``inflation: surface'', which are almost the same  for $H_{\mathrm{ex}}^{z}/H_{D} \gtrsim 5.2$. We write these two instability fields as $\vec{H}_{\mathrm{inf,sol}}$ and $\vec{H}_{\mathrm{inf,sur}}$, respectively. The instability field $\vec{H}_{\mathrm{inf,sol}}$ is the limit of metastability of an isolated soliton obtained by decreasing $H_{\mathrm{ex}}^{x}$ under the PBC with $N_{z} = 4000$. The field $\vec{H}_{\mathrm{inf,sur}}$ is the limit of metastability of the uniform state with the surface modulation. This is also obtained by decreasing $H_{\mathrm{ex}}^{x}$ but under the OBC with $N_{z} = 4000$. The system size is sufficiently large so that the surface modulations in both surfaces do not interfere. Below the instability field, the distorted conical state appears, and either the soliton or the surface modulation is a trigger of the conical instability. 

The two instability fields $\vec{H}_{\mathrm{inf,sur}}$ and $\vec{H}_{\mathrm{inf,sol}}$ are slightly different, and we show the difference of their $x$ component, $\Delta H_{\mathrm{inf}}^{x} = H_{\mathrm{inf,sur}}^{x} - H_{\mathrm{inf,sol}}^{x}$, as a function of $H_{\mathrm{ex}}^{z}/H_{D}$ in Fig.~\ref{fig-inflation1}(b). The horizontal axis shows the absolute values of the difference in logarithmic scale, and the square symbols stand for $\Delta H_{\mathrm{inf}}^{x} > 0$ while the circle symbols do  $\Delta H_{\mathrm{inf}}^{x} < 0$. $\Delta H_{\mathrm{inf}}^{x}$ becomes zero at around $H_{\mathrm{ex}}^{z}/H_{D} \sim 5.5$ as speculated from Fig.~\ref{fig-inflation1}(b), but it does not become zero at around $H_{\mathrm{ex}}^{z}/H_{D} \sim 5.2$. Instead, $\vec{H}_{\mathrm{inf,sol}}$ crosses the curve of $\vec{H}_{\bb}$ as shown in the inset of Fig.~\ref{fig-inflation1}(a), which is discussed in Appendix~\ref{sect-turning-point}. We see in the energy landscape that the inflation of the surface modulation is different from the surface instability. At $H_{\mathrm{ex}}^{z}/H_{D} = 5.7$, the energy landscapes for several values of $H_{\mathrm{ex}}^{x}/H_{D}$ are shown in Fig.~\ref{fig-inflation2}(a). The field values are displayed in the panel, and the red curve denotes the energy profile at $\vec{H}_{\mathrm{c}1}$. Down to $H_{\mathrm{ex}}^{x}/H_{D} = 0.207$, the local minimum and maximum structure near the surface exist, and $l_{\min} (<0)$, which gives the local minimum near the surface, does not approaches to the surface. Instead, the oscillation behavior in the energy profile becomes remarkable in decreasing field. Such a behavior implies that the instability is not the soliton entry from the surface. We remark that the isolated soliton becomes unstable between $H_{\mathrm{ex}}^{x}/H_{D} = 0.206$ and $0.207$ because of its inflation instability, consistent with Fig.~\ref{fig-inflation1}(a).

Since the energy landscape does not give a useful information about the inflation instability, i.e., the instability is different from the surface instability,
we investigate the instability of the surface modulation also in terms of the excitation spectrum. The dependence on $H_{\mathrm{ex}}^{x}$ of the excitation energy is shown in Fig.~\ref{fig-inflation2}(b). We use the same value of $H_{\mathrm{ex}}^{z}/H_{D}$ as in Fig.~\ref{fig-inflation2}(a).  
The eigenenergies of the bound state around the surface are indicated by the blue symbols. The inset shows the spatial profile of $M_{\mathrm{s},l}^{x}$ by the black solid curve, and that of $\tilde{A}_{n=1,l}$ defined by Eq.~\eqref{eq-amp-wf}, by the blue broken curve, where $n=1$ stands for the first excited state. We use $H_{\mathrm{ex}}^{x}/H_{D} = 0.2064$, 
which gives $\epsilon_{n} \simeq 1.5 \times 10^{-2}$. 
There is a clear oscillation in $M_{\mathrm{s},l}^{x}$ since the decaying length related to the soliton size increases for small $\epsilon_{n}$. The first excited state shows the largest amplitude at the position of the first dip from the surface in  $M_{\mathrm{s},l}^{x}$ rather than at the surface as shown in the inset of Fig.~\ref{fig-surface-barrier-1}(a). This excitation promotes the oscillation structure, and the size of the surface modulation becomes larger. 
The inflation instability corresponds to the process for increase of the winding number by pointing a spin at around the first dip to the helical axis, as indicated by the dashed curves in the middle panel of Fig.~\ref{fig-hb-h0-phase}(c). The process around the first dip is also explained by the change from the black dashed loop to the red solid one on the unit sphere in Fig.~\ref{fig-hb-h0-phase}(c). This process is followed by a similar process around the second dip. 
The instability of this mode leads to the distorted conical structure which spreads into the whole system.

We see that the inflation instability occurs not only around the surface modulation but also around the isolated soliton. 
In general, the inflation instability to the distorted conical order is triggered by the inhomogeneity such as the presence of the surface or the isolated soliton. This feature is different from that of the surface instability which occurs only around the surface. 
The case where both the surface modulation and an isolated soliton exist can be understood within the above discussion when the soliton is far from the surface. An example is demonstrated in Appendix~\ref{sect-inf-surface-soliton}.

Note that $\vec{H}_{\mathrm{inf,sol}}$ terminates at the end point of $\vec{H}_{0}$ discussed in the following subsection. The parameter region for the existence of an isolated soliton is surrounded by these two instability fields. 

\begin{figure}[t]
\begin{center}
\includegraphics[width=\hsize]{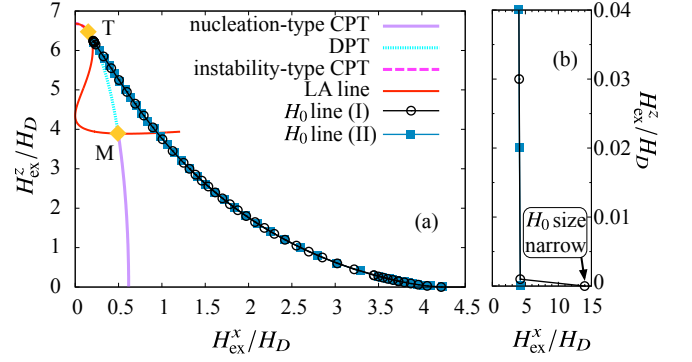}
\caption{(a) $H_{0}$ line in the whole phase diagram. (b) $H_{0}$ line around $H_{\mathrm{ex}}^{z}/H_{D} =0$. Note the range of the horizontal axis. 
 In these panels, the $H_{0}$ lines are plotted from the several aspects.
See the text for the difference between (I) and (II).
The value of the instability field indicated by ``$H_{0}$ size narrow'' in (b), is around $14H_{D}$, for which the out-of-plane component of spin is not taken into account. 
}
\label{fig-h0-1}
\end{center}
\end{figure}
\subsection{$H_{0}$ line}\label{sect-h0-line}
Finally we discuss the $H_{0}$ line, which is defined
by the limit of the metastability of an isolated soliton in high field side, i.e., an isolated soliton no longer exists in the high field side of this line. The $H_{0}$ line was originally considered for skyrmions in a two-dimensional chiral ferromagnetic system by Leonov {\it et al.}\cite{Leonov2010}. This line gives the upper bound for the field for possible observation of the remnant solitons.  
We remark that the necessary condition for the existence of an isolated soliton is  given by $\mathrm{Re}(\kappa a) > 0$. On the other hand, the sufficient condition is given by two instability fields: the inflation instability (the lower bound), and the $H_{0}$ line (the upper bound).

The $H_{0}$ line of an isolated soliton has three origins:  
(i) The first one is the spin motion towards the helical axis.  The soliton can be unwound  via the zero amplitude of the in-plane components. The method to seek for the $H_{0}$ line in this origin is discussed in detail later. 
(ii) The second one is that, when the temperature is finite,  thermal fluctuations reduce the moment locally, which unwinds the soliton through a vanishing moment at the soliton center. 
This was originally discussed for skyrmion unwinding, and the $H_{0}$ line was introduced for that system\cite{Leonov2010}. 
(iii) The third  origin is specific to the lattice model; at sufficiently large field, the width of the soliton becomes narrow. No matter how narrow the soliton is, it can exist in the continuum model. 
In the lattice model, however, an extremely narrow soliton with the width shorter than the lattice constant is neither stable nor well defined. 
In the present case of zero temperature and small $D$, we will see that the origin (i) is dominant. Thus we study the $H_{0}$ line from this viewpoint.

The whole structure of the $H_{0}$ line in the phase diagram is shown in Fig.~\ref{fig-h0-1}(a).
In Fig.~\ref{fig-h0-1}(a), the $H_{0}$ line is calculated in two ways for which we use the labels ``$H_{0}$-line ($k$)'' ($k= $ I, II ), and we see that they are consistent.  A simple way is to seek the field where an isolated soliton disappears by increase of the magnetic field with small steps, which we refer to as $H_{0}$ line~(I). The obtained $H_{0}$ line is shown by open circles in Fig.~\ref{fig-h0-1}.  A (meta) stable soliton region in the parameter space is bounded by $\vec{H}_{0}$ and $\vec{H}_{\mathrm{inf,sol}}$ and they meet at around $(H_{\mathrm{ex}}^{x},H_{\mathrm{ex}}^{z}) \sim (0.197, 6.287)H_{D}$[see Fig.~\ref{fig-inflation1}(a)]. The difference between the $x$ components of the $H_{0}$ line and the phase boundary becomes larger with decreasing $H_{\mathrm{ex}}^{z}$. At $H_{\mathrm{ex}}^{z} = 0$, $H_{0}^{x}/H_{D}$ marks the pretty high value about $14H_{D}$ in the panel~(c), which is different from the limiting value, $H_{\mathrm{ex}}^{x}/H_{D}\simeq 4.3$ for $H_{\mathrm{ex}}^{z} \to 0$.
\begin{figure}
\begin{center}
\includegraphics[width=0.85\hsize]{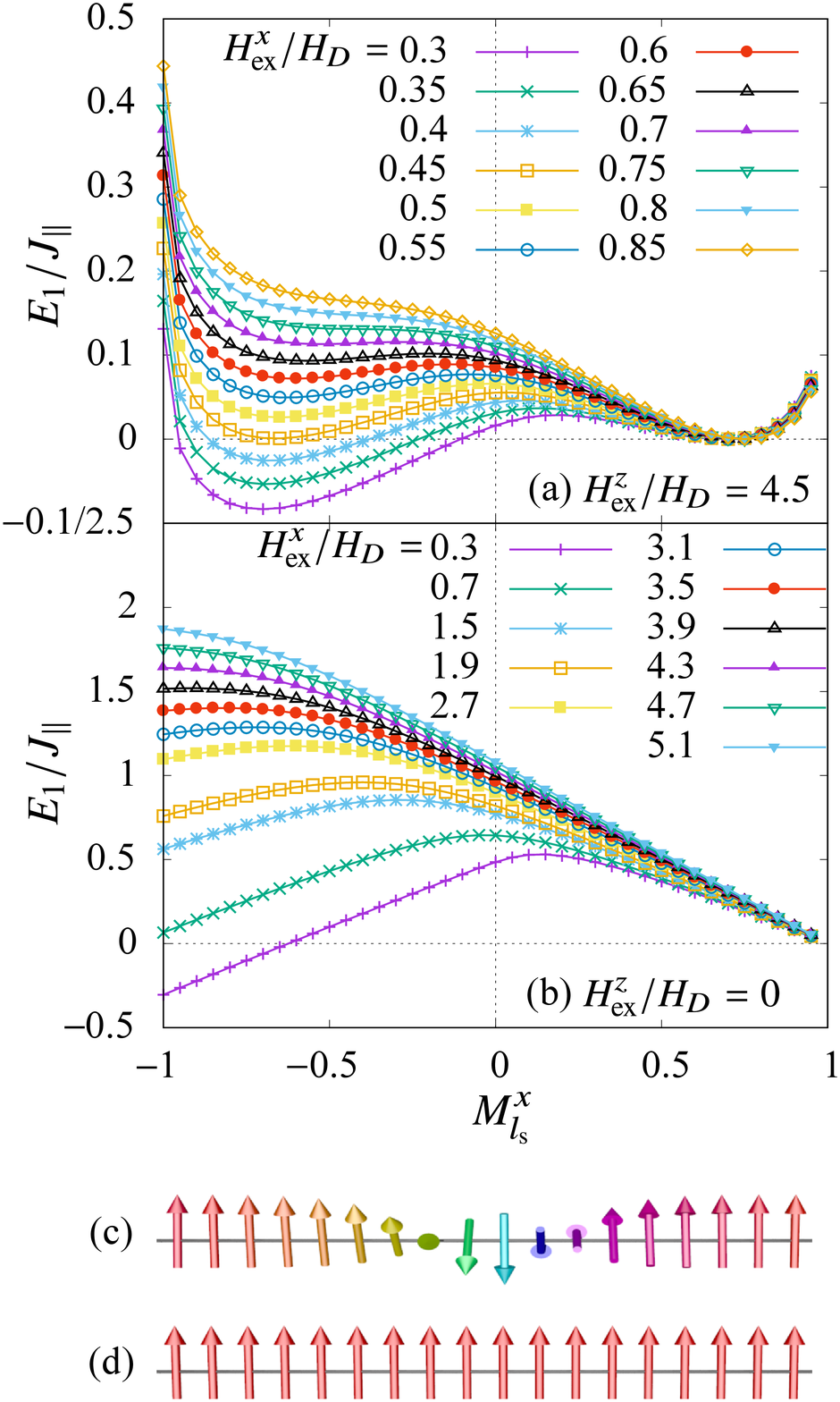}
\caption{Energy landscapes as functions of $M_{l_{\mathrm{s}}}^{x}$ for (a) $H_{\mathrm{ex}}^{z} = 4.5 H_{D}$ and (b) $H_{\mathrm{ex}}^{z} = 0$ . Schematic images of spin configurations at $H_{\mathrm{ex}}^{z} = 0$ for (c) $M_{l_{\mathrm{s}}}^{x} = -1.0$ and (d) $M_{l_{\mathrm{s}}}^{x} = 1.0$.}
\label{fig-h0-2}
\end{center}
\end{figure}

We also consider the $H_{0}$ line by using the energy landscape on the basis of mechanism (i) and refer to this scheme as $H_{0}$ line~(II).  
The condition for a (meta) stable soliton is that its total energy should be either globally or locally minimized for some value $M_{l_{\mathrm{s}}}^{x} < 0$. We let $M_{l_{\mathrm{s}}}^{y} = 0$, and now the problem is the variation of the energy with respect to one parameter $M_{l_{\mathrm{s}}}^{x}$ since $M_{l_{\mathrm{s}}}^{z} = [1 - (M_{l_{\mathrm{s}}}^{x})^{2}]^{1/2}$. In numerical calculations, we repeat iteration with $M_{l_{\mathrm{s}}}^{x}$ fixed in the update of $\vec{M}_{l}$. Figures~\ref{fig-h0-2}(a) and \ref{fig-h0-2}(b) show the energy landscapes as functions of $M_{l_{\mathrm{s}}}^{x}$ for $H_{\mathrm{ex}}^{z}/H_{D} = 4.5$ and $0$, respectively. Different curves correspond to different values of $H_{\mathrm{ex}}^{x}/H_{D}$. We find two local minima for low $H_{\mathrm{ex}}^{x}$: one is for positive $M_{l_{\mathrm{s}}}^{x}$ and the other for negative $M_{l_{\mathrm{s}}}^{x}$. The local minimum in the positive side corresponds to the uniform state, and hence it is taken at $M_{l_{\mathrm{s}}}^{x} = M_{\uu}^{x}$. The local minimum in the negative side corresponds to an isolated soliton solution. For $H_{\mathrm{ex}}^{z} = 0$, we show the sketches of the spin profiles corresponding to these two local minima in Figs.~\ref{fig-h0-2}(c) and \ref{fig-h0-2}(d). 
We see that the local minimum  in the negative side disappears for high $H_{\mathrm{ex}}^{x}$, which means that the soliton loses the stability against the motion of the moment at the center in the $M_{l_{\mathrm{s}}}^{x}$-$M_{l_{\mathrm{s}}}^{z}$~plane.
The instability field $H_{0}$ is determined by the disappearance of the local minimum: $\partial E_{1}/\partial M_{l_{\mathrm{s}}}^{x} = 0$ and $\partial^{2} E_{1}/\partial (M_{l_{\mathrm{s}}}^{x})^{2}= 0$. In Fig.~\ref{fig-h0-1}, the $H_{0}$ line obtained by the scheme (II) is shown by solid squares, which overlaps with the line of $H_{0}$ line~(I) except at $H_{\mathrm{ex}}^{z} = 0$; the origin of the soliton instability is confirmed as origin (i).

For $H_{0}$ line (II), $H_{0}^{x}$ at $H_{\mathrm{ex}}^{z} = 0$ is the same as the limiting value for $H_{\mathrm{ex}}^{z} \to 0$. As shown in Fig.~\ref{fig-h0-2}(b), the local minimum structure collapses at around $H_{\mathrm{ex}}^{x}/H_{D} \simeq 4.3$. Below this field, $E_{1}$ has two local minima at $M_{l_{\mathrm{s}}}^{x} = -1$ and $1$. On the other hand, $H_{0}$ line~(I) gives the different value $H_{\mathrm{ex}}^{x}/H_{D}\sim 14$ at $H_{\mathrm{ex}}^{z} = 0$.
This is because scheme~(I) restricts the spherical spin space to the $xy$ plane at $H_{\mathrm{ex}}^{z} = 0$, and the soliton cannot unwind by the mechanism (i). In this case, the increasing field squeezes the soliton width and leads to an instability which is possible only in the lattice model [mechanism (iii)]. The rather high $H_{0}$ field is attributed to the constraint of the spin space onto the $xy$plane, and this artificial constraint is eliminated by applying infinitesimal $H_{\mathrm{ex}}^{z}$. 
For realistic parameters of Cr$_{1/3}$NbS$_{2}$, the dominant instability mechanism of an isolated soliton is the mechanism (i), i.e., the spin motion towards the helical axis. The possibility of mechanism (iii) is discussed in Appendix.~\ref{sect-h0-Kdep-csg}.

\begin{figure}[t]
\begin{center}
\includegraphics[width = 0.95\hsize]{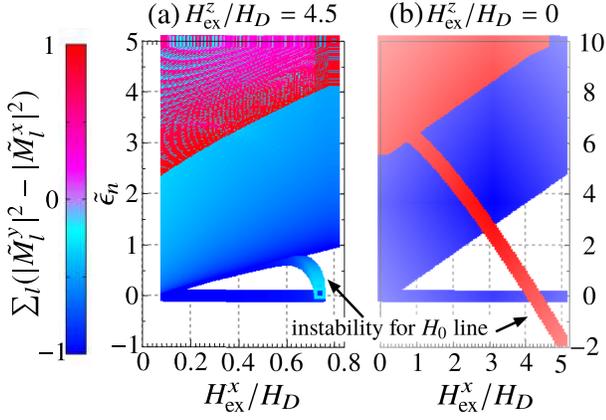}
\caption{
$H_{\mathrm{ex}}^{x}/H_{D}$ dependence of the eigenvalues of the Hessian for 
(a)$H_{\mathrm{ex}}^{z}/H_{D}= 4.5$ and (b)$H_{\mathrm{ex}}^{z}/H_{D}= 0$. The color represents 
$\sum_{l}(|\tilde{M}_{l}^{y}|^{2} - |\tilde{M}_{l}^{x}|^{2})$, where $\tilde{M}_{l}^{y}$ is related to the $\theta_{l}^{\prime}$ component of the spin wave. The eigenvalue of one localized mode different from the translational zero mode becomes zero, which causes the instability at $\vec{H}_{\mathrm{ex}} = \vec{H}_{0}$.
} 
\label{fig-e-h}
\end{center}
\end{figure}
\begin{figure}[t]
\begin{center}
\includegraphics[width = 0.95\hsize]{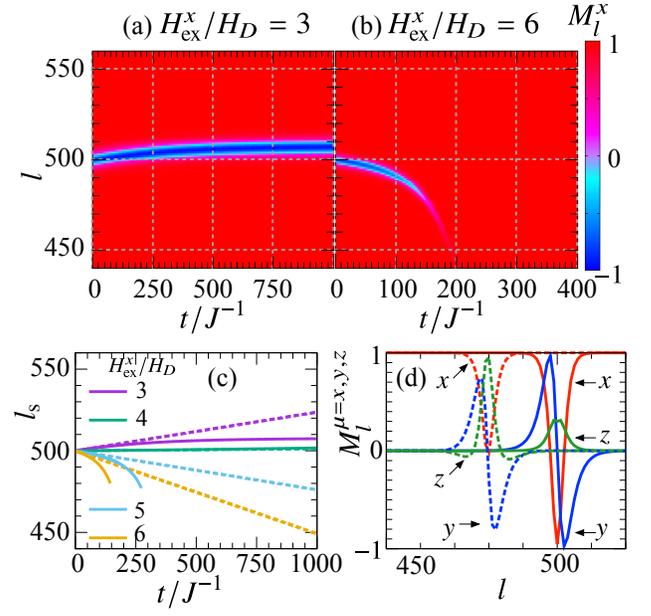}
\caption{
Time evolution of $M_{l}^{x}$ for (a) $H_{\mathrm{ex}}^{x}/H_{D} = 3$ and (b) $H_{\mathrm{ex}}^{x}/H_{D} = 6$. We set $H_{\mathrm{ex}}^{z} = 0 $. (c) The time evolution of the soliton center $l_{\mathrm{s}}$, and the values of $H_{\mathrm{ex}}^{x}/H_{D}$ are indicated in the caption. The solid curve (dashed line) stands for $\alpha = 0.05$ ($\alpha = 0$). The dashed lines describe the sliding motion of the soliton and give the initial slopes of the solid curves. For $H_{\mathrm{ex}}^{x}/H_{D}= 5$ and $6$, the solid curves terminate when soliton is destabilized. (d) The spin profiles for $H_{\mathrm{ex}}^{x}/H_{D} = 6$ at $t = 0$ and $t/J^{-1} = 145$, indicated by the solid curves and the dashed curves, respectively. The $M_{l_{\mathrm{s}}}^{z}$ evolves from 0.3 to 1, and the soliton vanishes.
}
\label{fig-sliding}
\end{center}
\end{figure}
In the remaining part, we clarify the $H_{0}$ line in terms of the spin wave excitation, i.e., the dynamics. 
The energy landscape in Fig.~\ref{fig-h0-2}(a) indicates that there are no unstable solution in $-1 \le M_{l_{\mathrm{s}}}^{x} \le 1$ for $M_{l_{\mathrm{s}}}^{z} > 0$ when the magnetic field is above the $H_{0}$ line ($H_{0}^{x}/H_{D}\sim0.75$ and $H_{0}^{z}/H_{D} = 4.5$). However, $M_{l_{\mathrm{s}}}^{x} = -1$ can be an unstable soliton when $H_{\mathrm{ex}}^{z} = 0$. Note that the single soliton profile 
$\varphi_{\mathrm{s}}(z) = 4\tan^{-1}[e^{m(z-z_{\mathrm{s}})}]$ 
for $\vec{M}_{\mathrm{s}}(z) = S(\cos\varphi_{\mathrm{s}},\sin\varphi_{\mathrm{s}},0)$ is always a stationary solution of the chiral sine-Gordon model. It is easily confirmed that there is still zero mode of the chiral soliton which stands for the translation of the soliton even above the $H_{0}$ line ($H_{0}^{x}/H_{D} \sim 4.3$ as discussed above).  Numerical diagonalization of the equation of motion \eqref{eq-eigen-csg} tells us there is no other localized mode, and thus there seems to be no instability. However, the instability of the $H_{0}$ line is complicated compared with the other instabilities ($\vec{H}_{\mathrm{b}}$ and $\vec{H}_{\mathrm{inf}}$), since it is combined with the translational zero mode, as seen  below. The combined dynamics corresponds to the sliding motion of the chiral soliton lattice\cite{KishineOvchinnikov2015}. First, we see the Hessian instead of the equation of motion itself. The Hessian is the quadratic form of the Hamiltonian with respect to the deviations (either a set of  $\tilde{M}_{l}^{x}$ and $\tilde{M}_{l}^{y}$ or a set of $\theta^{\prime}(z)$ and $\varphi^{\prime}(z)$) from the stationary point. Note $\theta^{\prime} = \theta - \theta_{\mathrm{s}}$ and $\varphi^{\prime} = \varphi - \varphi_{\mathrm{s}}$ and see Appendix~\ref{sect-h0-Kdep-csg} for further details. The relation between the Hessian analysis and the analysis for the equation of motion is discussed in the Appendix of  Ref.~\onlinecite{Wu2003}. Here the eigenequation is given by 
$\tilde{\omega}_{n} 
\tilde{M}_{n,l}^{\mu} 
=\sum_{m}\sum_{\nu=x,y}
\mathcal{K}_{l,m}^{\mu\nu}
\tilde{M}_{n,m}^{\nu}
$,
where $\tilde{\omega}_{n}$ is real. The dimensionless eigenvalues of the Hessian, $\tilde{\epsilon}_{n} = \tilde{\omega}_{n}/H_{D}$ are shown in Figs.~\ref{fig-e-h}(a) and \ref{fig-e-h}(b) for $H_{\mathrm{ex}}^{z}/H_{D} = 4.5$ and $H_{\mathrm{ex}}^{z}/H_{D} = 0$, respectively. The color shows $\sum_{l}(|\tilde{M}_{n,l}^{y}|^{2}-|\tilde{M}_{n,l}^{x}|^{2})$, where the information about the $z$ component in the original spin space is included only in $\tilde{M}_{l}^{y}$. When $H_{\mathrm{ex}}^{z}/H_{D} = 0$, $\tilde{M}_{l}^{x}$ and $\tilde{M}_{l}^{y}$ are decoupled, corresponding to $\varphi_{l}^{\prime}$ and $\theta_{l}^{\prime}$, respectively. In each case, there is zero mode which stands for the translation of the soliton. In addition, the low eigenvalue mode appears from the continuum region with increasing $H_{\mathrm{ex}}^{x}/H_{D}$, and its eigenvalue goes to zero. This field is $\vec{H}_{0}$, and the soliton is destabilized for $H_{\mathrm{ex}}^{x} > H_{0}^{x}$. The instability fields in Figs.~\ref{fig-e-h}(a) and \ref{fig-e-h}(b) are consistent with $H_{0}^{x}/H_{D} \sim 0.75$ and $4.3$ estimated from Figs.~\ref{fig-h0-2}(a) and \ref{fig-h0-2}(b), respectively. We show the Hessian eigenvalues for the uniform state for $H_{\mathrm{ex}}^{x} > H_{0}^{x}$ in Fig.~\ref{fig-e-h}(a). On the other hand, as discussed above, there is an unstable solution for $H_{\mathrm{ex}}^{x} > H_{0}^{x}$ and $H_{\mathrm{ex}}^{z} = 0$, for which the eigenvalue becomes negative and indicates instability. Its eigenvector stands for the $\tilde{M}_{l}^{y}\simeq\theta_{l}^{\prime}$ mode for $H_{\mathrm{ex}}^{z}=0$, describing the spin motion pointing to the $z$ direction. Although $\tilde{M}_{l}^{y}$ and $\tilde{M}_{l}^{x}$ are strongly coupled for finite $H_{\mathrm{ex}}^{z}$, the physical meaning of the instability mode is basically the same.

Then we see how this instability appears in dynamics by using the Landau Lifshitz Gilbert (LLG) equation 
$\frac{\dd \vec{M}_{l}}{\dd t} = -\vec{M}_{l}\times  \vec{H}_{l}^{\mathrm{eff}}- \alpha \vec{M}_{l} \times \frac{\dd \vec{M}_{l}}{\dd t}$ with $\vec{H}_{l}^{\mathrm{eff}} = -\frac{\partial \mathcal{H}}{\partial \vec{M}_{l}}$ and the Gilbert damping $\alpha$ since the instability does not appear in the linearized equation of motion Eq.~\eqref{eq-bog}. We set $\alpha = 0$ or $0.05$, $N_{z} = 1000$, and $H_{\mathrm{ex}}^{z} = 0$. The LLG equation is numerically solved by fourth order Runge--Kutta method with the time step $\Delta t = 0.01J^{-1}$. We construct the initial profile by setting $M_{l_{\mathrm{s}}}^{y} = 0$ and $M_{l_{\mathrm{s}}}^{z} = 0.3$ with $l_{\mathrm{s}} = 500$. Figures~\ref{fig-sliding}(a) and \ref{fig-sliding}(b) show the time evolution of $M_{l}^{z}$ for $H_{\mathrm{ex}}^{x}/H_{D} = 3$ and $6$. The soliton goes forward, i.e., $l_{\mathrm{s}}$ increases, for $H_{\mathrm{ex}}^{x} < H_{0}^{x}$, while it goes backward and vanishes for $H_{\mathrm{ex}}^{x} > H_{0}^{x}$.  In Fig.~\ref{fig-sliding}(c), the time evolution of the soliton center $l_{\mathrm{s}}$ is shown using solid curves for $\alpha = 0.05$ and dashed ones for $\alpha = 0$. The values of $H_{\mathrm{ex}}^{x}/H_{D}$ are shown in the panel. When $\alpha = 0$, the sliding motion is linear as shown by the dashed lines. The velocity corresponds to the eigenvalue of the Hessian for the localized mode indicated by the red symbols in Fig.~\ref{fig-e-h}(b)\cite{KishineOvchinnikov2015}, and the transport direction changes at around $H_{\mathrm{ex}}^{x} = H_{0}^{x}$.

When $\alpha = 0.05$, the sliding motion relaxes to either the static soliton profile or the uniform state owing to the instability. The former relaxation is seen in Figs.~\ref{fig-sliding}(a) and \ref{fig-sliding}(c), as indicated by the violet solid curve. In the case of $H_{\mathrm{ex}}^{x} = 4 H_{D} < H_{0}^{x}$, the soliton for $\alpha = 0.05$ moves  faster than it does for $\alpha=0$, which implies that the instability occurs though the relaxation time is long. This is due to the relatively large $M_{l_{\mathrm{s}}}^{x} = 0.3$ in the initial condition, and we confirm that the sliding motion relaxes to the static soliton structure when the initial condition is given by $M_{l_{\mathrm{s}}}^{x} = 0.2$. 
The latter relaxation to the uniform state is seen in Figs.~\ref{fig-sliding}(b) and  \ref{fig-sliding}(c), as indicated by the solid curves for $H_{\mathrm{ex}}^{x}/H_{D} = 5$ and $6$. The snapshots of the spin profiles at $t /J^{-1}= 0$ and $145$ are shown for $H_{\mathrm{ex}}^{x}/H_{D} = 6$ in Fig.~\ref{fig-sliding}(d) using the solid and dashed curves, respectively, where we clearly see the unwinding process by the evolution of $M_{l}^{z}$ from $0.3$ to $1$.

The above analyses of the $H_{0}$ line are consistent and give an interpretation of the instability mechanism, which is the unwinding process by pointing the spin at $l_{\mathrm{s}}$ to the helical axis. The simulation based on the LLG equation actually shows the unwinding process occurs with the sliding motion of the soliton. The instability occurs when the damping is finite, and thus it may be regarded as the Landau instability, although the signature does not appear in the excitation spectrum. 

Related to this instability, it should be noted whether the surface modulation above the $H_{0}$ line is described by a soliton since we showed that the surface modulation is interpreted as a virtual soliton outside the system in Sect.~\ref{sect-surface-instability}. We demonstrate the construction of the surface modulation above the $H_{0}$ line in Appendix~\ref{sect-surface-twist-h0}.

\section{Summary and Discussion}\label{sect-summary}
In this manuscript, we studied the characters of the phase transitions and instabilities of modulated structures in the monoaxial chiral magnet in the tilted magnetic field. 
In the first part, we performed the linearization to separate the parameter region into three parts. In one region, a wave picture of the distorted conical order works, while the particle picture of solitons is useful in the other two regions.  The ground state phase diagram, which was originally studied by Laliena {\it et al.}\cite{Laliena2016a}, consists of two CPTs and one DPT. Following de Gennes's classification, two CPTs are identified as the instability-type and the nucleation-type.
We emphasize that the nucleation-type transitions cannot be characterized by a small and local order parameter with a single $q$, but the winding number can be a generalized order parameter.
 The linearization clarified the mechanisms of the phase transitions and connected them to the soliton picture or the single $q$ helical wave picture. 
From this point of view, around the instability-type CPT, we constructed the Landau energy for the conical order parameter, which described the tricritical point precisely. 
On the other hand,  we performed several analyses based on the soliton picture to study the multicritical point and the DPT. 
We clarified that the two-soliton interaction can be repulsive or attractive, which is consistent with the linearization approach. The multicritical point can be assigned to the turning point of the interaction type on the phase boundary. We also demonstrated that how the soliton picture worked to describe the DPT line and the critical property of the nucleation-type CPT on the basis of the two-soliton interaction. 

We make a few remarks on the relation between the phase diagrams of $H_{\mathrm{ex}}^{x}$-$H_{\mathrm{ex}}^{z}$ and $H_{\mathrm{ex}}^{x}$-$T$.\cite{Laliena2017} These phase diagrams share the case of $\vec{H}_{\mathrm{ex}} = (H_{\mathrm{ex}}^{x}, 0,0)$ at $T = 0$, and $H_{\mathrm{ex}}^{z}$ or $T$ have the same effects of softening the in-plane amplitude $[(M_{l}^{x})^{2} + (M_{l}^{y})^{2}]^{1/2}$. The soft modulus effects  cause the DPT or the instability-type CPT and appear in the high temperature regions. It remains unclear whether the thermal fluctuations change the scenario of the phase transitions or not\cite{Nishikawa2016}. On the other hand, in phase diagram $H_{\mathrm{ex}}^{x}$-$H_{\mathrm{ex}}^{z}$, the thermal fluctuation does not matter. In addition, the quantum fluctuation is not essential either since Cr$_{1/3}$NbS$_{2}$ has a spin length of $3/2$.

In the second part of this manuscript, we investigated three instabilities of the modulated structures: the surface instability, the inflation instability, and the $H_{0}$ line. The inflation instability is sorted into two types, the soliton and the surface modulation.  
These instabilities were confirmed in the excitation spectrum of the localized mode, and we also added some interpretations of these instabilities using the energy landscapes for proper parameters, such as soliton coordinate and the spin moment at the soliton center. The surface instability is understood as the disappearance of the surface barrier and causes the penetration of solitons. The energy landscape for the surface barrier also gives an importance of the virtual soliton for describing not only the surface modulation itself but also the interaction between the surface modulation and the soliton (Appendix~\ref{sect-int-sol-surf}).  The landscape for the spin moment at the center clarifies the origin of the $H_{0}$ line, which is the unwinding process via the spin motion of the soliton center towards the helical axis. This is also confirmed in the dynamics using the LLG equation and may be regarded as a kind of the Landau instability. The $H_{0}$ line gives the upper bound for the magnetic field where remnants of solitons are observed. The inflation instability, on the other hand, gives the lower bound for the existence of an isolated soliton and can be understood as the instability of the particle picture towards the wave picture.  The winding process of the inflation instability around the surface modulation and/or the isolated soliton is regarded as the inverse process of the unwinding process at the $H_{0}$ line as seen in Fig.~\ref{fig-hb-h0-phase}(c). The inflation instability of the surface modulation is the spreading of the oscillating structure to the bulk, and clearly different from the surface instability.

The tilted magnetic field clarifies the importance of the $z$ component for the defect of the $xy$-spin plane. A finite $z$ component softens the amplitude in the $xy$-spin plane and results in the oscillation of the interaction and the winding/unwinding process important to the low temperature instability. In particular, the field region surrounded by the inflation instability and the $H_{0}$ line gives the sufficient condition for the existence of the soliton, and it is important to the manipulation of an isolated soliton. 

It is not easy to detect the DPT because the hysteresis appears frequently owing to the topological stability of solitons. Instead there are several indirect evidences related to the DPT. The attractive interaction can be confirmed through the observation of the cluster structure of solitons. Related to this, the observation of the soliton bound to the surface can be another clue and the oscillation structure in the tail of the soliton is also a possible clue. We note that the skyrmions forming their cluster or positioning near the edges are actually observed, which support the attractive interaction of skyrmions\cite{Leonov2016,Muller2017,Loudon2018,Du2018}. 

\begin{acknowledgments}
The author would like to thank H.~Tsunetsugu for critical comments, and Y.~Kato for reading the manuscript carefully. 
He also thanks Alex Bogdanov for his introduction of nucleation-type phase transition during his stay in Komaba in Tokyo in early 2017 and thanks M.~Kunimi and R.~Regan for their helpful comments. He  acknowledges support under Japan Society for the Promotion of Science (JSPS) KAKENHI Grants No.~JP16J03224, No.~JP19K14662, No.~JP25220803, and JSPS KAKENHI on Innovative Areas ``Topological Materials Science" (Grant No.~JP15H05855). This work was also supported by Chirality Research Center (Crescent) in Hiroshima University, the Mext program for promoting the enhancement of research universities, Japan, JSPS, Russian Foundation for Basic Research (RFBR) under the Japan - Russian Research Cooperative Program, and JSPS Core-to-Core Program, A. Advanced Research Networks, and
the Program for Leading Graduate Schools, the Ministry of Education, Culture, Sports, Science and Technology, Japan.
\end{acknowledgments} 

\appendix
\section{Landau expansion}\label{sect-landau-expansion}
In this appendix, we show the calculation details of the Landau expansion in the form of $E(\xi)/N_{z} = a_{0} + a_{2} \xi^{2} + a_{4}\xi^{4} + O(\xi^{6})$.
\subsection{Coefficients of $z$ component}
First, we show the coefficients of the $z$ component in Eq.~\eqref{eq-expand-3} by using those of the $x$ and the $y$ component. 
\begin{align}
\sigma_{1,z} &= -\dfrac{M_{\uu,\perp}}{M_{\uu,\parallel}} \sigma_{1,x}, 
\\
\alpha_{M,z} &= -\dfrac{1}{4M_{\uu,\parallel}}\left( 1 + 4\alpha_{M} M_{\uu,\perp} + \dfrac{M_{\uu,\perp}^{2}}{M_{\uu,\parallel}^{2}}\sigma_{1,x}^{2}\right), 
\\
\sigma_{2,z} &= -\dfrac{1}{4M_{\uu,\parallel}} \left(\dfrac{\sigma_{1,x}^{2}}{M_{\uu,\parallel}^{2}} - \sigma_{1,y}^{2} +4M_{\uu,\perp}\sigma_{2,x}\right), 
\\
\sigma_{1,z}^{\prime} &= 
-\dfrac{2\alpha_{M} \sigma_{1,x} + 2\alpha_{M,z}\sigma_{1,z} 
+\sum_{\mu=x,y,z}\sigma_{1,\mu}\sigma_{2,\mu}}{2M_{\uu,\parallel}},
\\
\sigma_{3,z} & = 
-\dfrac{\sigma_{1,x}\sigma_{2,x} -\sigma_{1,y}\sigma_{2,y} + \sigma_{1,z}\sigma_{2,z} + 2M_{\uu,\perp}\sigma_{3,x}}{2M_{\uu,\parallel}},
\\
\beta_{M,z} &= - \dfrac{\alpha_{M}^{2}  +\alpha_{M,z}^{2}+ 2M_{\uu,\perp}\beta_{M} + \sigma_{1,z}\sigma_{1,z}^{\prime} - \sigma_{2}^{2}/2}{2M_{\uu,\parallel}},
\end{align}
where $\sigma_{2}^{2} = \sum_{\mu = x,y,z}\sigma_{2,\mu}^{2}$.

\subsection{Expansion}
Here we expand each term in the energy functional up to the second order with respect to $\xi$ for discussion of $a_{2} = 0$ and write down in the following: 
\begin{widetext}
\noindent the exchange interaction
\begin{align}
-\dfrac{J_{\parallel}}{N_{z}}\sum_{l}\vec{M}_{l}\cdot\vec{M}_{l+1}
&\approx
-J_{\parallel}
\left\{
|\vec{M}_{\uu}|^{2}
+\left[
2(M_{\uu,\perp}\alpha_{M} + M_{\uu,\parallel}\alpha_{M,z}) + \dfrac{1}{2}\left( 1 + \sigma_{1,z}^{2}\right)\cos qa
\right]
\xi^{2}\right.\nonumber\\
&\hspace{-8em}\left.
+\left[
\alpha_{M}^{2} + \alpha_{M,z}^{2} + 2(M_{\uu,\perp} \beta_{M} +M_{\uu,\parallel} \beta_{M,z}) +\dfrac{1}{2}
(\sigma_{2,x}^{2} + \sigma_{2,y}^{2} + \sigma_{2,z}^{2})\cos 2qa+\sigma_{1,z}\sigma_{1,z}^{\prime}\cos qa
\right]\xi^{4}
\right\} \\
&\approx -J_{\parallel}\left[M_{\uu}^{2} -\dfrac{\xi^{2}}{2}\left(1 + \dfrac{M_{\uu,\perp}}{M_{\uu,\parallel}}\sigma_{1,x}^{2}\right)(1-\cos q_{c} a)\right] + O(\xi^{4}),
\end{align}
the DMI
\begin{align}
-\dfrac{D}{N_{z}}\sum_{l}\left(\vec{M}_{l}\times\vec{M}_{l+1}\right)^{z}
\approx - D
\left(
\xi^{2}\sigma_{1,x}\sigma_{1,y}\sin qa + \xi^{4} \sigma_{2,x} \sigma_{2,y} \sin 2qa
\right) \approx -D\xi^{2}\sigma_{1,x}\sigma_{1,y}\sin q_{c}a + O(\xi^{4}),
\end{align}
the hard-axis anisotropy
\begin{align}
\dfrac{K}{2N_{z}}\sum_{l}\left(M_{l}^{z}\right)^{2}
&\approx\dfrac{K}{2}\left[M_{\uu,\parallel}^{2} + \left(2 M_{\uu,\parallel}\alpha_{M,z} + \dfrac{1}{2}\sigma_{1,z}^{2}\right)\xi^{2}
+ \left(\alpha_{M,z}^{2} + 2M_{\uu,\parallel}\beta_{M,z} + \dfrac{1}{2}\sigma_{2,z}^{2}+ \sigma_{1,z}\sigma_{1,z}^{\prime}\right)\xi^{4}\right]\\
&\approx  \dfrac{K}{2}\left[ M_{\uu,\parallel}^{2}-\dfrac{\xi^{2}}{2}\left(1 + 4 \alpha_{M} M_{\uu,\perp}\right)\right]+ O(\xi^{4}),
\end{align}
and the Zeeman coupling
\begin{align}
-\dfrac{1}{N_{z}}\sum_{l}\vec{H}_{\mathrm{ex}}\cdot\vec{M}_{l} 
&\approx -\vec{H}_{\mathrm{ex}} \cdot \vec{M}_{c} 
\approx-\vec{H}_{\mathrm{ex}}\cdot \vec{M}_{\uu} 
+\xi^{2}\left[\dfrac{H_{\mathrm{ex}}^{z}}{4M_{\uu,\parallel}}\left(1 + \dfrac{M_{\uu,\perp}^{2}}{M_{\uu,\parallel}^{2}}\sigma_{1,x}^{2}\right) + \alpha_{M} \left(\dfrac{M_{\uu,\perp}}{M_{\uu,\parallel}}H_{\mathrm{ex}}^{z}-H_{\mathrm{ex}}^{z}\right)\right]+ O(\xi^{4}).
\end{align}
\end{widetext}

\subsection{Derivation of $\mathsf{D}=0$ from $a_{2} = 0$}
Coefficient  $a_{2} $ includes parameters $q_{c} a$ and $\sigma_{1,x}$ to be determined by the stationary condition of $a_{2}$.
The derivatives of $a_{2} $ with respect to $q_{c} a$ and $\sigma_{1,x}$ are, respectively, given using $\sigma_{1,x}^{2} + \sigma_{1,y}^{2} = 1$ by
\begin{align}
\dfrac{\partial a_{2}}{\partial (q_{c} a)}
&= \dfrac{J_{\parallel}}{2M_{\uu,\parallel}^{2} }\left({M_{\uu,\parallel}^{2}} +{M_{\uu,\perp}^{2}\sigma_{1,x}^{2}}\right) \sin q_{c} a \nonumber \\
&- D\sigma_{1,x}\sigma_{1,y} \cos q_{c} a = 0,\label{eq-a2-1}\\
\sigma_{1,y} \dfrac{\partial a_{2}}{\partial \sigma_{1,x}} &=
\left(-2J_{\parallel}\cos q_{c} a + \dfrac{H_{\uu}^{z}}{M_{\uu,\parallel}}  +  K\right)\dfrac{M_{\uu,\perp}^{2}}{2M_{\uu,\parallel}^{2}}\sigma_{1,x}\sigma_{1,y}\nonumber \\
&-D\left(\sigma_{1,y}^{2}- \sigma_{1,x}^{2} \right)\sin q_{c} a = 0. \label{eq-a2-2}
\end{align}
Usually $|q_{c}a| < \pi/2$ for ferromagnetic systems, and $q_{c} a > 0 $ owing to a positive $D$. 
By considering $\sin q_{c} a >0$ and  $\cos q_{c} a > 0$, we see that $\sigma_{1,x}\sigma_{1,y} > 0$ and $\sigma_{1,x}^{2} -\sigma_{1,y}^{2} < 0$ from Eqs.~\eqref{eq-a2-1} and \eqref{eq-a2-2}, respectively. Note  a useful relation $H_{\uu} = H_{\uu}^{x}/M_{\uu,\perp} = H_{\uu}^{z}/M_{\uu,\parallel}$.

Combining $a_{2} = 0$ and Eq.~\eqref{eq-a2-2},  we obtain the following equation
from $2a_{2}\sigma_{1,x} M_{\uu,\parallel}^{2} + M_{\uu,\parallel}^{2} \sigma_{1,y}^{2}\partial a_{2}/\partial \sigma_{1,x} =0$ and $2a_{2}\sigma_{1,y} - \sigma_{1,x}\sigma_{1,y}\partial a_{2}/\partial \sigma_{1,x} = 0$:
\begin{align}
\begin{pmatrix}
2J_{\parallel} \cos q_{c} a - H_{\uu} - K & 2 DM_{\uu,\parallel}^{2}\sin q_{c} a \\
2D\sin q_{c} a & 2J_{\parallel}\cos q_{c} a -H_{\uu}
\end{pmatrix}
\begin{pmatrix}
\sigma_{1,x} \\ \sigma_{1,y}
\end{pmatrix} = \begin{pmatrix} 0 \\ 0 \end{pmatrix}.
\end{align}
The condition for the existence of the nontrivial set of $\sigma_{1,x}$ and $\sigma_{1,y}$ is given by for the null determinant of the coefficient matrix:
\begin{align}
A\cos^{2} q_{c} a  + B\cos q_{c} a  + C = 0, \label{eq-zero-determinant}
\end{align}
where $A$, $B$, and $C$ are defined below Eq.~\eqref{eq-quadratic}, and we find that  
\begin{align}
(\sigma_{1,x},\sigma_{1,y}) = (2J_{\parallel}\cos q_{c} a -H_{\uu}, -2D \sin q_{c} a)/W \label{eq-sigma-xy}
\end{align} with 
\begin{align}W=\sqrt{(2J_{\parallel}\cos q_{c} a -
H_{\uu})^{2} + 4D^{2} \sin^{2} q_{c} a}.
\end{align}
Using Eqs.~\eqref{eq-zero-determinant} and \eqref{eq-sigma-xy}, Eq.~\eqref{eq-a2-1} is reduced to
\begin{align}
&\mp (J_{\parallel} B + A H_{\uu})\dfrac{\sqrt{B^{2}-4AC}}{2A}\nonumber \\
&+\dfrac{J_{\parallel}B^{2}}{2A} + \dfrac{H_{\uu}B}{2} +J_{\parallel}(-C  + J_{\parallel}H_{\uu}^{2} +4D^{2} M_{\uu,\parallel}^{2}) = 0. \nonumber 
\end{align}
Note that $\cos q_{c}a$ is a solution to Eq.~\eqref{eq-zero-determinant}.
The terms other than the first one can be summarized as $J_{\parallel}[B^{2}/(2A) -2C]$.
Therefore using $\mathsf{D} = B^{2}/4 - AC$, we have
\begin{align}
\pm\sqrt{\mathsf{D}}\left[2J_{\parallel}\dfrac{-B \pm 2\sqrt{\mathsf{D}} }{2A} -H_{\uu}\right] 
=\pm\sqrt{\mathsf{D}}\left[2J_{\parallel}\cos q_{c} a -H_{\uu}\right] 
=0.
\end{align}
Noting that $J_{\parallel} \cos q_{c} a -H_{\uu} = J_{\parallel} \cos q_{c} a -H_{\uu}^{x}/M_{\uu,\perp} = - J_{\parallel}(1-\cos q_{c} a) - H_{\mathrm{ex}}^{x}/M_{\uu,\perp} < 0$, we can conclude that $\mathsf{D}  = 0$, i.e., $a_{2} = 0 $ is equivalent to $\mathsf{D} = 0$. This is reasonable because the expansion Eqs.~\eqref{eq-expand-1} and \eqref{eq-expand-3} up to order $\xi$ is the same as the expansion \eqref{eq-asymptotic-form}.

\subsection{Fourth order terms in $\xi^{4}$}
In this subsection, we derive the coefficient $a_{4}$ in the Landau energy. 
Here $\xi$ is a scalar order parameter associated with this phase transition ($a_{2} = 0$), and $a_{4}\xi^{4}$ is the only form of the fourth-order invariants in the Landau expansion.
Note that $q$ has a $\xi^{2}$ term. The fourth order term in each part of the energy is given as follows:
the exchange interaction
\begin{align}
\dfrac{J_{\parallel}}{2}&\Bigl[ \alpha_{q} a (1+\sigma_{1,z}^{2}) \sin q_{c} a +\sum_{\mu = x,y,z}\sigma_{2,\mu}^{2}  (1-\cos 2q_{c} a) \nonumber \\
&\hspace{5em}
+2\sigma_{1,z}\sigma_{1,z}^{\prime}(1-\cos q_{c} a)\Bigr]\xi^{4},
\end{align}
the DMI
\begin{align}
-D\left[ \alpha_{q} a \sigma_{1,x}\sigma_{1,y} \cos q_{c} a 
+
\sigma_{2,x}\sigma_{2,y}\sin 2q_{c}a
\right]\xi^{4},
\end{align}
the hard-axis anisotropy
\begin{align}
\dfrac{K}{2}\left(\alpha_{M,z}^{2} + 2M_{\uu,\parallel} \beta_{M,z} + \dfrac{1}{2}\sigma_{2,z}^{2} +\sigma_{1,z}\sigma_{1,z}^{\prime}\right)\xi^{4},
\end{align}
and Zeeman coupling
\begin{align}
-\left( H_{\mathrm{ex}}^{x} \beta_{M}+ H_{\mathrm{ex}}^{z} \beta_{M,z}  \right)\xi^{4}.
\end{align}
We sum up them and obtain $a_{4}$ using $1-\cos 2q_{c} a = 2\sin^{2}q_{c}a$ and $\sigma_{2}^{2} = \sigma_{2,x}^{2} + \sigma_{2,y}^{2} + \sigma_{2,z}^{2}$ as 
\begin{align}
a_{4} &= J_{\parallel}\left[\left(\sigma_{2,x}^{2} + \sigma_{2,y}^{2}+ \sigma_{2,z}^{2}\right)\sin^{2} q_{c}a  + \sigma_{1,z}\sigma_{1,z}^{\prime} ( 1- \cos q_{c}a)\right] \nonumber \\
&-D \sigma_{2,x} \sigma_{2,y} \sin 2q_{c} a 
 -\dfrac{K}{2}\left(
\alpha_{M}^{2} + \dfrac{\sigma_{2,x}^{2} + \sigma_{2,y}^{2}}{2}
\right) \nonumber \\
&+ \dfrac{H_{\mathrm{ex}}^{z} }{2M_{\uu,\parallel}}\left(\alpha_{M}^{2} + \alpha_{M,z}^{2}  + \dfrac{\sigma_{2}^{2}}{2}  + \sigma_{1,z}\sigma_{1,z}^{\prime}\right). 
\end{align}
\begin{widetext}
The derivative with respect to $\alpha_{M}$ leads to
\begin{align}
\dfrac{\partial a_{4}}{\partial \alpha}
 =  - \left(K  - \dfrac{H_{\mathrm{ex}}^{z}}{M_{\uu,\parallel}^{3}}\right)\alpha_{M} + \dfrac{H_{\mathrm{ex}}^{z}}{M_{\uu,\parallel}}\dfrac{M_{\uu,\perp}}{4M_{\uu,\parallel}^{2}}\left(1 + \dfrac{M_{\uu,\perp}^{2}}{M_{\uu,\parallel}^{2}}\sigma_{1,x}^{2}\right)-\left[
J_{\parallel} (1 - \cos q_{c}a )+ \dfrac{1}{2}\dfrac{H_{\mathrm{ex}}^{z}}{M_{\uu,\parallel}}\right] 
\dfrac{\sigma_{1,x}\sigma_{1,z}}{M_{\uu,\parallel}^{3}}= 0.
\end{align}
The derivatives with respect to $\sigma_{2,x}$ and $\sigma_{2,y}$ are given by 
\begin{align}
&\hspace{2em}\begin{pmatrix}
\dfrac{2J_{\parallel}}{M_{\uu,\parallel}^{2}}\sin^{2}q_{c}a - \dfrac{K}{2} + \dfrac{H_{\mathrm{ex}}^{z}}{2M_{\uu,\parallel}^{3}} & - D\sin 2q_{c}a \\
-D\sin 2q_{c} a  &  2J_{\parallel}\sin^{2}q_{c}a - \dfrac{K}{2} + \dfrac{H_{\mathrm{ex}}^{z}}{2M_{\uu,\parallel}}
\end{pmatrix}
\begin{pmatrix}
\sigma_{2,x} \\\sigma_{2,y}
\end{pmatrix}
=\begin{pmatrix}
y_{x} \\ y_{y}
\end{pmatrix},
 \\
y_{x}&=\left[
-\dfrac{M_{\uu,\perp}}{4M_{\uu,\parallel}^{2}}\left(\dfrac{\sigma_{1,x}^{2}}{M_{\uu,\parallel}^{2}} - \sigma_{1,y}^{2}\right)
\! \!
\left(
2J_{\parallel}\sin^{2}q_{c}a + \dfrac{H_{\mathrm{ex}}^{z}}{2M_{\uu,\parallel}}
\right) +\dfrac{\sigma_{1,x}\sigma_{1,z}}{2M_{\uu,\parallel}^{3}}
\left(
J_{\parallel}(1 - \cos q_{c}a) + \dfrac{H_{\mathrm{ex}}^{z}}{2M_{\uu,\parallel}}
\right)\right],
\\
y_{y}&= \dfrac{\sigma_{1,y}\sigma_{1,z}}{2M_{\uu,\parallel}}
\left(
J_{\parallel}2\sin^{2}q_{c}a + \dfrac{H_{\mathrm{ex}}^{z}}{2M_{\uu,\parallel}}
\right).
\end{align}
\end{widetext}

\section{Numerical calculation}\label{sect-numerics}
We summarize the detail of numerical calculations to solve a set of  Eqs.~\eqref{eq-moment} and \eqref{eq-field} on the finite-sized chain with $N_{z}$ spins to obtain the phase boundary and to study the properties of isolated solitons. We impose either the periodic boundary condition (PBC) or the open boundary condition (OBC)
\begin{align}
\begin{cases}
\vec{M}_{l = N_{z}} = \vec{M}_{l = 0} &\text{PBC}, \\
\vec{M}_{l = N_{z}} = \vec{M}_{l = -1} = 0 & \text{OBC}.
\end{cases}\label{eq-bc}
\end{align}
Numerical calculations are performed in the following iterative manner:
First we choose an initial condition and calculate the effective field from Eq.~\eqref{eq-field}. Then we update the spin configuration by Eq.~\eqref{eq-moment}. When studying an isolated soliton,  we sometimes impose a condition to fix the soliton position at site $l_{\mathrm{s}}$ in this updating process. For multiple isolated solitons, we fix their positions in a similar way when they are well separated.
We repeat to this procedure until achieving a desirable precision. 
Our convergence condition is $\max_{l}|\Delta \vec{M}_{l}| < 5.0\times10^{-15}$ with 
the change of the moment at site $l$ after one iteration, $\Delta \vec{M}_{l}$, except for the case of calculating the phase diagram. In this case, the condition is 
$\max_{l}|\Delta \vec{M}_{l}| < 1.0\times10^{-18}$, and $\Delta \vec{M}_{l}$ is the change after $1000$ iterations.

We choose two methods to fix the soliton position. In one update, we fix the direction of the in-plane component of $\vec{M}_{l = l_{\mathrm{s}}}$ to the $-x$ direction: for $H_{\mathrm{ex}}^{x} > 0$,
\begin{align}
\vec{M}_{l_{\mathrm{s}}} = (-|{H}_{l_{\mathrm{s}}}^{\mathrm{eff},xy}|, 0, H_{l_{\mathrm{s}}}^{\mathrm{eff},z})/ |{H}_{l_{\mathrm{s}}}^{\mathrm{eff}}| \label{eq-fix-condition-1}
\end{align}
with $|{H}_{l_{\mathrm{s}}}^{\mathrm{eff},xy}| = [({H}_{l_{\mathrm{s}}}^{\mathrm{eff},x})^{2} + ({H}_{l_{\mathrm{s}}}^{\mathrm{eff},y})^{2}]^{1/2}$, and in the other update we completely fix the direction of $\vec{M}_{l = l_{\mathrm{s}}}$: 
\begin{align}
\vec{M}_{l_{\mathrm{s}}} = (M_{l_\mathrm{s}}^{x,*}, 0, M_{l_{\mathrm{s}}}^{z,*}) \equiv \vec{M}_{l_{\mathrm{s}}}^{*}~\text{with}~|\vec{M}_{l_{\mathrm{s}}}^{*}|  = 1, \label{eq-fix-condition-2}
\end{align}
where $\vec{M}_{l_{\mathrm{s}}}^{*}$ is a given vector to fix the direction of the moment at the soliton center. In contrast to Eq.~\eqref{eq-fix-condition-2}, Eq.~\eqref{eq-fix-condition-1} imposes the condition that   the $x$ component is nonpositive and does not fix the direction of the moment. 
Note that the solutions obtained under these constraints are not guaranteed to be stable if the constraints are absent. 

We also mention how to choose initial configurations. 
There are a lot of solutions which are stable against all the local and small perturbations when it is related to the nucleation-type phase transition. The different solutions are characterized by different topological indices, although such a stability is lost in some region, for example near the instability-type phase transition. We obtain the most stable state from these states through the comparison of their (free) energies. 
We define the following quantity with the parametrization  $\vec{M}_{l} =  (\cos\varphi_{l}\sin \theta_{l},\sin\varphi_{l}\sin\theta_{l},\cos\theta_{l})$:
\begin{align}
w = \dfrac{1}{2\pi}\sum_{l}  (\varphi_{l+1} - \varphi_{l})~\text{where}~-\pi < \varphi_{l+1} - \varphi_{l} \le \pi,
\end{align}
which is a topological index of this system under the PBC and represents that the winding number stands for how many times in-plane spins rotate about the helical axis in the spin space along the chain direction. 
We choose the following state in which spins rotate with a constant angle as the initial state of iteration process to obtain the solution with  winding number $w$:
\begin{align}
\vec{M}_{l}  = 
(\cos \varphi_{l}, \sin \varphi_{l},0)~\text{with}~\varphi_{l} = 2\pi w l/N_{z}. \label{eq-initial-sinusoidal}
\end{align}
When the field is applied to the perpendicular direction, this single-harmonic state evolves into a state with its higher harmonics but the winding number does not change. Note that this does not necessarily hold. For a field particularly in the region of the phase diagram where $\kappa a$ is pure imaginary, a state with a winding number far from $w_{0} = (2\pi)^{-1}N_{z}\tan^{-1}(D/J_{\parallel})$ may change its winding number to an integer value closer to $w_{0}$. As another case, consider a state with $w=1$ when $H_{\mathrm{ex}}^{x}$ is not very weak. 
For sufficiently large $N_{z}$ compared with an isolated soliton size, 
we could not obtain a well localized state by an iterative evolution from the initial sinusoidal state described by Eq.~\eqref{eq-initial-sinusoidal} with $w=1$. 
In this case, we choose a trial single soliton solution in a different way.
For soliton center $l_{\mathrm{s}}$, one possibility is 
\begin{align}
\varphi_{l}  = \pi \{ 1 + \tanh [(l -l_{\mathrm{s}}) /\Delta l] \},
\end{align}
or under the PBC
\begin{align}
\varphi_{l}  = \pi \{  2+ \tanh [(l - l_{\mathrm{s}}) /\Delta l]+ \tanh [(l -l_{\mathrm{s}}-N_{z}+1)/\Delta l]\}.
\end{align}
Here $\Delta l$ denotes the width of soliton and is taken as, for example of order $50$. This trial state can be generalized to an initial state with $N_{\mathrm{s}}$ solitons with or without the surface configuration which contributes to the noninteger part of $w$ when it exists. The number of solitons $N_{\mathrm{s}}$ is the same as the integer part of $w$. We also use the initial condition
\begin{align}
\vec{M}_{l}  = 
(M_{\uu}^{x}\cos \varphi_{l}, M_{\uu}^{x}\sin \varphi_{l},M_{\uu}^{z})~\text{with}~\varphi_{l} = 2\pi w l/N_{z},
\end{align}
in large $H_{\mathrm{ex}}^{z}/H_{D}$ under the PBC. We have adopted $\vec{M}_{\uu}$ for the coefficients as an example.

\begin{widetext}
\section{Matrix elements of $\mathcal{K}$}\label{sect-matrix-elements}
\begin{description}
\item[Exchange interaction]~\\
For convenience, we use the notations $\cos \theta_{\sss,l}  \rightdef\ccc \theta_{\sss,l}$, $\sin \theta_{\sss,l}  \rightdef\sss \theta_{\sss,l}$, and $\varphi_{\sss,l} -\varphi_{\sss,l+1} \rightdef \Delta\varphi_{\sss,l}$.
The exchange term is transformed using $\tilde{M}$ as
$\vec{M}_{ l }\cdot\vec{M}_{ l + 1 } = \sum_{\mu,\nu=x,y,z}\tilde{M}_{ l }^{\mu} \vec{\tilde{e}}_{l}^{\ \!\!\ \!\mu}\cdot \vec{\tilde{e}}_{l+1}^{\ \!\!\ \!\nu}\tilde{M}_{ l + 1 }^{\mu} $, where
\begin{align}
\vec{\tilde{e}}_{l}^{\ \!\!\ \!\mu}\cdot \vec{\tilde{e}}_{l+1}^{\ \!\!\ \!\nu}
=
\begin{pmatrix}
\cos \Delta\varphi_{\sss,l}&
\ccc \theta_{\sss,l+1} \sin \Delta\varphi_{\sss,l} &
-\sss \theta_{\sss,l+1} \sin \Delta\varphi_{\sss,l}\\
-\ccc \theta_{\sss,l} \sin \Delta\varphi_{\sss,l} & 
\ccc \theta_{\sss,l}\ccc\theta_{\sss,l+1} \cos\Delta\varphi_{\sss,l}+\sss \theta_{\sss,l}\sss\theta_{\sss,l+1} &
-\ccc \theta_{\sss,l}\sss \theta_{\sss,l+1}\cos\Delta\varphi_{\sss,l} + \sss \theta_{\sss,l}\ccc \theta_{\sss,l+1} \\
\sss \theta_{\sss,l} \sin \Delta\varphi_{\sss,l} &
-\sss \theta_{\sss,l}\ccc \theta_{\sss,l+1}\cos \Delta \varphi_{\sss,l}+ \ccc\theta_{\sss,l}\sss \theta_{\sss,l+1} &
\sss \theta_{\sss,l}\sss\theta_{\sss,l+1} \cos\Delta\varphi_{\sss,l}+\ccc \theta_{\sss,l}\ccc\theta_{\sss,l+1}
\end{pmatrix}^{\mu\nu}.
\end{align}

\item[Dzyaloshinskii--Moriya interaction]~\\
The second term is written as $\vec{e}^{\ \!\!\ \!z}\cdot(\vec{M}_{ l }\times\vec{M}_{ l + 1 }) 
= \sum_{\mu,\nu=x,y,z}\tilde{M}_{ l }^{\mu}
  [\vec{e}^{z}\cdot(\vec{\tilde{e}}_{l}^{\ \!\!\ \!\mu}\times
\vec{\tilde{e}}_{l+1}^{\ \!\!\ \!\nu})^{z}]\tilde{M}_{ l + 1 }^{\nu}
$ and we calculate the matrix element as follows:
\begin{align}
 [(\vec{\tilde{e}}_{l}^{\ \!\!\ \!\mu}\times
\vec{\tilde{e}}_{l+1}^{\ \!\!\ \!\nu})^{z}]
=\begin{pmatrix}
-\sin\Delta\varphi_{\sss,l} & 
\ccc \theta_{\sss,l+1} \cos\Delta\varphi_{\sss,l}&
-\sss \theta_{\sss,l+1} \cos\Delta\varphi_{\sss,l}\\
-\ccc \theta_{\sss,l} \cos\Delta\varphi_{\sss,l}&
-\ccc\theta_{\sss,l}\ccc\theta_{\sss,l+1}\sin\Delta\varphi_{\sss,l} &
\ccc \theta_{\sss,l}\sss \theta_{\sss,l+1} \sin\Delta\varphi_{\sss,l}\\
\sss \theta_{\sss,l} \cos\Delta\varphi_{\sss,l}&
\sss \theta_{\sss,l}\ccc \theta_{\sss,l+1} \sin\Delta\varphi_{\sss,l}&
-\sss\theta_{\sss,l}\sss\theta_{\sss,l+1}\sin\Delta\varphi_{\sss,l} 
\end{pmatrix}^{\mu\nu}.
\end{align}
\item[Zeeman coupling]~\\
The third term is given by $\vec{H}_{\mathrm{ex}}\cdot \vec{M}_{ l } = \sum_{\mu}\vec{H}_{\mathrm{ex}}\cdot \vec{\tilde{e}}_{l}^{\ \!\!\ \!\mu}\tilde{M}_{ l }^{\mu}\to \vec{H}_{\mathrm{ex}}\cdot\vec{\tilde{e}}_{l}^{\ \!\!\ \!z} \tilde{M}_{ l }^{z}$. In the final transformation, we retain the term contributing the equilibrium state energy and the second order expansion. 
\begin{align}
 \vec{H}_{\mathrm{ex}}\cdot\vec{\tilde{e}}_{l}^{\ \!\!\ \!z}
 =H_{\mathrm{ex}}^{x}\cos\varphi_{\sss,l}\sss\theta_{\sss,l}
 +H_{\mathrm{ex}}^{z}\ccc\theta_{\sss,l}.
\end{align}
\item[Anisotropy]~\\
The fourth term is given by $(\vec{M}_{ l }\cdot \vec{e}^{\ \!\!\ \!z} )^{2}= \sum_{\mu,\nu=x,y,z}\tilde{M}_{ l }^{\mu} (\vec{\tilde{e}}_{l}^{\ \!\!\ \!\mu}\cdot\vec{e}^{\ \!\!\ \!z})(\vec{\tilde{e}}_{l}^{\ \!\!\ \!\nu}\cdot\vec{e}^{\ \!\!\ \!z})\tilde{M}_{ l }^{\nu}$.
\begin{align} 
(\vec{\tilde{e}}_{l}^{\ \!\!\ \!\mu}\cdot\vec{e}^{\ \!\!\ \!z})(\vec{\tilde{e}}_{l}^{\ \!\!\ \!\nu}\cdot\vec{e}^{\ \!\!\ \!z})
=\begin{pmatrix}
0 & 0 & 0\\
0 & 
\sss^{2} \theta_{\sss,l}&
\sss \theta_{\sss,l}\ccc \theta_{\sss,l}\\
0 &
\sss \theta_{\sss,l}\ccc \theta_{\sss,l}&
\ccc^{2} \theta_{\sss,l}
\end{pmatrix}.
\end{align}
\end{description}
We summarize the above expressions.
Defining $\tilde{J} = \sqrt{J_{\parallel}^{2} + D^{2}}$ and $\tan\alpha_{0}= D/J_{\parallel}$, we  obtain the explicit forms of $E$ and $\mathcal{K}$ in Eq.~\eqref{eq:hessian} as follows:
\begin{align}
E(\{\varphi_{\sss,l}\},\{\theta_{\sss,l}\})
=-N_{2\dd}\sum_{l}&\Biggl\{\tilde{J}(
\sss \theta_{\sss,l}\sss\theta_{\sss,l+1} \cos(\Delta\varphi_{\sss,l} + \alpha_{0} )+J_{\parallel}\ccc \theta_{\sss,l}\ccc\theta_{\sss,l+1}
\nonumber\\
&\hspace{5em}
+H_{\mathrm{ex}}^{x}\cos\varphi_{\sss,l}\sss\theta_{\sss,l}
+H_{\mathrm{ex}}^{z}\ccc\theta_{\sss,l}
-\dfrac{K}{2}\ccc^{2}\theta_{\sss,l}
+ (J_{x}+J_{y})\Biggr\} 
\end{align}
and 
\begin{align}
\mathcal{K}_{l,l+1}^{xx}
&=-\tilde{J}\cos(\Delta\varphi_{\sss,l} + \alpha_{0} ) \\
\mathcal{K}_{l,l+1}^{xy}
&=-\tilde{J}\ccc\theta_{\sss,l+1}\sin(\Delta\varphi_{\sss,l} + \alpha_{0} )\\
\mathcal{K}_{l,l+1}^{yx}
&=+\tilde{J}\ccc\theta_{\sss,l}\sin(\Delta\varphi_{\sss,l}+\alpha_{0} )\\
\mathcal{K}_{l,l+1}^{yy}
&=-\tilde{J}
\ccc \theta_{\sss,l}\ccc\theta_{\sss,l+1}\cos(\Delta\varphi_{\sss,l}+\alpha_{0} ) -J_{\parallel}\sss \theta_{\sss,l}\sss\theta_{\sss,l+1}\\
\mathcal{K}_{l,l}^{xx}
&=
\tilde{J}
\sss \theta_{\sss,l}[\sss\theta_{\sss,l+1} \cos(\Delta\varphi_{\sss,l}+\alpha_{0} )
+\sss \theta_{\sss,l-1}\cos(\Delta\varphi_{\sss,l-1}+\alpha_{0} )]
+J_{\parallel}
\ccc \theta_{\sss,l}(\ccc\theta_{\sss,l+1}
+\ccc \theta_{\sss,l-1}) 
\nonumber \\
&+H_{\mathrm{ex}}^{x}\cos\varphi_{\sss,l}\sss\theta_{\sss,l}
+H_{\mathrm{ex}}^{z}\ccc\theta_{\sss,l} - K \ccc^{2}\theta_{\sss,l}\\
\mathcal{K}_{l,l}^{yy}
&=
\tilde{J}
\sss \theta_{\sss,l}[\sss\theta_{\sss,l+1} \cos(\Delta\varphi_{\sss,l}+\alpha_{0} )
+\sss \theta_{\sss,l-1}\cos(\Delta\varphi_{\sss,l-1}+\alpha_{0} )]
+J_{\parallel}
\ccc \theta_{\sss,l}(\ccc\theta_{\sss,l+1}
+\ccc \theta_{\sss,l-1})
\nonumber \\
&+H_{\mathrm{ex}}^{x}\cos\varphi_{\sss,l}\sss\theta_{\sss,l}
+H_{\mathrm{ex}}^{z}\ccc\theta_{\sss,l} - K (\ccc^{2}\theta_{\sss,l}-\sss^{2}\theta_{\sss,l}).
\end{align}
The other components are zero. 
\end{widetext}

\section{Interaction between soliton and surface}\label{sect-int-sol-surf}
\begin{figure}[t]
\begin{center}
\includegraphics[width = 0.95\hsize]{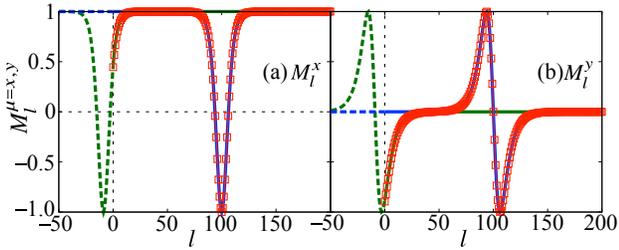}
\caption{Comparison between the spin profiles of Method~I and Method~II for $H_{\mathrm{ex}}^{x}= 0.8 H_{D}$.  The left panel (a) shows the $x$ component, while the right panel (b) does the $y$ component. Note that $M_{l}^{z}= 0$ since we set $H_{\mathrm{ex}}^{z}=0$. $l_{\mathrm{s}} = -9$ is a point which gives the local minimum of the soliton energy as shown in Fig.~\ref{fig-surface-barrier-1}(b). The red squares stand for the $x$ and $y$ components of the spin profile obtained using Method~II when the soliton center is fixed at $l_{\mathrm{s}} = 100$. The blue and green curves are the spin profiles with $l_{\mathrm{s}} = 100$ and  $-9$, respectively, using Method~I. The dashed curves describe the virtual spin structures outside the system.}
\label{fig-spin-profile}
\end{center}
\end{figure}
Here, we discuss the interaction between the soliton and the surface modulation in detail. 
We introduce two methods for study of the energy barrier.
In Refs.~\onlinecite{Shinozaki2018,Masaki2018}, the energy of an isolated soliton is calculated as follows: (i) For a given field value, we first construct an isolated soliton solution with its center at $l_{\mathrm{s}} = 0$ for either an infinite system or sufficiently large system with PBC. (ii) Let the obtained spin profiles be $\vec{M}_{l,1}$. Then we describe the soliton profile with the center at the position $l_{\mathrm{s}}$ by $\vec{M}_{l} = \vec{M}_{l-l_{\mathrm{s}},1}$. (iii) Finally we evaluate the energy of the system with open boundary at $l = 0$ by summing up in Eq.~\eqref{eq-energy-chain} for $l \ge 0$ and subtracting the energy of the uniform state. We write the energy in this scheme as $E_{1}^{\mathrm{OB}}$ and  call this calculation scheme Method~I. The advantage of this method is that we can consider the case of the soliton outside the system ($l_{\mathrm{s}} < 0 $), and also the case of a single soliton for magnetic field lower than the barrier field $H_{\bb}$. The disadvantage of this method is that the solution does not satisfy the boundary condition, and we cannot describe the surface modulation which appears also when a soliton exists in the system. 

Instead, we directly construct an isolated soliton solution using the OBC.  The system in this case has a surface twisted structure for the field $H > H_{\bb}$. We fix the position of the soliton center at $l = l_{\mathrm{s}}$ by imposing the condition that $\varphi_{l_{\mathrm{s}}} = \pi$ in updating. Then we calculate the energy of this system using Eq.~\eqref{eq-energy-chain} by summing up from $l=0$ to $l \le N_{z}/2$ to eliminate the effects of the opposite surface. We call this method Method~II. 
This method is more realistic to consider surface effects, since the surface twisted structure is present and interacts with the soliton in the system. On the other hand, we cannot consider negative values of $l_{\mathrm{s}}$, and the energy profile of a single soliton for the field lower than the barrier field. This type of calculation was also done by Iwasaki {\it et al.}\cite{IwasakiMochizukiNagaosa2013a}

\begin{figure*}[t]
\begin{center}
\includegraphics[width = 0.95\hsize]{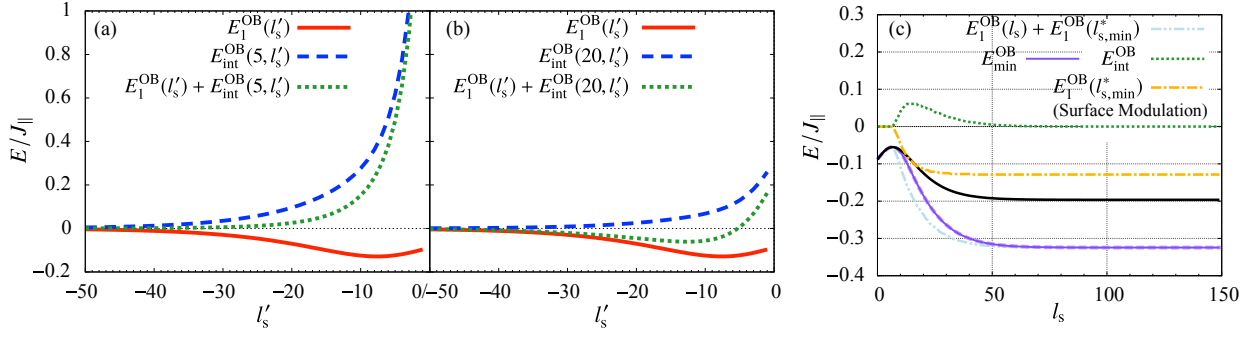}
\caption{Energies of the virtual soliton as functions of its center $l_{\mathrm{s}}^{\prime}$ when an isolated soliton exists at (a) $l_{\mathrm{s}} = 5$ and (b) $20$. The solid curves represent the energy owing to the tail of the virtual soliton, while the broken curves do its interaction with the isolated soliton with $l_{\mathrm{s}} = 5$ or $20$. The dotted curves are the sum of these two energies.
(c) Position ($l_{\mathrm{s}}$) dependence of several energies of the isolated soliton when the virtual soliton is fixed at $l_{\mathrm{s, min}}^{*}$. The black solid curve and the purple broken curve are the same as in Fig.~\ref{fig-surface-barrier-1}(b) at $(H_{\mathrm{ex}}^{x}, H_{\mathrm{ex}}^{z}) = (0.4, 0) H_{D}$. See the text for other curves. 
}
\label{fig-E-comp}
\end{center}
\end{figure*}

For the following discussion, let us consider the chain with $2N_{z} + 1$ sites, where $l = -N_{z}, \cdots, N_{z}$ with PBC $\vec{M}_{l = - N_{z} - 1 } = \vec{M}_{l=N_{z}} $. We consider the two-soliton configuration; the soliton centers are set to $l_{\mathrm{s}} \ge 0$ and $l_{\mathrm{s}}^{\prime} < 0$. For convenience, we call the soliton with its center at $l_{\mathrm{s}}^{\prime}$ the virtual soliton. We define the energy for the state with winding number $w$ in $[-N_{z},N_{z}]$, $E_{w}^{\mathrm{OB}}$ by taking the summation of Eq.~\eqref{eq-energy-chain} only for $0 \le l \le N_{z}$, which is measured from $E_{\uu}$. In the present case, $E_{2}^{\mathrm{OB}}$ is a function of $l_{\mathrm{s}}$ and $l_{\mathrm{s}}^{\prime}$. By fixing $l_{\mathrm{s}}$, and minimizing $E_{2}^{\mathrm{OB}}$ with respect to $l_{\mathrm{s}}^{\prime} < 0 $, we obtain the energy of the soliton interacting with the surface modulation, $E_{2}^{\mathrm{OB}}(l_{\mathrm{s}}) \equiv \min_{l_{\mathrm{s}}^{\prime}<0} E_{2}^{\mathrm{OB}}(l_{\mathrm{s}},l_{\mathrm{s}}^{\prime})$.

First we confirm that the surface modulation can be described by the virtual soliton and the interaction can be neglected when the soliton is far from the surface.
We show a spin profile at $H_{\mathrm{ex}}^{x} = 0.8H_{D}$ in Fig.~\ref{fig-spin-profile}, in which the left (right) panel shows the $x(y)$ component. Open squares are obtained using Method~II by imposing a soliton with  $l_{\mathrm{s}} = 100$. On the other hand, green and blue dashed curves are the profiles obtained by setting the center of the soliton which is constructed under PBC, to $l_{\mathrm{s}} =  -9 \simeq l_{\mathrm{s},\min}$ and $l_{\mathrm{s}}=100$. We use dashed curves for $l < 0 $ and solid curves for $l\ge 0$. First of all, the tail part of the soliton with $l_{\mathrm{s}} = l_{\mathrm{s},\min}$ coincides with the surface modulation inside the system very well. It is confirmed that the surface modulation is described as a soliton that virtually exists at $l_{\mathrm{s},\min}$, which gives the local minimum of the energy profile outside the system. Second, the intersoliton distance is sufficiently large, and the soliton inside the system is not affected by the surface. Hence the soliton with $l_{\mathrm{s}} = 100$ is  well-described by $\vec{M}_{l-l_{\mathrm{s}},1}$, and the total energy is given by the sum of the energies of the surface modulation and the isolated soliton: $E_{\min}^{\mathrm{OB}}(l_{\mathrm{s}}) \simeq \min_{l_{\mathrm{s}}^{\prime}}E_{1}^{\mathrm{OB}}(l_{\mathrm{s}}^{\prime}) + E_{1}^{\mathrm{OB}}(l_{\mathrm{s}}) = E_{1}^{\mathrm{OB}}(l_{\mathrm{s},\min}) + E_{1}^{\mathrm{OB}}(l_{\mathrm{s}})$.

Then we consider the case where an isolated soliton is near the surface. The soliton center inside the system is denoted by $l_{\mathrm{s}}$ and $H_{\mathrm{ex}}^{x}/H_{D} = 0.4$.  Figures~\ref{fig-E-comp}(a) and \ref{fig-E-comp}(b) show, for $l_{\mathrm{s}} = 5$ and $20$, respectively, the energies, $E_{1}^{\mathrm{OB}}$, $E_{\mathrm{int}}^{\mathrm{OB}}(l_{\mathrm{s}},l_{\mathrm{s}}^{\prime})$, and the sum of these two, as a function of the virtual-soliton center $l_{\mathrm{s}}^{\prime} < 0$. The first two energies are, respectively, the energy of the virtual soliton when the soliton is absent in the system, and the two-soliton interaction energy when the centers of two solitons are at $l_{\mathrm{s}} > 0$ and $l_{\mathrm{s}^{\prime}} < 0$. The interaction energy is defined by $E_{2}^{\mathrm{OB}}(l_{\mathrm{s}},l_{\mathrm{s}}^{\prime}) - E_{1}^{\mathrm{OB}}(l_{\mathrm{s}}) - E_{1}^{\mathrm{OB}}(l_{\mathrm{s}}^{\prime})$. $E_{1}^{\mathrm{OB}}(l_{\mathrm{s}}^{\prime})$ has the minimum structure at $l_{\mathrm{s}}^{\prime}\sim -8$ which corresponds to the surface modulation.  When we take account of the interaction, such a  minimum structure disappears in the total energy when $l_{\mathrm{s}} = 5$, while it shifts to the negative side $l_{\mathrm{s}}^{\prime} <  -8$ when $l_{\mathrm{s}} = 20$. Figure~\ref{fig-E-comp}(c) shows the energies as functions of the soliton center $l_{\mathrm{s}} > 0$. The energies are minimized with respect to $l_{\mathrm{s}}^{\prime}$ for each $l_{\mathrm{s}}$, and $l_{\mathrm{s}}^{\prime}$ determined by the minimization is denoted by $l_{\mathrm{s},\min}^{*}$. Here the solid black and dashed purple curves are the same as those shown in Fig.~\ref{fig-surface-barrier-1}(b) at $H_{\mathrm{ex}}^{x} = 0.4 H_{D}$. The solid purple curve labeled $E_{\min}^{\mathrm{OB}}(l_{\mathrm{s}}) = \min_{l_{\mathrm{s}}^{\prime}< 0}E_{2}^{\mathrm{OB}}(l_{\mathrm{s}},l_{\mathrm{s}}^{\prime})$, is in remarkably good agreement with the dashed purple curve. 
The soliton interacting with the surface modulation, which is obtained by Method~II, 
is well described by using the soliton with its center $l_{\mathrm{s}}$ and the virtual soliton with $l_{\mathrm{s}}^{\prime}$.
We decompose $E_{\min}^{\mathrm{OB}}$ into 
$E_{1}^{\mathrm{OB}}(l_{\mathrm{s}}) + E_{1}^{\mathrm{OB}}(l_{\mathrm{s},\min}^{*}) + E_{\mathrm{int}}^{\mathrm{OB}}(l_{\mathrm{s}},l_{\mathrm{s},\min}^{*})$ to see further details. They are indicated in Fig.~\ref{fig-E-comp}(c) by the black-solid, the yellow-dotted-dashed, and the green-dotted curve, respectively. The light blue curve, which is the sum of the yellow and the black curve, is also shown for reference. From Fig.~\ref{fig-E-comp}(c), we see that the interaction effects appear in the range of  $8 \lesssim l_{\mathrm{s}} \lesssim 50$; the gain due to the binding energy of the virtual soliton near the surface is smaller than the loss due to the interaction energy for $l_{\mathrm{s}} \lesssim 8$ as seen in panel (a). Hence the virtual soliton is pushed to $-\infty$ and the surface modulation is absent. When the soliton is deeply inside the system ($l_{\mathrm{s}} \gtrsim 50$), the interaction effects are not important, and $l_{\mathrm{s,min}}^{*} \simeq l_{\mathrm{s,min}}$.  The energy gain appears from $l_{\mathrm{s}}\sim 8$ and saturates to $E_{\min}$ for $l_{\mathrm{s}} \to \infty$ as shown by the yellow curve in panel (c). 
Since the local maximum value $E_{\max}$ of the black curve is not affected very much by the surface modulation, and thus the surface barriers of the virtual soliton and the soliton deeply inside the system are evaluated, respectively, by $E_{\max} - E_{\min}$ and $E_{\max} - E_{1}^{\mathrm{OB}}(l_{\mathrm{s,\min}}^{*}) =  - 2 E_{\min} > 0$. This analysis also demonstrates the validity to study the surface instability without considering the surface modulation, i.e., by Method~I.

\begin{figure*}[t]
\begin{center}
\includegraphics[width = 0.95\hsize]{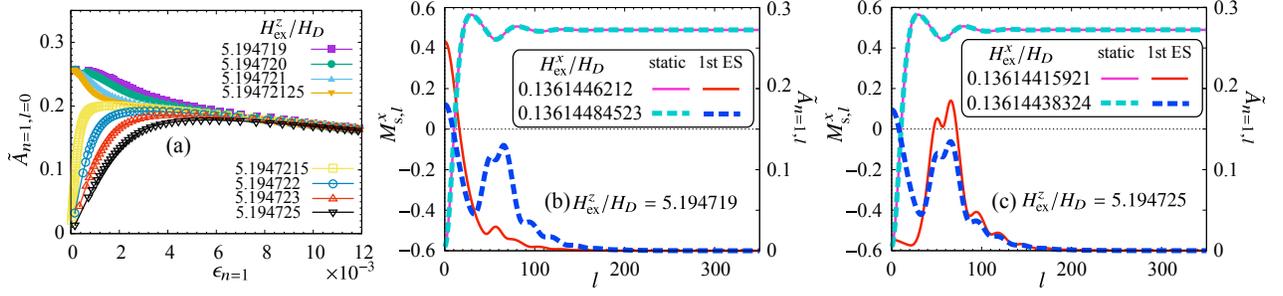}
\caption{(a) The amplitudes of the spin wave at the origin as a function of its eigenenergy. The different symbols represent the different values of $H_{\mathrm{ex}}^{z}/H_{D}$. (b)(c) Spatial profiles of $M_{\mathrm{s},l}^{x}$ on the left-side axis and $\tilde{A}_{n=1,l}$ on the right side axis for (b) $H_{\mathrm{ex}}^{z}/H_{D} = 5.194719$ and (c) $5.194725$. The values of $H_{\mathrm{ex}}^{x}/H_{D}$ are shown in the panels.}
\label{fig-inflation3}
\end{center}
\end{figure*}
\section{Boundary between inflation instability and surface instability}\label{sect-turning-point}
In this appendix, we investigate the boundary between the inflation instability and the surface instability.
The instability character for the surface modulation changes from the entry of the soliton to its size inflation with increasing $H_{\mathrm{ex}}^{z}$. In both cases, the number of the low energy state is one, but its spatial profile changes. In order to check this change, we have a look at the behavior of the amplitude at the origin. Figure~\ref{fig-inflation3}(a) shows the amplitude for the first excited state, $\tilde{A}_{n=1,l=0}$, for several values of $H_{\mathrm{ex}}^{z}/H_{D}$ as a function of the eigenenergy, where we practically change $H_{\mathrm{ex}}^{x}/H_{D}$ instead of specifying the eigenenergy in the calculation. In the limit of $\epsilon_{n=1} \to 0$, the solid symbols go to the finite value of $\tilde{A}_{n=1,l=0}$, while the open symbols seem to become zero. It is remarkable that for relatively large excitation energies $\epsilon_{n} \ge 5 \times10^{-3}$, the amplitudes
$\tilde{A}_{n=1,l=0}$ for different $H_{\mathrm{ex}}^{z}/H_{D}$ take similar values, and whether they evolve or go to zero for $\epsilon_{n} \to 0$ depends on $H_{\mathrm{ex}}^{z}/H_{D}$. We see from Fig.~\ref{fig-inflation3}(c) the evolution of $\tilde{A}_{n=1,l\sim60}$ in addition to $\tilde{A}_{n=1,l=0} \to 0$; the amplitude $\tilde{A}_{n=1,l}$ 
evolves around the region where the instability occurs. 
Figures~\ref{fig-inflation3}(b) and \ref{fig-inflation3}(c) demonstrate this property for $H_{\mathrm{ex}}^{z}/H_{D} = 5.194719$ and $5.194725$, respectively. In these panels, the curves labeled ``static'' show the spatial profiles of $M_{\mathrm{s},l}^{x}$ for the static state to calculate the excited state. We remark that their differences are fairly slight.  The smaller value of $H_{\mathrm{ex}}^{x}/H_{D}$ in each panel stands for the case where the applied field is closer to the instability field. For larger $H_{\mathrm{ex}}^{x}/H_{D}$, the spatial profiles of $\tilde{A}_{n=1,l}$ indicated by the broken blue curves in Figs.~\ref{fig-inflation3}(b) and \ref{fig-inflation3}(c) are similar. 
The field decrease makes clear the difference between these first excited states: In Fig.~\ref{fig-inflation3}(b), 
the amplitude at around the origin increases and the peak structure of $\tilde{A}_{n=1,l}$ around $l\sim 60$ decreases, while in Fig.~\ref{fig-inflation3}(c), 
the opposite behavior can be seen.

The turning point from the surface instability to the inflation instability of the surface modulation seems to exist between $H_{\mathrm{ex}}^{z}/H_{D} = 5.19472125$ and $5.1947215$ in the panel~Fig.~\ref{fig-inflation3}(a). 
As mentioned in Sect.~\ref{sect-div-inst} (see the inset of Fig.~\ref{fig-inflation1}(a)), $\vec{H}_{\mathrm{inf,sol}}$ crosses $\vec{H}_{\bb}$. We remark  the instability occurs by the following mechanism for $H_{\mathrm{ex}}^{z}$ such that $H_{\mathrm{inf,sol}}^{x} > H_{\bb}^{x}$. In the presence of the surface modulation only, first the surface instability occurs and a soliton starts penetrating into the system at $H_{\mathrm{ex}}^{x} = H_{\mathrm{b}}^{x}$. Following this instability, the inflation instability of the soliton occurs before the whole structure of the soliton appearing in the system.

\begin{figure}[t]
\begin{center}
\includegraphics[width = \hsize]{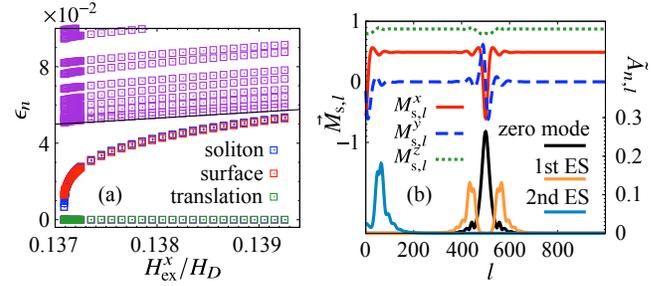}
\caption{(a) Energy spectra at $H_{\mathrm{ex}}^{z}/H_{D} = 5.2$. There are the surface modulation and an isolated soliton at $l = 500$. The blue (red) symbols show the eigenenergies of the bound state around the soliton (surface). The black curve shows the lowest eigenvalue of the excited states for the completely uniform state. (b) Spatial profiles of $\vec{M}_{\mathrm{s},l}$ on the left axis and $\tilde{A}_{n=0,1,2,l}$ on the right axis. $n = 0$, $1$, $2$ represent the zero mode, and the first and the second excited state, respectively. The magnetic field is set  to  $(H_{\mathrm{ex}}^{x}, H_{\mathrm{ex}}^{z})  =  (0.137088, 5.2)H_{D}$.}
\label{fig-inflation-surface-soliton}
\end{center}
\end{figure}
\section{Inflation instability in the presence of  surface modulation and isolated soliton}\label{sect-inf-surface-soliton}
In this appendix, we discuss the case where both the surface modulation and the isolated soliton are present in the static spin profile. These structures are sufficiently distant from each other: $l_{\mathrm{s}} = 500$. We set $H_{\mathrm{ex}}^{z}/H_{D} = 5.2$, for which $H_{\mathrm{inf,sol}}^{x} > H_{\mathrm{inf,sur}}^{x}$, as seen in Fig.~\ref{fig-inflation1}(b). The static profile can be obtained using Eqs.~\eqref{eq-bc} and \eqref{eq-fix-condition-1} and is shown in Fig.~\ref{fig-inflation-surface-soliton}(b) on the left axis at $(H_{\mathrm{ex}}^{x}, H_{\mathrm{ex}}^{z})  =  (0.137088, 5.2)H_{D}$. The dependence of the energy spectrum on $H_{\mathrm{ex}}^{x}/H_{D}$ is shown in Fig.~\ref{fig-inflation-surface-soliton}(a). We see that the three localized modes exist below the black solid curve, which represents the bottom edge of the extended modes in the bulk. The curve is obtained by calculating the lowest eigenvalues for the completely uniform state $\vec{M}_{\mathrm{s},l} = \vec{M}_{\mathrm{u}}$ under the PBC, as
$\omega = \min_{k} [-K_{1,k}^{yx} +
(K_{0}^{xx} + K_{1,k}^{xx})^{1/2}(K_{0}^{yy} + K_{1,k}^{yy} )
^{1/2}]$. Here $k$ is a wave number and $K_{1,k}^{yx} = 2D M_{\uu,\parallel}^{2}\sin ka$, $K_{1,k}^{xx} = -2J_{\parallel}\cos ka$, $K_{1,k}^{yy} = -2J_{\parallel}\cos ka$, $K_{0}^{xx} = 2J_{\parallel} + \vec{M}_{\uu}\cdot \vec{H}_{\mathrm{ex}} - K M_{\uu,\parallel}^{2}$, and $K_{0}^{yy} = 2J_{\parallel} + \vec{M}_{\uu}\cdot \vec{H}_{\mathrm{ex}} - K (M_{\uu,\parallel}^{2} - M_{\uu,\perp}^{2})$.
The eigenstates indicated by the blue, the red, and the green symbols stand for the localized modes around the soliton and the surface, and the translation of the soliton, respectively. In particular the translation mode is the zero mode. 
It depends on $H_{\mathrm{ex}}^{x}/H_{D}$ whether the first excited state with nonzero excitation energy is the localized mode around the soliton or the surface. The amplitudes of the spatial profiles for these localized modes are shown in Fig.~\ref{fig-inflation-surface-soliton}(b) for the right axis at $(H_{\mathrm{ex}}^{x}, H_{\mathrm{ex}}^{z})  =  (0.137088, 5.2)H_{D}$, where the bound state around the soliton has lower energy than that around the surface. For $H_{\mathrm{ex}}^{x}/H_{D} \gtrsim 0.13715$, the bound state around the surface is the first excited state. Both of the first and the second excited states scarcely have their amplitudes at each modulation center, which means they are inflation-type excited states. 
The translation mode becomes unstable at $\vec{H}_{\mathrm{ex}} = \vec{H}_{0}$ via the sliding motion of the soliton.

\begin{figure}[t]
\begin{center}
\includegraphics[width = 0.95\hsize]{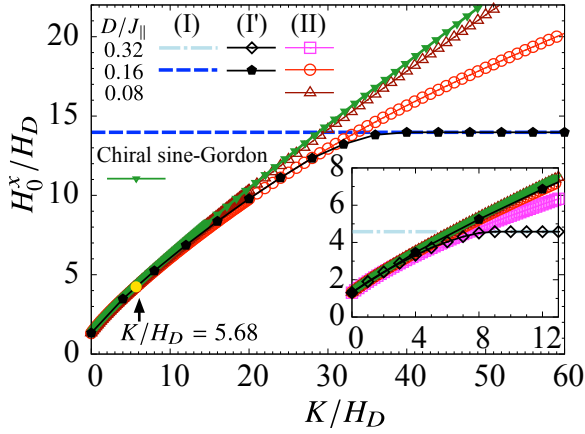}
\caption{
	$K$ dependence of $H_{0}^{x}$ at $H_{\mathrm{ex}}^{z} = 0$ for $D/J_{\parallel} = 0.16$ and $0.08$. 
	We add the data for $D/J_{\parallel} = 0.32$ in the inset, which shows the small anisotropy region.
	The scheme and the values of the DMI are shown in the main plot.  
	The values labeled ``(I$^{\prime}$)'' are obtained by the same method as for $H_{0}$ line~(I)
	but with nonzero $H_{\mathrm{ex}}^{z}/H_{D}= 10^{-10}$.
	The solid circle is for the realistic value of $K/H_{D}$ of Cr$_{1/3}$NbS$_{2}$ with $D/J_{\parallel} = 0.16$. 
	The green inverted triangles show $H_{0}^{x}$ using the chiral sine-Gordon model.
}
\label{fig-csg-h0}
\end{center}
\end{figure}
\section{Continuum limit and $H_{0}$ line due to size narrowing}\label{sect-h0-Kdep-csg}
In the case of $H_{\mathrm{ex}}^{z} = 0$, the chiral sine-Gordon model is also useful because we have analytic solutions for static states. The model is given by 
\begin{align}
E&=J_{\parallel} S^{2} a \int \dd z \left[ 
\dfrac{1}{2}\left(\dfrac{\dd \theta}{\dd z}\right)^{2}
+\dfrac{1}{2}\sin^{2}\theta\left(\dfrac{\dd \varphi}{\dd z}\right)^{2}
-Q_{0} \sin^{2}\theta \dfrac{\dd \varphi}{\dd z} \right.\nonumber \\
&\hspace{6em}\left.
- m^{2}\sin \theta \cos \varphi +\dfrac{\gamma^{2}}{2}\cos^{2}\theta
\right].
\end{align}
Here $\vec{M}(z) = S(\cos\varphi(z)\sin\theta(z),\sin\varphi(z)\sin\theta(z),\cos\theta(z))$, with spin length $S$.  Parameters $Q_{0}$ and $m^{2}$ are, respectively, the magnitude of the DMI and the magnetic field. They are related to the parameters of the lattice model as $Q_{0} = D/(J_{\parallel}a)$ and $m^{2} = H_{\mathrm{ex}} /(J_{\parallel}Sa^{2})$, respectively. 
The last term describes the hard axis anisotropy along the helical axis: $\gamma^{2} = K/(J_{\parallel}a^{2})$.
Small deviations from the static solution $\varphi_{\mathrm{s}}(z)$ and $\theta_{\mathrm{s}} = \pi/2$ are denoted by  $\varphi^{\prime}(z)$ and $\theta^{\prime}(z)$, respectively.  The energy functional is expanded up to second order in $\theta^{\prime}$ and $\varphi^{\prime}$ as follows:
\begin{align}
E &\approx E_{0} + J_{\parallel}S^{2}a \int \dd z\left\{\dfrac{1}{2}\left( \dfrac{\dd \theta^{\prime}}{\dd z}\right)^{2} + \dfrac{1}{2} \left(\dfrac{\dd\varphi^{\prime}}{\dd z}\right)^{2} +\dfrac{\gamma^{2}}{2}\theta^{\prime2}\right.\nonumber \\
&\hspace{-2em}\left.-\left[\dfrac{1}{2}\left(\dfrac{\dd \varphi_{\mathrm{s}}}{\dd z}\right)^{2} -Q_{0}\dfrac{\dd \varphi_{\mathrm{s}}}{\dd z} \right] \theta^{\prime2}
+\dfrac{m^{2}}{2}\cos\varphi_{\mathrm{s}} \left(\varphi^{\prime 2} + \theta^{\prime 2}\right)
\right\},
\\
E_{0} &= J_{\parallel}S^{2}a\int \dd z \left[\dfrac{1}{2}\left(\dfrac{\dd \varphi_{\mathrm{s}}}{\dd z}\right)^{2} -Q_{0}\dfrac{\dd \varphi_{\mathrm{s}}}{\dd z} - m^{2}\cos\varphi_{\mathrm{s}}\right].
\end{align}
By taking the Berry phase into account, the action may be constructed as
\begin{align}
A&= \int \dd t  L  = \int \dd t\left[ S\int \dfrac{\dd z}{a} (1 -  \cos \theta)\dfrac{\dd \varphi}{\dd t} - E \right]\\
&\approx \int \dd t\left[
S\int \dfrac{\dd z}{a} \theta^{\prime} \dfrac{\dd \varphi^{\prime}}{\dd t} -  E\right].
\end{align}
The Euler--Lagrange equation $(\dd/\dd t) \delta L/\delta \dot{q} = \delta L /\delta q$ with $\dot{q} = \dd q /\dd t$ for  $q = \varphi^{\prime}$, $\theta^{\prime}$, and the Fourier transform in time leads to
\begin{align}
-\ii \omega_{n}
\begin{pmatrix}
\varphi_{n}^{\prime} \\
\theta_{n}^{\prime}
\end{pmatrix}
=J_{\parallel} S a^{2}\begin{pmatrix}
0 & \hat{\mathcal{L}}_{\theta} \\
-\hat{\mathcal{L}}_{\varphi} & 0
\end{pmatrix}
\begin{pmatrix}
\varphi_{n}^{\prime}\\
\theta_{n}^{\prime}
\end{pmatrix},\label{eq-eigen-csg}\\
\hat{\mathcal{L}}_{\theta} = -\dfrac{\dd^{2}}{\dd z^{2}} - \left[\left(\dfrac{\dd \varphi_{\mathrm{s}}}{\dd z}\right)^{2} - 2Q_{0}\dfrac{\dd \varphi_{\mathrm{s}}}{\dd z}-m^{2}\cos\varphi_{\mathrm{s}} -\gamma^{2}\right],\\
\hat{\mathcal{L}}_{\varphi} = -\dfrac{\dd^{2}}{\dd z^{2}} + m^{2}\cos\varphi_{\mathrm{s}}.
\end{align}
Equation~\eqref{eq-eigen-csg} is equivalent to Eq.~\eqref{eq-bog} as seen by substituting $\theta_{\mathrm{s},l} = \pi/2$ into Eq.~\eqref{eq-bog}, assigning $\tilde{M}_{n,l}^{x} = \varphi_{n,l}^{\prime}$ and $\tilde{M}_{n,l}^{y} = -\theta_{n,l}^{\prime}$, and taking the continuous limit. We remark that $\varphi^{\prime}$ and $\theta^{\prime}$ are coupled. We use a notation $\epsilon_{n}$ for the $n$th dimensionless eigenvalue $\omega_{n}/(J_{\parallel} S Q_{0}^{2} a^{2}) = \omega_{n }/(D^{2}S/J_{\parallel})$.

For a realistic parameter set for Cr$_{1/3}$NbS$_{2}$, the instability mechanism for the $H_{0}$ line is the spin motion at the soliton center towards the helical axis. Since the instability of the narrow soliton is not forbidden as a mechanism of the $H_{0}$ line in principle, we consider its possibility for different parameters.
If the spins are restricted in the plane, the motion to the out-of-plane is forbidden. This corresponds to the infinitely large anisotropy. On the other hand, the narrow soliton is unstable only in the lattice model, and thus the continuum limit ($D \to 0$ or $a \to 0$) never leads to the instability of  the narrow soliton. From this brief consideration,  the motion towards the helical axis is the dominant instability mechanism for small $D$ or $K$, while the narrowing width is the dominant mechanism in the opposite case. Usually the information about the hard axis anisotropy $K$ is obtained by measuring the parallel critical field $(0, 0, H_{\cc}^{z})$ in the monoaxial chiral magnet. However, we remark that the $H_{0}$ field also depends on $K$. 

We show the $K$ dependence of $\vec{H}_{0}$ for $D/J_{\parallel} = 0.32$, $0.16$, and $0.08$ in Fig.~\ref{fig-csg-h0} for $H_{\mathrm{ex}}^{z} = 0$. The result for $D/J_{\parallel} = 0.32$ is shown only in the inset, which is for small $H_{0}^{x}/H_{D}$ and $K/H_{D}$. 
The labels (I) and (II) means $H_{0}$ line~(I) and $H_{0}$ line~(II) discussed in Sect.~\ref{sect-h0-line}, respectively.
 To take account of the origins (i) and (iii) together, we also calculate $\vec{H}_{0}$ in the scheme (I) with an infinitesimally small value $H_{\mathrm{ex}}^{z}/H_{D} = 10^{-10}$. They are shown for $D/J_{\parallel} = 0.16$ and 0.32 by black solid and open symbols, respectively,  with label ``(I$^{\prime})$''. The $H_{0}$ field~(I) at $H_{\mathrm{ex}}^{z} = 0$ represents the size narrow instability (iii) and is independent of $K$. We note that $H_{0}^{x}/H_{D}$ based on scheme (I) is about $49$ for $D/J_{\parallel} = 0.08$ (not shown).

As considered above, the instability mechanism changes from the origin (i) to the origin (iii) with increasing $K/H_{D}$. The changes occur at $K/H_{D} \sim 34$ for $D/J_{\parallel} = 0.16$ and at $K/H_{D} \sim 8$ for $D/J_{\parallel}$ = 0.32.  It should be mentioned that the anisotropy of Cr$_{1/3}$NbS$_{2}$ is $K/H_{D} = 5.68$ as indicated by the solid circle in Fig.~\ref{fig-csg-h0}. 
The black symbols stand for the continuous curve, which is lower than both the symbols for (I) and (II). The difference is due to whether the soliton center is on-site or between two nearest neighbor sites: When the $H_{0}$ field of the origin (i) is close to that of (iii), the soliton center is between the two sites, and the assumption in the scheme~(II) is not valid.  

It is also confirmed that the size narrowing instability is hard to occur for smaller $D/J_{\parallel}$. In this limit, the system is described by the continuum model, in which the instability of the narrow width of the soliton never occurs. 
We further study the $H_{0}$ field of the origin (i) on the basis of the chiral sine-Gordon model. The calculated values of $H_{0}^{x}/H_{D}$ are shown in Fig.~\ref{fig-csg-h0} by green inverted-triangles. We see that the deviation between the lattice model and the chiral sine-Gordon model is smaller for smaller $D$.

\begin{figure}[t!]
\begin{center}
\includegraphics[width = 0.9\hsize]{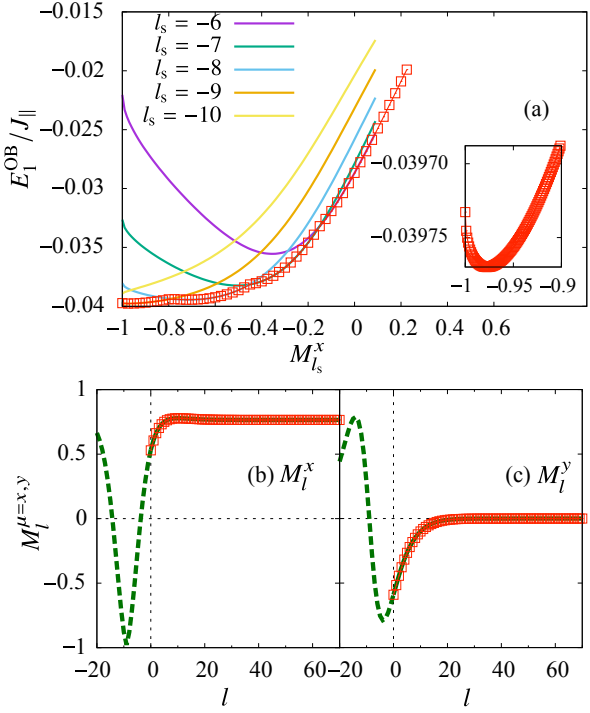}
\caption{Energy profiles.  
Solid lines are the energies as functions of $M_{l_{\mathrm{s}}}^{x}$ when the virtual soliton center is fixed at $l_{\mathrm{s}}$, 
and squares are the energy minimized with respect to $l_{\mathrm{s}}$ for each $M_{l_{\mathrm{s}}}^{x}$. 
The global minimum is at around $M_{l_{\mathrm{s}}}^{x} = -0.9775$ and $l_{\mathrm{s}} = -9$. The right panel shows the enlarged image around the global minimum. Spin profiles describing the surface twisted state for the field $\vec{H}_{\mathrm{ex}} = (1.0,0.0,4.5) H_{D}$, which is higher than the $H_{0}$ line. 
The red squares with line describe the result obtained through Method~II, while the line without symbol is the solution for the condition $(M_{l_{\mathrm{s}}}^{x},M_{l_{\mathrm{s}}}^{y}) \simeq (-0.9775, 0)$. The minimization condition gives that $l_{\mathrm{s}} = -9$.
}
\label{fig-twist}
\end{center}
\end{figure}

\section{Surface modulation above $H_{0}$ line}\label{sect-surface-twist-h0}
Here, we discuss the surface modulation for the field higher than the $H_{0}$ line, at which an isolated soliton is unstable. Since we showed that the surface modulation is interpreted as a virtual soliton outside the system, we explain how the surface modulation at such a high field is described by a soliton. In the calculation of the $H_{0}$ line based on the energy landscape, we imposed the fixed magnetic moment at the soliton center, $\vec{M}_{l_\mathrm{s}}$. Using this method, we construct the solution for the field above the $H_{0}$ line. For such a field, the energy landscape for $\vec{M}_{l_\mathrm{s}}$ in the bulk has a single  minimum structure, which denotes the uniform state. The energy landscapes for negative values of $l_{\mathrm{s}}$ at $\vec{H}_{\mathrm{ex}} = (1.0,0.0,4.5) H_{D}$, are shown in Fig.~\ref{fig-twist}(a) with  solid curves. Here $l_{\mathrm{s}}$ takes only integer values. In contrast to the case for positive $l_{\mathrm{s}}$, there exists the minimum structure for each negative $l_{\mathrm{s}}$. The curve with red squares show the energy minimized with respect to $l_{\mathrm{s}}$ for fixed $\vec{M}_{l_\mathrm{s}}$, and the inset is the enlarged one. We obtain $l_{\mathrm{s}}$ and $M_{l_{\mathrm{s}}}^{x}$ which minimize the energy, as $l_{\mathrm{s}} = -9$ and $M_{l_{\mathrm{s}} = - 9 }^{x} = -0.9775$. Then we demonstrate that the surface modulation is described by the soliton solution obtained in this way. In Figs.~\ref{fig-twist}(b) and \ref{fig-twist}(c), the spatial profiles of $M_{l}^{x}$ and $M_{l}^{y}$ are shown; the dashed lines indicate the virtual soliton with its center at $l_{\mathrm{s}} = -9$, while the open squares the surface modulation obtained using Method~II, and they are in good agreement.

\bibliographystyle{apsrev4-1}
\bibliography{library}
\end{document}